\documentclass{aa}  
\usepackage{natbib}
\bibpunct{(}{)}{;}{a}{}{,} 

\usepackage{color}
\usepackage{placeins}
\usepackage{hyperref}
\hypersetup{colorlinks, citecolor=blue, urlcolor=black}

\usepackage{graphicx}
\usepackage{txfonts}

\usepackage[squaren]{SIunits}
\usepackage{makecell}
\usepackage{natbib}
\usepackage{booktabs}
\usepackage{tikz}
\usepackage{rotating}
\usepackage{ulem}
\usepackage{multirow}
\usepackage[switch]{lineno}
\usepackage{titlesec}
\usepackage[percent]{overpic}
\usepackage{orcidlink}

\begin{document} 

   \title{Torsionally excited methanol in massive young stellar objects: The  role of the mid-IR radiation field.}

   \author{
   {C.~Sanna\orcidlink{0009-0006-5480-1219}}\inst{1,2}
   \and {S.~Leurini\orcidlink{0000-0003-1014-3390}}\inst{2}
   \and {A.~Giannetti\orcidlink{0000-0003-3869-6501}}\inst{3}
   \and {E.~Schisano\orcidlink{0000-0003-1560-3958}}\inst{4}
   \and {L.~Testi\orcidlink{0000-0003-1859-3070}}\inst{5,6}
   \and {S.~Molinari\orcidlink{0000-0002-9826-7525}}\inst{4}
   \and {J.~S.~Urquhart\orcidlink{0000-0002-1605-8050}}\inst{7}
   \and {T.~Pillai\orcidlink{0000-0003-2133-4862}}\inst{8}
   \and {F.~Massi\orcidlink{0000-0001-6407-8032}}\inst{6}
   \and {K.~Immer\orcidlink{0000-0003-4140-5138}}\inst{9}
   \and {E.~Molinari\orcidlink{0000-0002-1742-7735}}\inst{10}
   \and {S.~Casu\orcidlink{0000-0002-0154-3571}}\inst{2}
   }       

   \institute{
        Dipartimento di Fisica, Università degli Studi di Cagliari, S.P.Monserrato-Sestu km 0,700, I-09042 Monserrato (CA), Italy
        \and
        INAF - Osservatorio Astronomico di Cagliari, Via della Scienza 5, I-09047 Selargius (CA), Italy
        \and
        INAF - Istituto di Radioastronomia di Bologna, Via Gobetti 101, 40129 Bologna, Italy
        \and
        INAF - Istituto di Astrofisica e Planetologia Spaziali, Via Fosso del Cavaliere 100, I-00133, Rome, Italy
        \and
        Alma Mater Studiorum – Università di Bologna, Dipartimento di Fisica e Astronomia “Augusto Righi”, Via Gobetti 93/2, I-40129, Bologna, Italy
        \and
        INAF-Osservatorio Astrofisico di Arcetri, Largo E. Fermi 5, I-50125 Firenze, Italy 
        \and
        Centre for Astrophysics and Planetary Science, University of Kent, Canterbury, CT2 7NH, UK
        \and
        Haystack Observatory, Massachusetts Institute of Technology, 99 Millstone Road, Westford, MA 01886, USA
        \and 
        European Southern Observatory, Karl-Schwarzschild-Straße 2, 85748 Garching bei München, Germany
        \and
        Osservatorio Astronomico di Brera, Via Brera 28, I–20121, Milano, Italy
   }

   \date{\today}
 
\abstract
{Observing the earliest stages of high-mass star formation in the mid-infrared (mid-IR) is challenging because the extinction is high and the resolution limited. Millimeter vibrationally and torsionally excited transitions of several molecules have been proposed as indirect probes of local mid-IR radiation fields around young stellar objects (YSOs).}
{We investigate the excitation and spatial distribution of torsionally excited methanol ($\mathrm{CH_3OH}$) in 12 high-mass star-forming clumps in different evolutionary stages, assessing their potential as tracers of the mid-IR luminosity of embedded protostars.}
{We analyzed rotational transitions of methanol within the same torsionally excited states $\varv_\mathrm{t}$$\geq$1 in the Atacama Large Millimeter/submillimeter Array (ALMA) 890\,$\mathrm{\mu m}$ observations and compared them with Spitzer 24\,$\mathrm{\mu m}$ data by plotting line luminosities normalized by methanol abundance against mid-IR luminosities.}
{The torsionally excited methanol emission is generally compact and systematically coincident with ALMA dust continuum peaks and 24\,$\mathrm{\mu m}$ emission, even in faint sources. 
The only clump without an IR counterpart also shows no $\varv_\mathrm{t}$$\geq$1 emission, supporting radiative pumping as the dominant excitation mechanism.
We found a promising power-law trend between $\varv_\mathrm{t}$$\geq$1 methanol lines and 24\,$\mathrm{\mu m}$ luminosities, largely independent of the methanol abundance.}
{$\mathrm{CH_3OH}$ $\varv_\mathrm{t}$$\geq$1 transitions appear to be promising tracers of local mid-IR radiation fields of embedded protostars, providing core-scale information even in highly extincted regions. 
As a pilot study, our results suggest that extending our study to a larger statistical sample is crucial for testing and improving this method.}


\maketitle
\nolinenumbers

\section{Introduction} \label{sec:intro}

High-mass stars ($M \geq 8M_\mathrm{\odot}$) play a fundamental role in the evolution of galaxies. 
Through intense radiative and mechanical feedback, they profoundly affect their environment, regulating the energy balance and chemical composition of the interstellar medium and thereby affecting the entire star formation process \citep[e.g.,][]{McKee2007, KennicuttEvans2012, Krumholz2014}.
Despite substantial progress over the past decades \citep[e.g.,][]{ZinneckerYorke2007, Tan2014, Motte2018, Beuther2025}, several fundamental stages, especially the earliest ones, in the formation of high-mass stars remain unclear.
This is primarily due to a combination of observational challenges, including their short lifetimes, large distances, and their formation within dense, clustered, and highly obscured regions.

A major challenge in high mass star formation (HMSF) studies is the lack of robust observational diagnostics to classify young stellar objects (YSOs) according to their evolutionary stage. 
Although low-mass YSOs classification schemes are well established \citep[e.g.,][]{Lada1984, Andre1993, Evans2009, Robitaille2006}, their application to high-mass YSOs is complicated due to their deeply embedded and clustered nature \citep[e.g.,][]{ZinneckerYorke2007, Motte2018}.
\citet{Molinari2008} proposed to extend the use of $L_\mathrm{bol}/M_\mathrm{clump}$, the bolometric luminosity-to-envelope mass ratio (abbreviated as $L/M$), a commonly accepted evolutionary indicator in the low-mass regime \citep[see][]{Bontemps1996, Saraceno1996, Andre2000, Andre2008}, to the high-mass regime as well. 
This ratio is expected to increase significantly over time, reflecting the rising central luminosity and the progressive dispersal of the surrounding envelope. 
The $L/M$ diagram has frequently been used in large-scale surveys \citep[e.g.,][]{Elia2010, Ma2013, Molinari2016, Elia2017, Koenig2017, Giannetti2017b, Urquhart2014, Urquhart2018, Molinari2025}, supporting its validity as a proxy for identifying the evolutionary stages of massive protostars.  
\citet{Giannetti2017b} further validated this trend using spectroscopic observations on a sample of selected sources from the APEX\footnote{12\,$\mathrm{m}$ Atacama Pathfinder Experiment (APEX) telescope \citep{Gusten2006}.} Telescope Large Area Survey of the Galaxy \citep[ATLASGAL;][]{Schuller2009, Urquhart2018, Giannetti2014}.

Although $L/M$ is a widely accepted evolutionary indicator in the HMSF regime, it mainly reflects properties on large spatial scales.
The bolometric luminosity, derived from spectral energy distributions (SED) of clustered regions, includes all radiation emitted by the sources.
Even for deeply embedded early-stage objects that are often invisible in the mid-IR due to severe extinction, the reprocessed radiation emitted in the range from far-IR to (sub-)millimeter still makes $L/M$ an effective indicator.
However, $L_\mathrm{bol}$ is mainly derived from Herschel Space Observatory data \citep{Pilbratt2010}, which have a limited angular resolution. 
As a result, $L/M$ typically traces clump-scale structures \citep[$\sim$1--10\,$\mathrm{pc}$; e.g.,][]{Motte2018, Motte2022}, reflecting the global emission budget of entire protoclusters.
Currently, no IR facilities can directly trace the evolutionary stages of deeply embedded cores on the typical small scale \citep[$\sim$0.01--0.1\,$\mathrm{pc}$; e.g.,][]{Motte2018, Motte2022}.
In recent decades, new telescopes have significantly improved the angular resolution, but much remains to be understood.
For example, the James Webb Space Telescope \citep[JWST;][]{McElwain2023} provides a higher resolution in the near- and mid-IR, but it cannot fully overcome the challenges of extinction or sensitivity limitations.

We explore an alternative diagnostic: we use rotational transitions of methanol ($\mathrm{CH_3OH}$) within the same torsionally excited states ($\varv_\mathrm{t}$$\geq$1) as indirect tracers of the mid-IR radiation field around individual high-mass YSOs.
Our approach is based on the study of \citet{CarrollGoldsmith1981} on the radiative pumping of vibrational transitions of molecules via IR radiation.
They were the first to describe that while rotational levels of the vibrational ground state can be populated by collisions and radiation for the $\mathrm{CS}$ molecule, the levels of the vibrational excited states are mainly populated by IR radiation. 
Infrared pumping has also been proposed for other molecules, including $\mathrm{HC_3N}$ \citep{Goldsmith1982} and $\mathrm{CH_3CN}$ \citep{Goldsmith1983}. 
Since this mechanism indirectly links the intensity of these transitions to the IR radiation field of the background source, vibrationally excited transitions have been increasingly used as tracers of deeply obscured IR emission, such as from $\mathrm{HC_3N}$ in star-forming regions \citep[e.g.,][]{Wyrowski1999, Chen2025}, and from $\mathrm{HCN}$ in luminous galaxies \citep{Aalto2015}. 

\citet{Menten1986} were the first to suggest that torsionally excited levels of methanol could also be radiatively excited by IR radiation. 
The same mechanism drives Class $\mathrm{II}$ $\mathrm{CH_3OH}$ maser emission \citep{Sobolev1994}.
Non-local thermal equilibrium (Non-LTE) radiative transfer modeling from \citet{Leurini2007a} confirmed that torsionally excited levels are radiatively populated by far-IR photons from the surrounding dust, while collisional excitation plays a negligible role under typical protostellar conditions.
Empirical evidence further corroborated this theoretical framework.  
\citet{Giannetti2017b} found a close correlation between the temperature derived from $\mathrm{CH_3OH}$ $\varv_\mathrm{t}$$\geq$1 lines and the 22\,$\mathrm{\mu m}$ luminosity of massive clumps, indicating that these transitions are sensitive to the embedded IR radiation field.
However, this correlation reflects integrated clump-scale properties, and further investigation at the core scale is essential to understand the relation between the radiation field and $\mathrm{CH_3OH}$ $\varv_\mathrm{t}$$\geq$1 lines in individual YSOs.

We investigate $\mathrm{CH_3OH}$ $\varv_\mathrm{t}$$\geq$1 lines as tracers of mid-IR radiation in individual YSOs, focusing on the core scale. 
Methanol is particularly well suited for this purpose because its $\varv_\mathrm{t}$$\geq$1 transitions are numerous within a narrow spectral range. This enables the simultaneous observation of multiple lines that are each potentially sensitive to photons of different mid-IR wavelengths and at high angular resolution.
To assess the reliability of these transitions as proxies for the IR radiation, we compare their emission with the available 22--24\,$\mathrm{\mu m}$ IR fluxes from YSOs in protoclusters in different evolutionary stages. 
We also explore correlations between line intensities in various excitation regimes and IR fluxes to evaluate the diagnostic potential of these transitions for probing the intensity of the local radiation field.

This paper is organized as follows. Sect.\,\ref{sec:observations} describes the transitions and sample selection, observations, and data reduction. Sect.\,\ref{sec:results} presents the morphology of the Atacama Large Millimeter/submillimeter Array (ALMA) continuum emission in the observed protoclusters and the detection of torsionally excited methanol lines. 
In Sect.\,\ref{sec:analysis} we describe the analysis of the spatial distribution and the luminosity correlation between $\mathrm{CH_3OH}$ $\varv_\mathrm{t}$$\geq$1 transitions and mid-IR emission. 
The implications for the evolutionary classification and the potential of torsionally excited transitions as diagnostic tools are discussed in Sect.\,\ref{sec:discussion}. 
Finally, we summarize in Sect.\,\ref{sec:conclusions} our main results and discuss the prospects for future work.

\section{Observations}\label{sec:observations}

\subsection{Line selection} \label{sec:lines_sel}

We selected torsionally excited transitions of methanol, specifically, the $7_\mathrm{k}$ -- $6_\mathrm{k}$ band in the vibrational states $\varv_\mathrm{t}$=1,2 (see the list of transitions in the Appendix Table\,\ref{tab:freq_blocks}). 
The transition frequencies and spectroscopic parameters were obtained from the Splatalogue\footnote{The Splatalogue
Database for Astronomical Spectroscopy (\url{http://www.splatalogue.net/}) \citep{Remijan2007}, from which we selected entries from the Cologne Database for Molecular Spectroscopy \citep[CDMS,][]{Muller2001, Muller2005} and the Jet Propulsion Laboratory catalog \citep[JPL,][]{Pickett1998}.} database, which provides accurate line lists for ground and torsionally excited states.

The targeted transitions fall within the $\sim$336.6--338\,$\mathrm{GHz}$ spectral window and include a dense set of lines with upper-energy levels $E_\mathrm{u}/k_\mathrm{B} \simeq$ 350--1000\,$\mathrm{K}$. 
These transitions were chosen because the energy gaps between the torsionally excited levels and the ground state correspond to photons in the $\sim$14--40\,$\mathrm{\mu m}$ range, consistent with the typical mid-IR emission of embedded protostars.
As a result, these transitions serve as sensitive tracers of local mid-IR radiation near the source.

\subsection{Sample selection} \label{sec:sample}

We selected clumps from the ATLASGAL\footnote{ATLASGAL compact source catalogs are based on \citet{Contreras2013, Urquhart2014, Csengeri2014}.} TOP100 sample (hereafter TOP100) \citep{Giannetti2014}, which includes approximately 110 massive clumps covering the full evolutionary sequence of high-mass star formation. 
Based on the classification scheme originally introduced by \citet{Giannetti2014} and \citet{Csengeri2016}, \citet{Koenig2017} refined it into four evolutionary stages: 70\,$\mathrm{\mu m}$ weak, mid-IR weak, mid-IR bright, and $\mathrm{H\,II}$ regions, using 70\,$\mathrm{\mu m}$ data from Herschel
Infrared Galactic Plane Survey \citep[Hi-GAL;][]{Molinari2010}, mid-IR emission at 21--24\,$\mathrm{\mu m}$ from the Midcourse Space Experiment \citep[MSX;][]{Price2001}, the Wide-Field Infrared Survey Explorer \citep[WISE;][]{Wright2010}, and the Multiband Infrared Photometer for
Spitzer survey of the inner Galactic Plane \citep[MIPSGAL;][]{Carey2009} with flux thresholds and radio-continuum observations at 4--8\,$\mathrm{GHz}$ from the Coordinated Radio “N” Infrared Survey for High-mass star formation \citep[CORNISH;][]{Hoare2012, Purcell2013, Urquhart2013a} and the Red MSX Source \citep[RMS;][]{Urquhart2007, Urquhart2009} surveys, complemented by targeted methanol maser observations \citep{Walsh1999}.
The TOP100 sample is designed to include the brightest representatives of each evolutionary class in the ATLASGAL catalog \citep{Contreras2013, Urquhart2014, Csengeri2014}.
Consequently, these clumps are among the best-characterized high-mass regions in the Galaxy through the extensive follow-up observations \citep[e.g.,][]{Csengeri2016, Koenig2017, Giannetti2017b, Wienen2021, Billington2019}.

This scheme was further refined by \citet{Urquhart2018, Urquhart2022}, who classified sources as quiescent, protostellar, YSO, and $\mathrm{H\,II}$ regions.
A key improvement over previous methods was that the arbitrary 24\,$\mathrm{\mu m}$ flux threshold was replaced with the presence or absence of an 8\,$\mathrm{\mu m}$ point source to distinguish protostellar objects from YSOs.
This approach overcomes the limitations of fixed flux thresholds and provides a more comprehensive view in different wavelengths, including 3--8\,$\mathrm{\mu m}$ from the Galactic
Legacy Infrared Mid-Plane Survey Extraordinaire \citep[GLIMPSE;][]{Benjamin2003, Churchwell2009}, 24\,$\mathrm{\mu m}$ MIPSGAL, and 70\,$\mathrm{\mu m}$ Hi-GAL images.
We adopted the \citet{Urquhart2022} classification.

In Table\,\ref{tab:sources} we present the 12 sources we analyzed, selected from the TOP100 sample.
As a small subset, they inherit the well-characterized properties of the TOP100, making them an ideal starting point for a detailed study that spans all evolutionary stages.
The TOP100 clumps were selected according to the following criteria:

\begin{enumerate}
    \item Coverage of the full evolutionary sequence.
    
    \item Distances ($d_\mathrm{ref}$) in the range 3--5\,$\mathrm{kpc}$ \citep[from][]{Giannetti2014, Urquhart2022}, which, given a typical ALMA angular resolution of $\sim$0.5$\arcsec$, correspond to spatial resolutions of $\sim$1500--2500\,$\mathrm{au}$. This is comparable to the typical sizes of dense protostellar cores \citep[see][]{Motte2007, Russeil2010, Motte2018, Coletta2025}.
    
    \item Clump masses ($M_\mathrm{clump}$) in the range 800--2000\,$M_\mathrm{\odot}$ \citep[from][]{Koenig2017}.
    
    \item $\mathrm{H_2}$ column densities ($N_\mathrm{H_2}$) in the range (8--30)\,$\times$\,$10^{22}\,\mathrm{cm^{-2}}$ \citep[from][]{Koenig2017}, excluding too low column densities to optimize the detections and too high values for optical depth issues.
\end{enumerate}

These criteria ensured a representative subset that covered all evolutionary stages of star formation while maintaining homogeneity within each class. 
After filtering the TOP100 sample according to these criteria, we initially identified 16 sources with an overabundance of $\mathrm{H\,II}$ regions. 
To preserve class homogeneity and for observational considerations, the selection was limited to 12 sources, restricting the number of $\mathrm{H\,II}$ regions, so that no class dominated the sample. 
The final distribution of sources is two quiescent sources, three protostellar sources, three YSOs, and four $\mathrm{H\,II}$ sources. 
In addition, nine of the sources in our subset were previously detected in torsionally excited methanol transitions by \citet{Giannetti2017b}, who used the same lines as we did (see Sect.\,\ref{sec:lines_sel}). 

\begin{table*}
    \caption{Properties of the selected sources.} 
    \label{tab:sources}
    \centering
    \begin{tabular}{lcccccrr}
    \hline\hline
    Name & ATLASGAL   & Class &  \multicolumn{2}{c}{Position (ICRS)} & $d_\mathrm{ref}$ & $\varv_\mathrm{LSR}$  & $\varv_\mathrm{ref}$ \\
    & CSC name &  & R.A. (J2000) & Dec. (J2000) & ($\mathrm{kpc}$) & ($\mathrm{km\,s^{-1}}$)& ($\mathrm{km\,s^{-1}}$)\\
    \hline
    G014.49 & AGAL014.492$-$00.139 & quiescent      & 18:17:22 & $-$16:25:01 & 3.1 &    40.5  &   39.5\\
    G030.89 & AGAL030.893$+$00.139 & quiescent      & 18:47:13 & $-$01:45:07 & 4.9 &    105.3 &   107.4\\
    G014.19 & AGAL014.194$-$00.194 & protostellar   & 18:16:58 & $-$16:42:16 & 3.1 &    38.7  &   39.4\\
    G008.68 & AGAL008.684$-$00.367 & protostellar   & 18:06:23 & $-$21:37:10 & 4.4 &    35.6  &   41.5\\
    G023.21 & AGAL023.206$-$00.377 & protostellar   & 18:34:54 & $-$08:49:19 & 4.6 &    77.5  &   77.4\\
    G335.78 & AGAL335.789$+$00.174 & YSO            & 16:29:47 & $-$48:15:51 & 3.3 & $-$50.5  & $-$49.4\\
    G019.88 & AGAL019.882$-$00.534 & YSO            & 18:29:14 & $-$11:50:26 & 3.3 &    43.2  &   44.5\\
    G337.92 & AGAL337.916$-$00.477 & YSO            & 16:41:10 & $-$47:08:04 & 2.9 & $-$40.6  & $-$38.9\\
    G305.21 & AGAL305.209$+$00.206 & $\mathrm{H\,II}$            & 13:11:13 & $-$62:34:38 & 4.0 & $-$42.9  & $-$41.6\\
    G301.14 & AGAL301.136$-$00.226 & $\mathrm{H\,II}$            & 12:35:35 & $-$63:02:30 & 4.3 & $-$39.6  & $-$39.3\\
    G337.40 & AGAL337.406$-$00.402 & $\mathrm{H\,II}$            & 16:38:50 & $-$47:27:59 & 3.0 & $-$42.5  & $-$40.2\\
    G343.13 & AGAL343.128$-$00.062 & $\mathrm{H\,II}$            & 16:58:17 & $-$42:52:08 & 2.7 & $-$31.5  & $-$35.5\\
\hline
\end{tabular}
\tablefoot{
The columns list the name: short source name used in this work; the 
ATLASGAL CSC name: source names from \citet{Contreras2013}; 
the evolutionary class, source velocities ($\varv_\mathrm{LSR}$), and distances ($d_\mathrm{ref}$) from \citet{Urquhart2022}; 
$\varv_\mathrm{ref}$: reference velocity derived in this work from the $\mathrm{CH_3OH}$ $7_{-1}$ -- $6_{-1}$, $A^{-}$, $\varv_\mathrm{t}$=1 line and used to convert the observed spectra from the sky frequency into the rest-frame frequency.
}
\end{table*}

\subsection{Observations and data reduction}\label{sec:obs}

\subsubsection{ALMA observations}\label{sec:almaobs}

Our study is based on ALMA Band 7 observations ($\sim$$338\,\mathrm{GHz}$, $\sim$$0.89\,\mathrm{mm}$) from project ID 2017.1.00377.S (PI: S. Leurini), carried out during Cycle 5 (2017--2018 session). 
The data were obtained in spectral line mode as single-pointing observations using the 12\,$\mathrm{m}$ main array and the 7\,$\mathrm{m}$ Atacama Compact Array (ACA), which were combined to recover compact and extended emission.
The 12\,$\mathrm{m}$ array observations were performed in the C-3 configuration with baselines of $\sim$15--680\,$\mathrm{m}$, corresponding to a largest angular scale (LAS) of $\sim$$5\arcsec$ and a synthesized beam of $\sim$0.6$\arcsec$\,$\times$\,0.4$\arcsec$.
The ACA observations cover baselines of $\sim$9--49\,$\mathrm{m}$, recovering spatial scales up to $\sim$17--19$\arcsec$ with a synthesized beam of $\sim$4.7$\arcsec$\,$\times$\,2.7$\arcsec$.
The resulting rms noise levels are $\sim$0.1--0.7\,$\mathrm{mJy\,beam^{-1}}$ for the 12\,$\mathrm{m}$ data and $\sim$10--20\,$\mathrm{mJy\,beam^{-1}}$ for the ACA data. 
The spectral setup consisted of four $\sim$$1\,\mathrm{GHz}$ spectral windows covering 336.4--340.2\,$\mathrm{GHz}$ with a spectral resolution of $\sim$0.488\,$\mathrm{MHz}$ ($\sim$0.86\,km\,s$^{-1}$), providing full coverage of the $\mathrm{CH_3OH}$ $7_\mathrm{k}$ -- $6_\mathrm{k}$ transitions in the torsional states $\varv_\mathrm{t}=0,1,2$.

\subsubsection{ALMA data reduction and imaging}\label{sec:reduction}

The data calibration and image processing were performed using the Common Astronomy Software Applications package \citep[CASA\footnote{\url{http://casa.nrao.edu}};][]{McMullin2007, CASA2022}.
The initial calibration followed the standard ALMA pipeline procedures (CASA~5.1.1\footnote{The pipeline version required for Cycle 5 is casa-release-5.1.1; see \url{https://almascience.eso.org/processing}.}), with additional manual flagging applied to correct antenna-specific errors. 
Subsequent data reduction steps were performed with CASA~6.4.1\footnote{Documentation CASA tasks: \url{https://casadocs.readthedocs.io/en/v6.4.1}.}, which provides updated analysis tools.

First, we combined the 12\,$\mathrm{m}$ and 7\,$\mathrm{m}$ data to recover compact and extended emission. We combined the visibilities using CASA tasks (see ALMA tutorials\footnote{\url{https://casaguides.nrao.edu/index.php/ALMA_Tutorials}}) such as \texttt{mstransform} and \texttt{concat} to ensure that the datasets were properly aligned and merged.
A critical step in the reduction was the identification of line-free channels, which is essential for an accurate continuum subtraction, especially given the high line-density typically observed toward massive YSOs. 
To address this, we used the ALMA pipeline \texttt{find\_continuum} task \citep{Hunter2023}, which analyzes the mean spectrum of a dirty cube (produced with \texttt{tclean}) to identify channels contaminated by spectral lines. 
This procedure produces a \texttt{cont.dat} file that lists the frequency ranges corresponding to line-free regions, which were then used for continuum subtraction.
The continuum subtraction was then performed in the $u\varv$ domain using the \texttt{uvcontsub} task, assuming a zeroth-order (constant) baseline fit.
The data imaging was performed with the \texttt{tclean} task using Briggs weighting (\texttt{robust} = 0.5), producing continuum images in \texttt{mfs} mode and spectral cubes in cube mode. 
The final combined cubes have a typical synthesized beam of $\sim$0.6--1$\arcsec$, a cell size of $\sim$0.1--0.2$\arcsec$, a field of view of $\sim$29$\arcsec$, and a channel width of $\sim$0.448\,$\mathrm{MHz}$ ($\sim$0.435\,$\mathrm{km\,s^{-1}}$).

While the data reduction process is generally robust, the measured fluxes can be slightly affected by systematic uncertainties that primarily arise from residual baseline contributions after continuum subtraction. 
In complex line spectra, blended or broad lines can affect the baseline determination. 
To account for these factors, we adopted a conservative 10\% uncertainty on the integrated fluxes, in agreement with the typical absolute flux calibration uncertainty of 5--10\% for ALMA Band 7 \citep[see, e.g.,][]{Remijan2019, Francis2020, Braatz2020}. 
We also examined the residuals in line-free channels after baseline fitting, which indicated that the uncertainties introduced by continuum subtraction are consistent with the adopted 10\%.

\subsection{Auxiliary infrared data}\label{sec:ir_data}

We obtained the IR data from the NASA/IPAC Infrared Science Archive \citep[IRSA\footnote{\url{https://irsa.ipac.caltech.edu/frontpage/}};][]{Alexov2005, Teplitz2018}). 
Specifically, we retrieved mid-IR Spitzer images at 24\,$\mathrm{\mu m}$ from the MIPSGAL survey \citep[MIPS, $\sim$6$\arcsec$;][]{Carey2009}, at 22\,$\mathrm{\mu m}$ from the WISE survey \citep[W4, $\sim$12$\arcsec$;][]{Wright2010}, and at 8\,$\mathrm{\mu m}$ from the GLIMPSE survey \citep[IRAC, $\sim$2$\arcsec$;][]{Benjamin2003, Churchwell2009}.

\section{Observational results} \label{sec:results}

\begin{figure*}
    \centering 
    \includegraphics[width=0.40\textwidth]{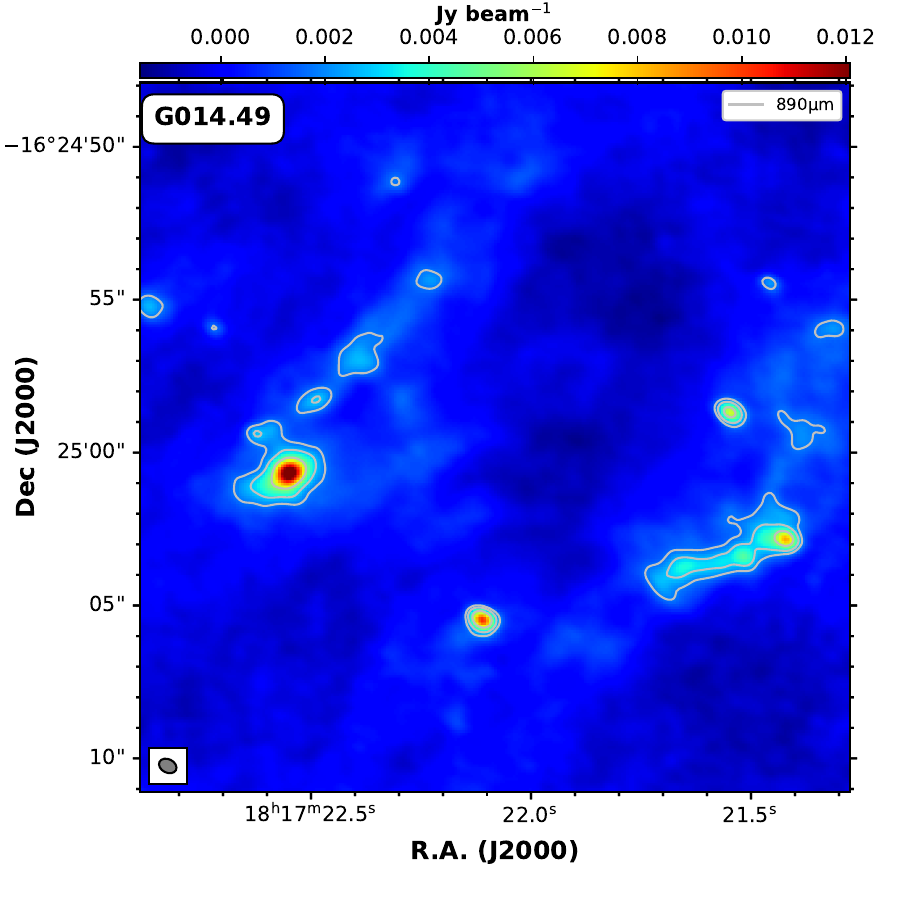}   
    \includegraphics[width=0.40\textwidth]{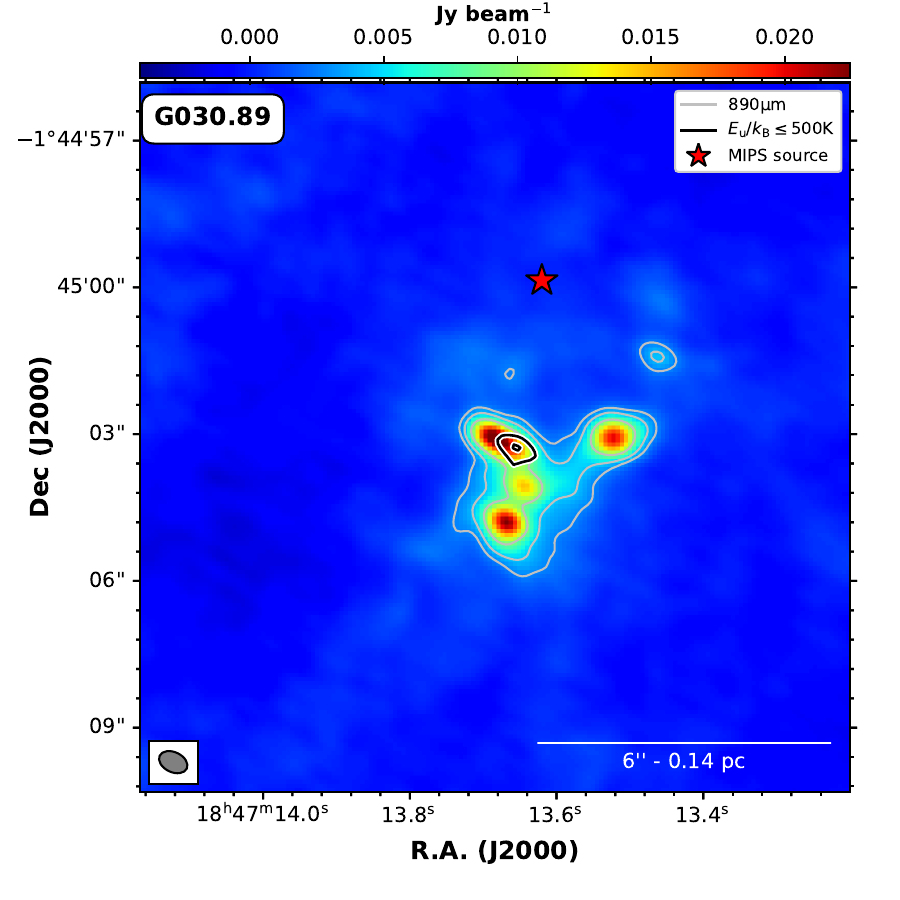}    
    \includegraphics[width=0.40\textwidth]{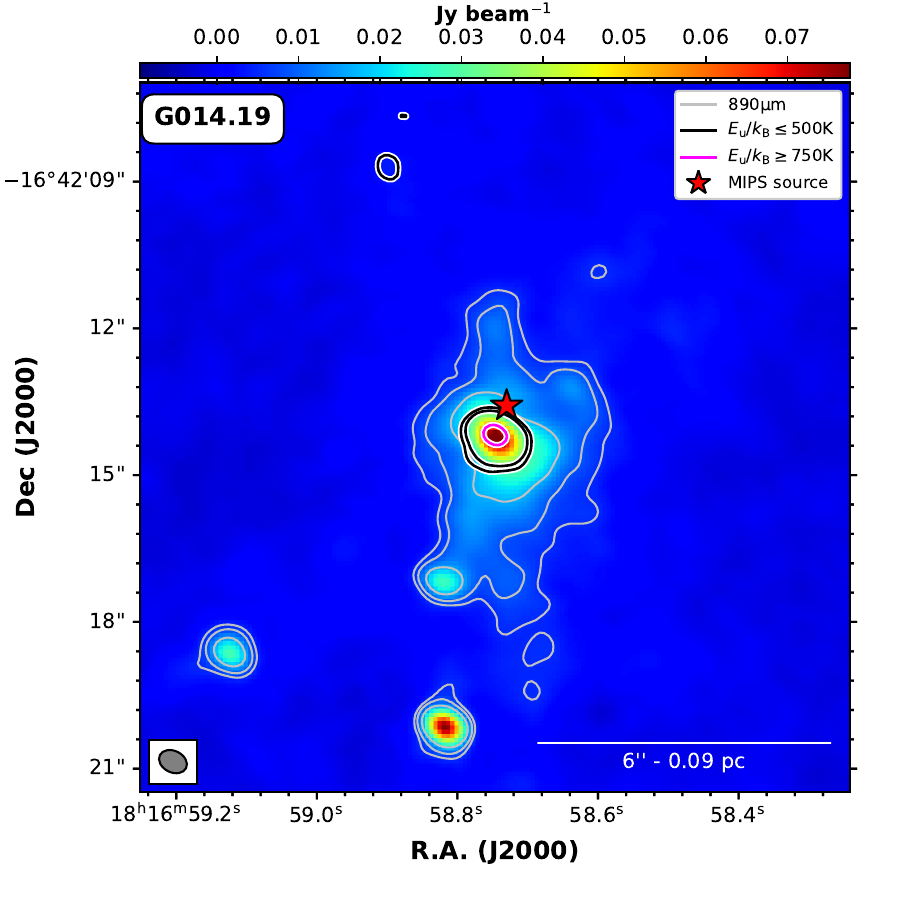}    
    \includegraphics[width=0.40\textwidth]{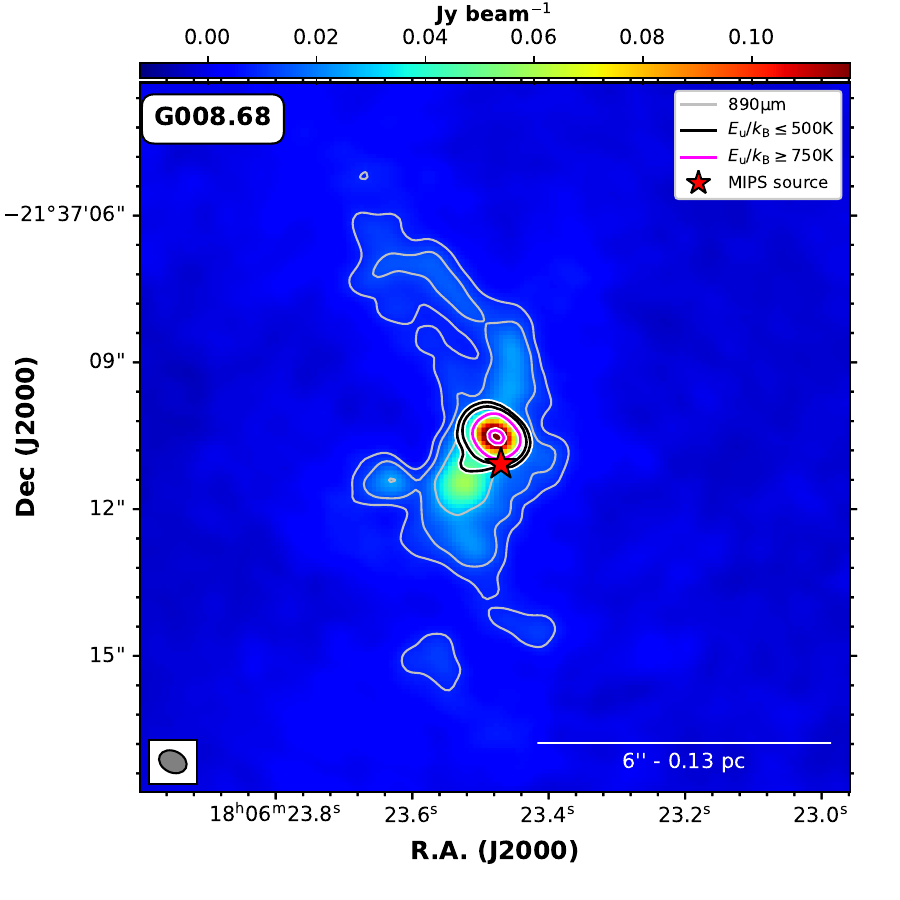}    
    \includegraphics[width=0.40\textwidth]{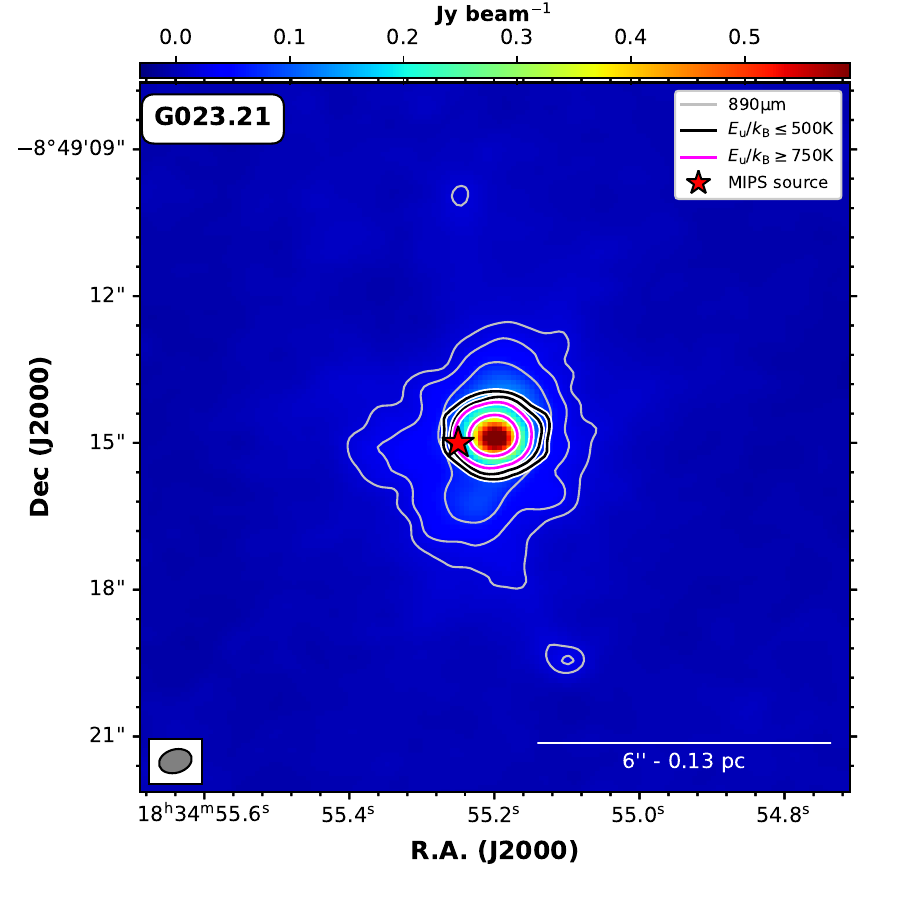}    
    \includegraphics[width=0.40\textwidth]{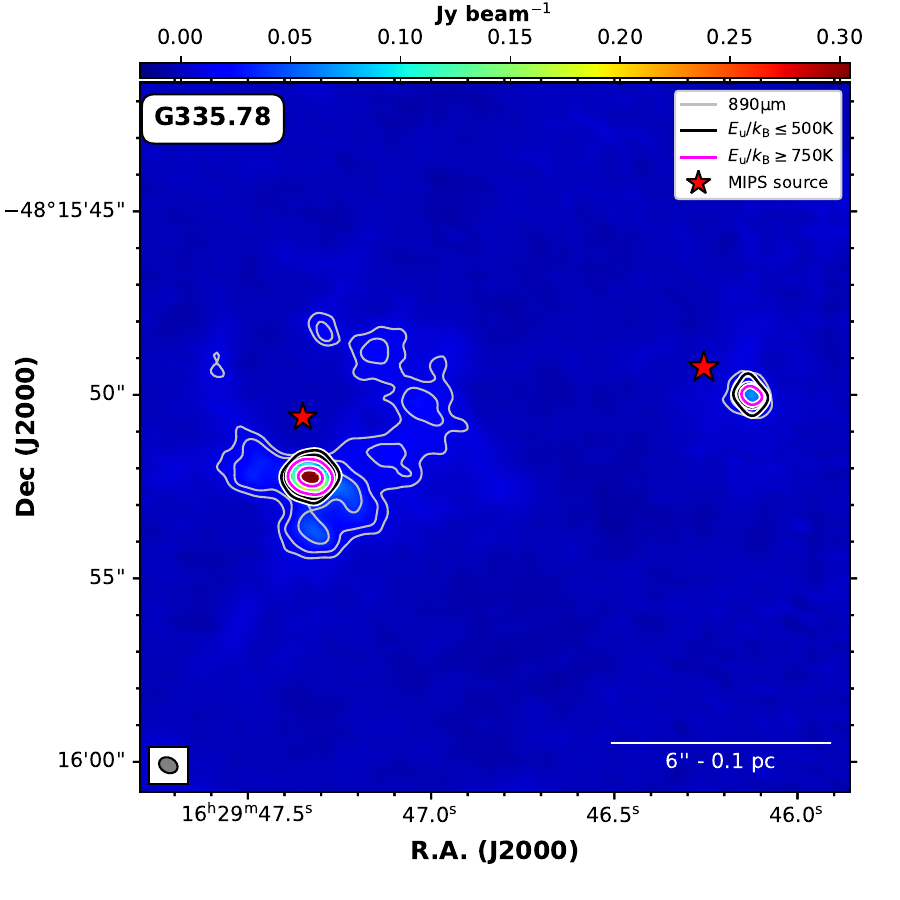}
    \caption{
    ALMA 890\,$\mathrm{\mu m}$ continuum images (background) of the younger clumps in the sample. 
    The silver contours show the continuum emission at [3, 5, 10]\,$\sigma$. 
    The black contours trace low-$\Delta E/k_\mathrm{B}$ emission at [3, 5]\,$\sigma$, and the magenta contours indicate high-$\Delta E/k_\mathrm{B}$ emission at [10, 50]\,$\sigma$. 
    The red star marks point sources from the MIPSGAL catalog \citep{Gutermuth2015}, with exceptions for G030.89, and secondary sources in G335.78, where centroids are determined in this work.
    For G014.49, the absence of line contours and IR emission indicates a non-detection. 
    The ALMA beam size ($\sim$0.6$\arcsec$) is shown in the bottom left corner, and the MIPSGAL ($\sim$6$\arcsec$) beam size is shown in the bottom right corner.
    }
    \label{fig:maps_b1}
    \end{figure*}
    
    \begin{figure*}
    \centering
    \includegraphics[width=0.40\textwidth]{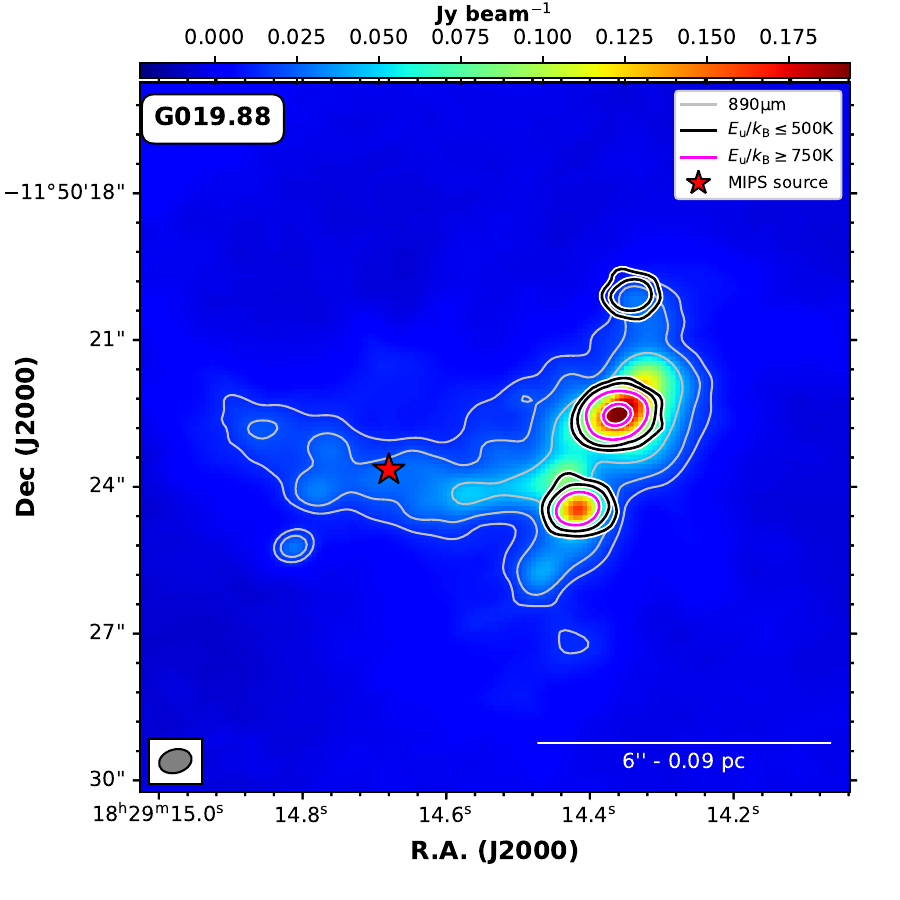}  
    \includegraphics[width=0.40\textwidth]{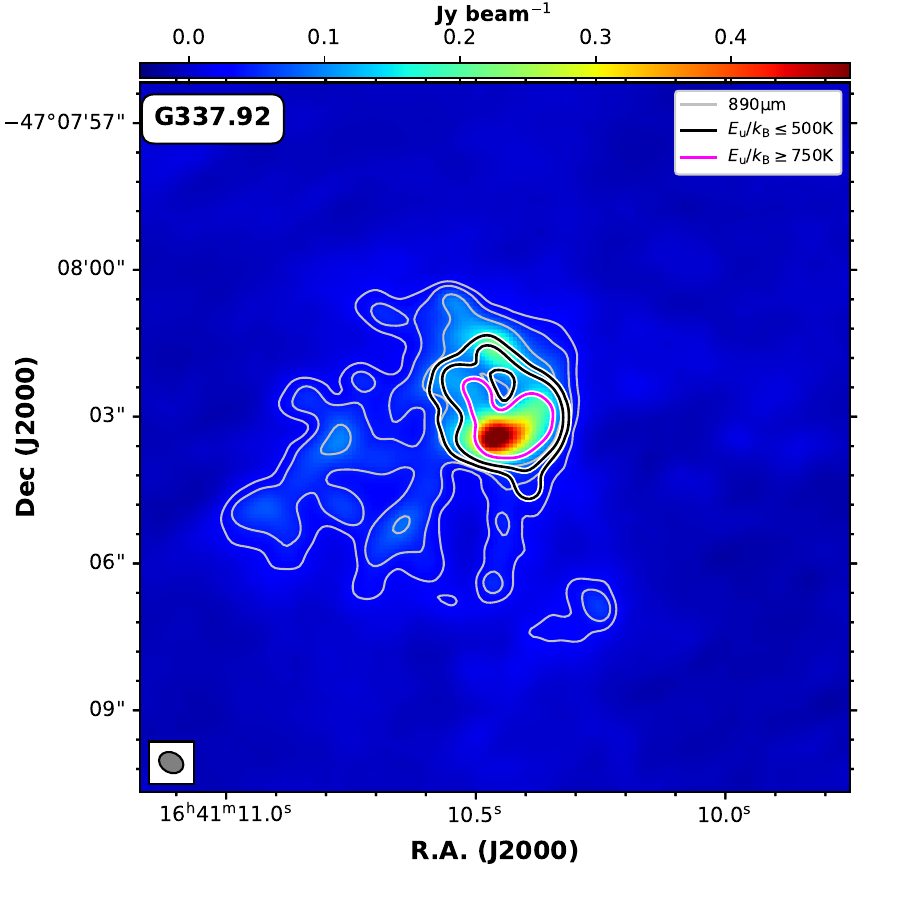} 
    \includegraphics[width=0.40\textwidth]{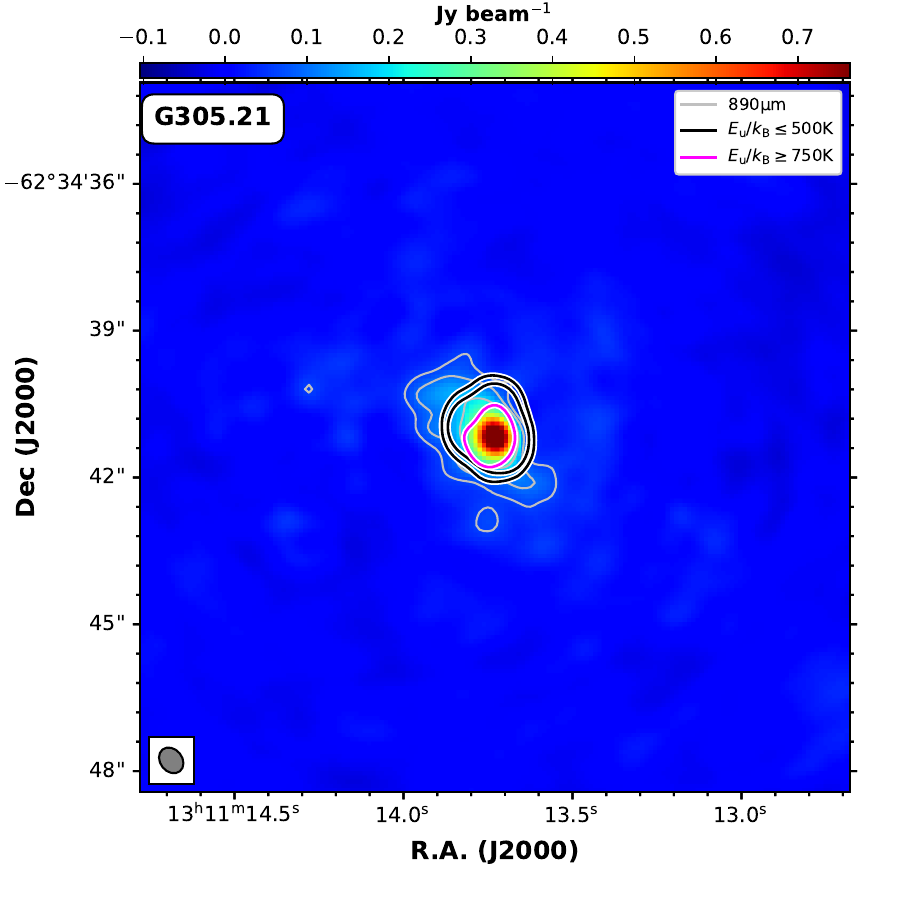}
    \includegraphics[width=0.40\textwidth]{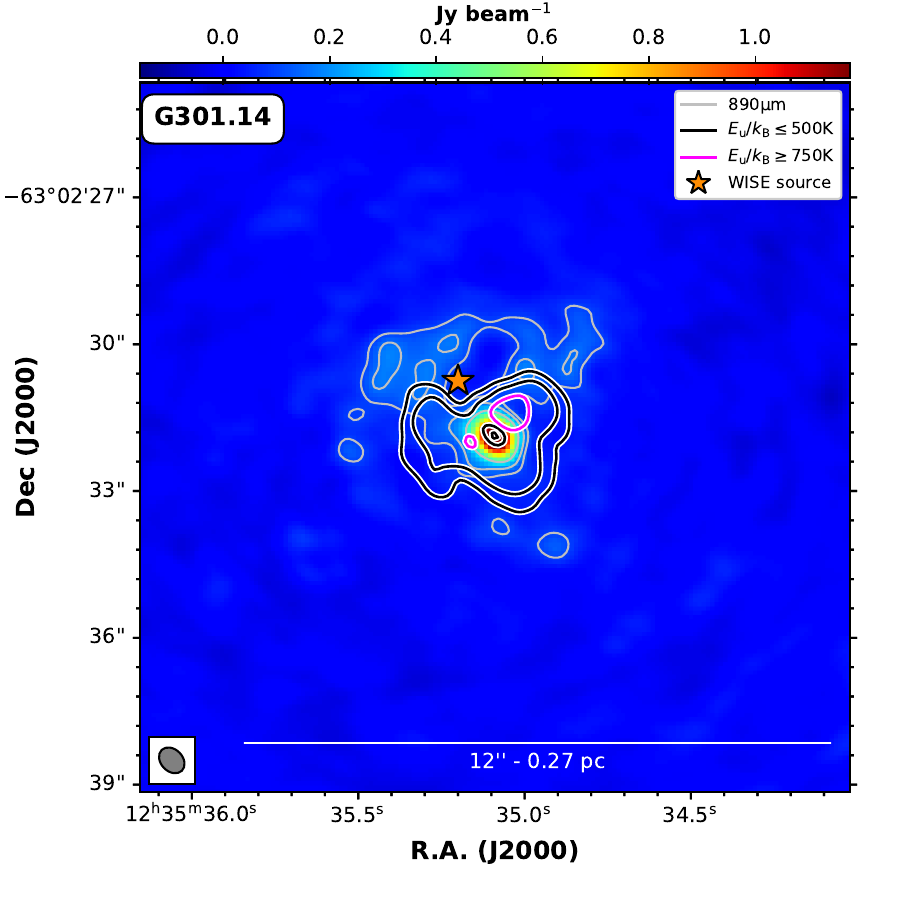}  
    \includegraphics[width=0.40\textwidth]{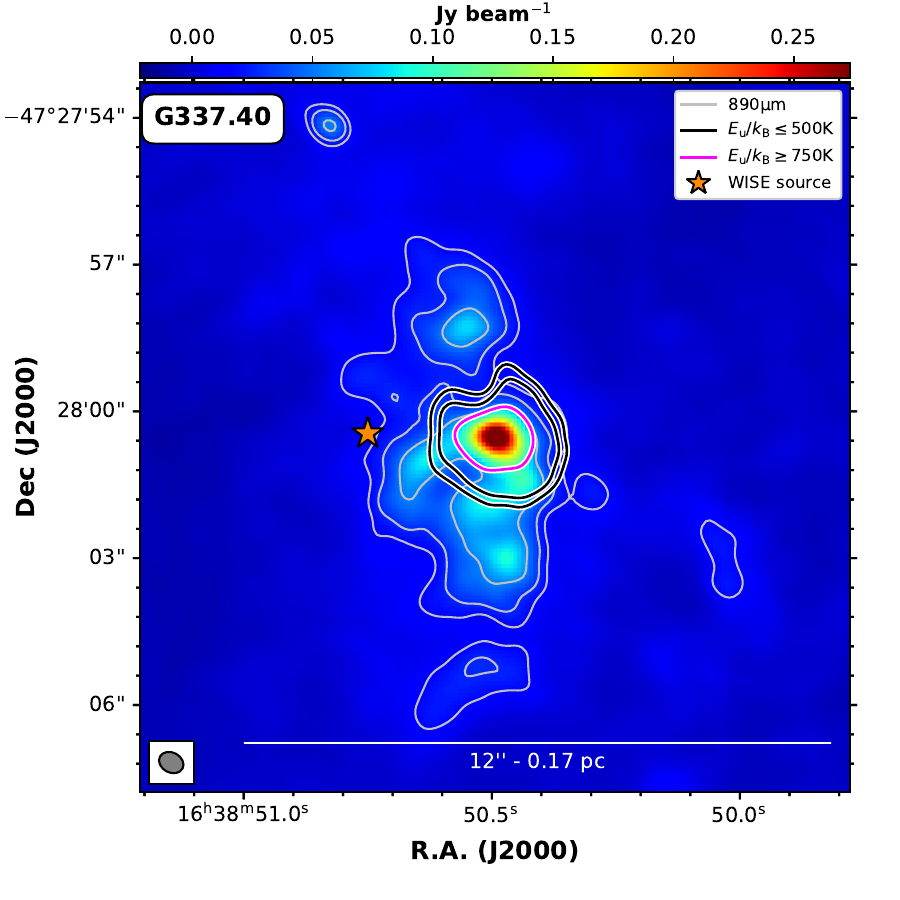}    
    \includegraphics[width=0.40\textwidth]{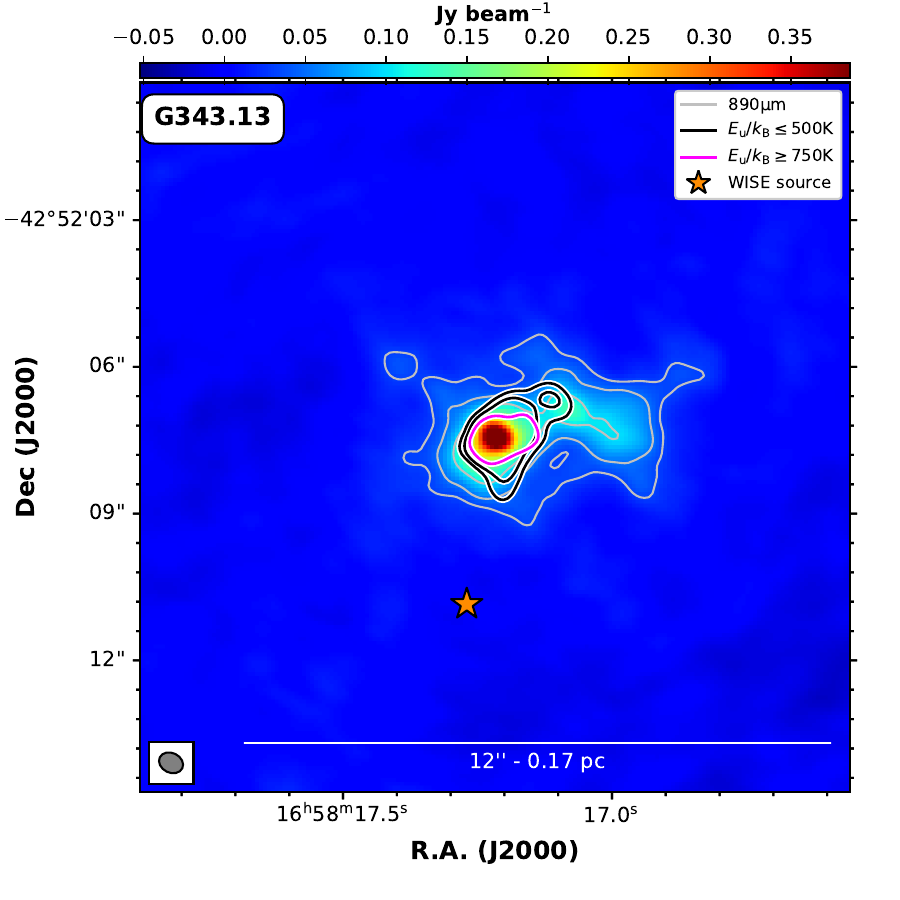}    
    \caption{
    ALMA 890\,$\mathrm{\mu m}$ continuum images (background) of the evolved clumps in the sample. 
    The silver contours show the continuum emission at [3, 5, 10]\,$\sigma$. 
    The black contours trace low-$\Delta E/k_\mathrm{B}$ emission at [3, 5]\,$\sigma$, and the magenta contours indicate high-$\Delta E/k_\mathrm{B}$ emission at [10, 50]\,$\sigma$. 
    All sources are saturated at 24\,$\mathrm{\mu m}$, except for G019.88, where the red star marks point sources from the MIPSGAL catalog \citep{Gutermuth2015}. 
    The orange star indicates the centroid of the WISE catalog \citep{Cutri2012} for G337.92, G301.14, G337.40, and G343.13. 
    Absence of stars (e.g., G305.21, G337.92) indicates complex diffuse emission not in the catalogs.
    The ALMA beam size ($\sim$0.6$\arcsec$) is shown in the bottom left corner, and the MIPSGAL ($\sim$6$\arcsec$) or WISE ($\sim$12$\arcsec$) beam size is shown in the bottom right corner.
    }
    \label{fig:maps_b2}
\end{figure*}

\subsection{Continuum emission} \label{sec:cores}

The ALMA 890\,$\mathrm{\mu m}$ continuum maps for the 12 clumps in our sample are shown in Figs.\,\ref{fig:maps_b1} (for the younger sources) and \,\ref{fig:maps_b2} (for the most evolved clumps). 
The sample exhibits a broad range of continuum morphologies, reflecting different degrees of fragmentation and structural complexity. 
The youngest clumps in our sample, classified as quiescent (e.g., G014.49, G030.89), tend to show significant fragmentation, with multiple cores often distributed along filamentary structures. 
The clumps classified as protostellar and YSO show a broader variety of core distributions.
Some (e.g., G023.21, G305.21) contain multiple compact cores embedded in clumpy or centrally concentrated structures, while others (e.g., G019.88, G337.92) display more elongated and irregular filamentary morphologies. 
In contrast, clumps classified as $\mathrm{H\,II}$ (e.g., G337.40, G301.14) generally present diffuse and more irregular continuum emission.

To characterize these structures, we identified dense cores using the 890\,$\mathrm{\mu m}$ thermal dust continuum emission. 
The core positions were determined using the algorithm \texttt{dendrogram} \citep{Rosolowsky2008}, implemented in the package \texttt{astrodendro}\footnote{\url{https://dendrograms.readthedocs.io/en/stable/}} \citep{Robitaille2019}. 
The algorithm was applied to maps without primary beam correction to preserve a uniform noise level across the entire field. 
We explored different combinations of dendrogram parameters and adopted \texttt{min\_value} = 5$\sigma$, \texttt{min\_delta} = 1$\sigma$, and \texttt{min\_npix} $\simeq$ 18~pixels (approximately half of the synthesized beam area).
The dendrogram algorithm segmented the emission into a hierarchical structure, with the smallest independent structure identified as leaves, which we considered to be dense cores.
A leaf is detected when it consists of at least a minimum number of connected pixels (\texttt{min\_npix}) above a fixed intensity threshold (\texttt{min\_value}).
Therefore, \texttt{min\_npix} is a segmentation parameter that defines the smallest area required to identify a leaf and does not represent the minimum size of a physical source.
For compact sources, the combination of a narrow emission profile and the adopted segmentation thresholds can result in a recovered area smaller than the synthesized beam area, especially for compact faint sources close to the detection limit or in crowded regions, as discussed by \citet{Rigby2024}.
Consequently, adopting a more conservative criterion, such as \texttt{min\_npix} equal to one synthesized beam area, can lead to the exclusion of significant compact detections.
Similar considerations motivated the adoption of comparable parameters in previous analogous studies \citep[e.g.,][]{Hatchfield2020, Rigby2024, Wallace2026}.

After the core identification, we measured the fluxes in the primary-beam-corrected maps to account for the antenna response. 
The fragmentation level in the 12 clumps is moderate, with approximately 3--11 cores per region, and all sources contain at least several detected cores.
A summary of the number of cores identified in each region is given in Table \ref{tab:n_cores}.

We focused on the analysis of the torsionally excited methanol lines and only considered the cores in which these lines are detected. A detailed study of the continuum properties and core characteristics will be presented in a forthcoming paper.

\begin{table*}
\caption{Number of continuum cores identified via the dendrogram analysis at 5$\sigma$ (see details in Sect.\,\ref{sec:cores}).}
\label{tab:n_cores}
\centering
\begin{tabular}{lclclclc}
\hline\hline
quiescent & N. cores & protostellar & N. cores & YSO   & N. cores & $\mathrm{H\,II}$   & N. cores\\
Clump     &          & Clump        &          & Clump &          & Clump & \\
\hline
G014.49 & 10 & G014.19 & 8  & G335.78 &  10 & G305.21   &  6\\
G030.89 &  7 & G008.68 & 7  & G019.88 &  9 & G343.13    &  3\\
        &    & G023.21 & 3  & G337.92 & 11 & G301.14    &  8\\
        &    &         &    &         &    & G337.40    &  7\\
\hline
\end{tabular}
\end{table*}

\subsection{Emission of \texorpdfstring{$\mathrm{CH_3OH}$ $\varv_\mathrm{t}$}{CH3OH vt}=1,2 rotational transitions} \label{sec:lines}

The spectra were extracted by integrating the ALMA data cubes over the areas of the dust continuum cores identified in Sect.\,\ref{sec:cores}.
Since the spectral axis of ALMA cubes is provided in sky frequency, an accurate reference velocity is required to convert the spectra into the rest frequency and to reliably identify molecular lines. 
The systemic velocities reported for ATLASGAL clumps, derived from single-dish observations \citep[e.g.,][]{Urquhart2022, Wienen2015}, trace the integrated emission of various extended components of the gas envelope and can therefore be inadequate to represent the kinematics at the compact core scale.
For this reason, we adopted the transition $\mathrm{CH_3OH}$ $7_{-1}$ -- $6_{-1}$, $A$ $\varv_\mathrm{t}$=1 at 337.969438\,$\mathrm{GHz}$ as a reference velocity ($\varv_\mathrm{ref}$).
This line was selected because it is relatively isolated compared to the other lines of the band, has a moderate upper-energy level for a $\varv_\mathrm{t}$=1 transition ($E_\mathrm{u}/k_\mathrm{B} \simeq 390\,\mathrm{K}$), and is consistently detected in all sources with significant torsionally excited methanol emission, making it a reliable local velocity reference.

The spectral range covered by our observations contains a high density of molecular transitions, as noted in previous studies using similar observational setups \citep[e.g.,][]{Sanchez2014, Beltran2014}. 
In particular, the interval between 336.6 and 339\,$\mathrm{GHz}$ contains not only $\mathrm{CH_3OH}$, but also several complex organic molecules (e.g., $\mathrm{C_2H_5OH}$, $\mathrm{CH_3OCH_3}$, and $\mathrm{CH_3COOH}$) and other important species such as $\mathrm{C^{17}O}$, $\mathrm{SO_2}$, $\mathrm{C^{34}S}$, and $\mathrm{HC_3N}$. 
This spectral complexity complicates the flux determination of individual torsionally excited features and the baseline determination and introduces residual uncertainties in the measured fluxes that must be considered (see Sect.\,\ref{sec:almaobs}).

\begin{figure*}
\centering

\includegraphics[width=\hsize]{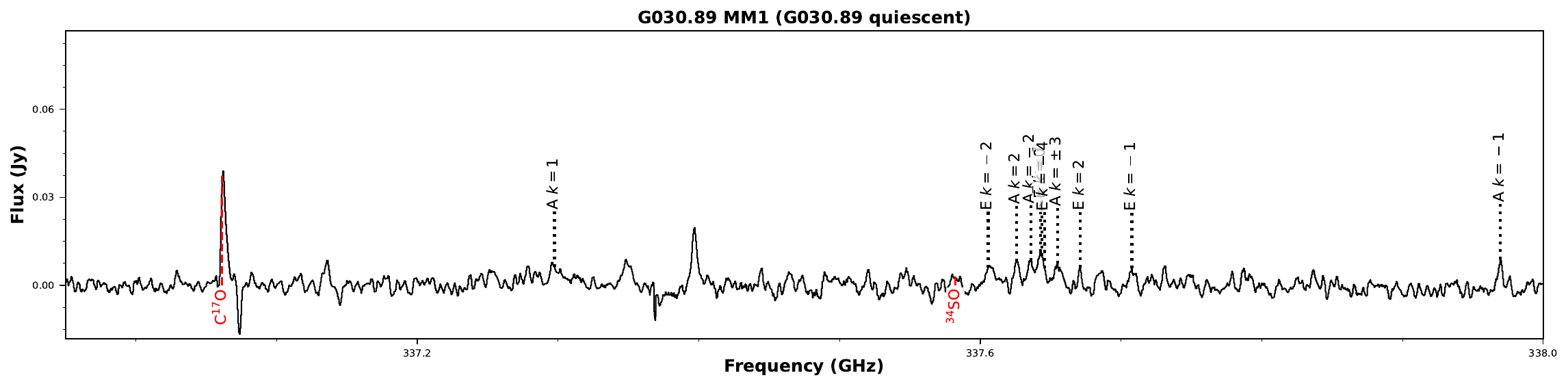}
\includegraphics[width=\hsize]{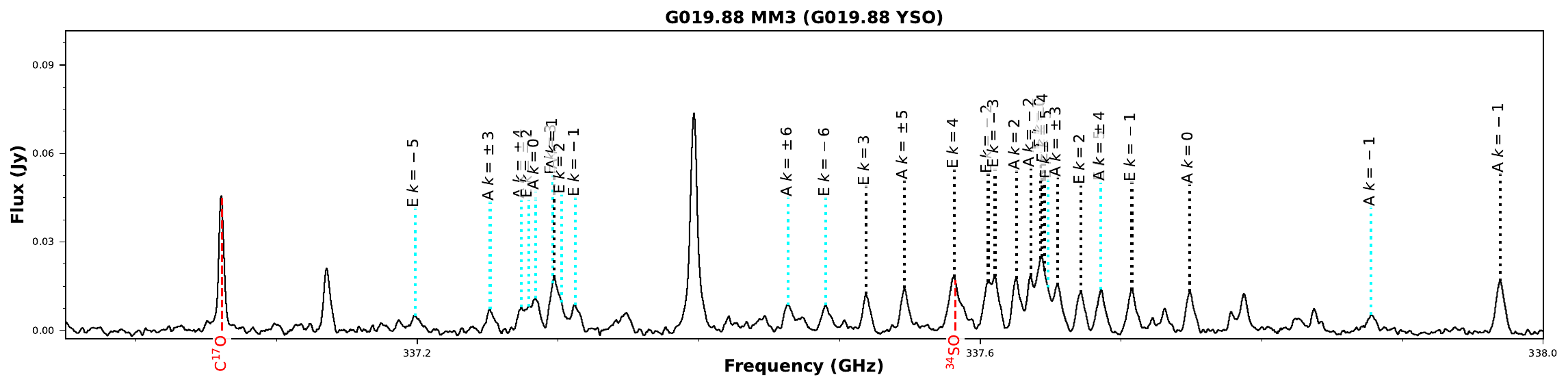}
\includegraphics[width=\hsize]{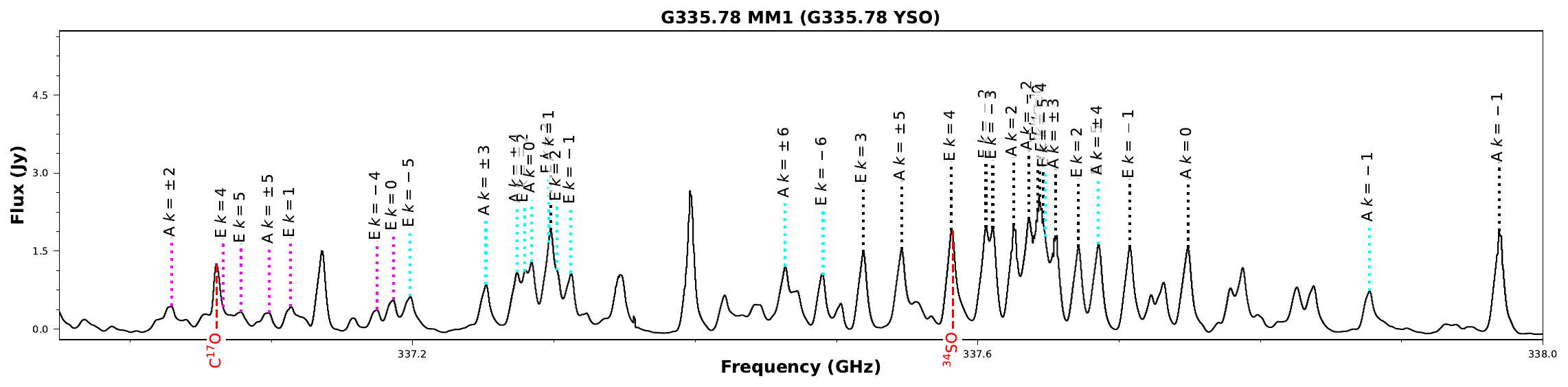}
\includegraphics[width=\hsize]{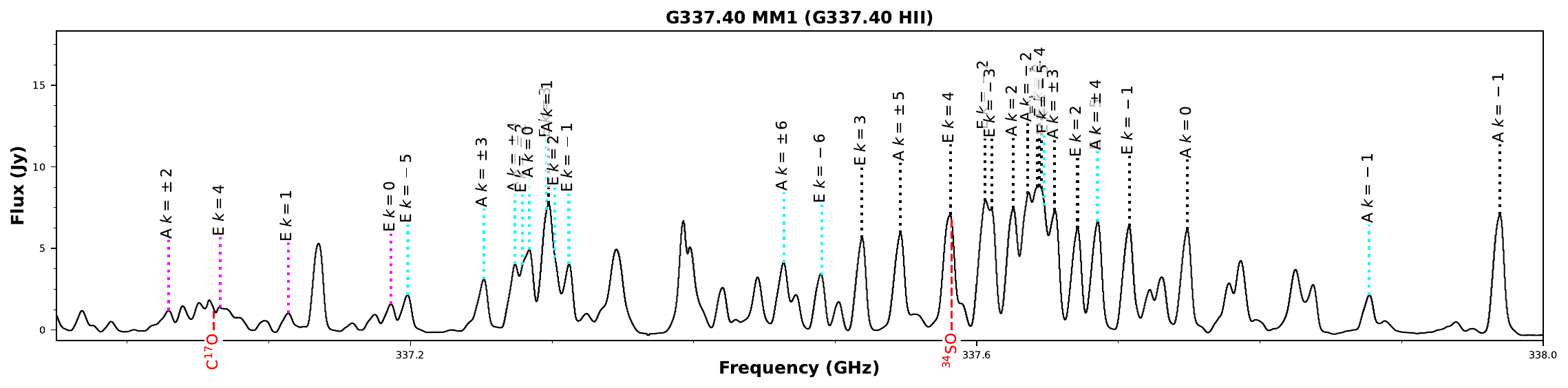}
\caption{
Four spectra covering 336.950 -- 338.000\,$\mathrm{GHz}$, showing the $\mathrm{CH_3OH}$ $7_\mathrm{k}$ -- $6_\mathrm{k}$ $\varv_\mathrm{t}$$\geq$1 transitions. 
All clumps are located at similar distances ($\sim$3--5\,$\mathrm{kpc}$), minimizing distance-related flux variations.
The panels illustrate the progressive detection of $\varv_\mathrm{t}$$\geq$1 transitions across the sample: 
G030.89~MM1 ($\sigma \simeq 0.002\,\mathrm{Jy}$) shows only low-$\Delta E/k_\mathrm{B}$ emission; 
G019.88~MM3 ($\sigma \simeq 0.001\,\mathrm{Jy}$) displays low- and mid-$\Delta E/k_\mathrm{B}$ lines but no significant high-$\Delta E/k_\mathrm{B}$ emission; 
G335.78~MM1 ($\sigma \simeq 0.1\,\mathrm{Jy}$) and G337.40~MM1 ($\sigma \simeq 0.3\,\mathrm{Jy}$) exhibit emission in all three $\Delta E/k_\mathrm{B}$ ranges, with the latter showing brighter lines. 
The transitions are labeled with the symmetry species (A or E) and the $k$ quantum number. 
The low-, mid-, and high-$\Delta E$ levels ($E_\mathrm{u}/k_\mathrm{B}\,\le\,500\,\mathrm{K}$, $500\,\mathrm{K}\,<\,E_\mathrm{u}/k_\mathrm{B}\,<\,750\,\mathrm{K}$, and $E_\mathrm{u}/k_\mathrm{B}\,\ge\,750\,\mathrm{K}$, respectively) are shown in black, cyan, and magenta.
}
\label{fig:Figlines}
\end{figure*}

Fig.\,\ref{fig:Figlines} shows representative spectra we analyzed, focusing on the $\mathrm{CH_3OH}$ $7_\mathrm{k}$ -- $6_\mathrm{k}$ $\varv_\mathrm{t}$=1,2 transitions, including approximately 40 lines in the narrow frequency range of $\sim$336.6--338\,$\mathrm{GHz}$. 
Most transitions cluster in two main spectral windows: one window around $\sim$337.6\,$\mathrm{GHz}$, and another window around $\sim$337.2\,$\mathrm{GHz}$, corresponding to $\varv_\mathrm{t}$=1 and 2, respectively.
The most prominent feature appears at $\sim$337.643\,$\mathrm{GHz}$, resulting from the blend of four $\varv_\mathrm{t}$=1 transitions ($\mathrm{k}$=0,1,$-$4,$-$5). 
This blended line is mainly dominated by transitions with $E_\mathrm{u}/k_\mathrm{B} \simeq 360\,\mathrm{K}$ and consistently appears as the brightest feature due to the efficient excitation of its relatively low-energy levels (which are more easily populated) and the convolution of multiple transitions, which enhances the overall signal-to-noise ratio, thereby facilitating detection.

The line intensities vary significantly in the clumps and individual cores, reflecting differences in the physical conditions of each source. 
Overall, the line morphology is dominated by compact emission, with brighter emission generally associated with the more evolved objects. 
The linewidths (full width at half maximum, FWHM) typically range from $\sim$4 to 10\,$\mathrm{km\,s^{-1}}$, and comparisons between transitions often reveal velocity gradients within or between individual cores.

Given the large number of torsionally excited lines, we first identified specific frequency intervals dominated by $\varv_\mathrm{t}$$\geq$1 transitions that were only marginally affected by contamination from other species.
Since the $\mathrm{CH_3OH}$ $\varv_\mathrm{t}$$\geq$1 levels are considered to be predominantly populated by IR radiative pumping \citep[e.g.,][]{Menten1986, Leurini2007a}, we grouped the lines according to their upper-energy levels ($E_\mathrm{u}/k_\mathrm{B}$) to explore possible trends as a function of excitation energy.
Using the relation $E = h\nu = hc/\lambda$, we associated each upper-energy level with a characteristic mid-IR wavelength.
Although this correspondence cannot be interpreted as a direct identification of the exciting pumping photons, it provides a useful empirical proxy to investigate the role of the mid-IR radiation field in the excitation.
The complete set of transitions spans a broad range of $E_\mathrm{u}/k_\mathrm{B}$, roughly corresponding to mid-IR wavelengths between $\sim14$ and 40\,$\mathrm{\mu m}$, sampling an essential portion of the mid-IR SED.
We divided the transitions into three empirical energy intervals ($\Delta E/k_\mathrm{B}$): low (356--500\,$\mathrm{K}$), mid (500--750\,$\mathrm{K}$), and high ($\geq$750\,$\mathrm{K}$), corresponding approximately to wavelength ranges of 29--40\,$\mathrm{\mu m}$, 19--27\,$\mathrm{\mu m}$, and 14--19\,$\mathrm{\mu m}$, respectively.
These intervals were only chosen to sample different portions of the excitation range and should not be interpreted as distinct physical pumping regimes.
These intervals were used to produce integrated moment-0 maps and compute the line luminosities presented in the following sections, excluding lines potentially contaminated by $\mathrm{C^{17}O}$\,(3 -- 2) and $\mathrm{^{34}SO}$\,($8_8$ -- $7_7$) (see Fig.\,\ref{fig:Figlines}). 

We applied a 3$\sigma$ threshold to each $\Delta E/k_\mathrm{B}$. For the following analysis, only continuum cores with positive low-$\Delta E/k_\mathrm{B}$ detection were considered.
The results are summarized in Table\,\ref{tab:cores_energy}.
The source G014.49 in the sample alone lacks detectable $\varv_\mathrm{t}$$\geq$1 emission for any of its cores, and all other clumps exhibit $\varv_\mathrm{t}$$\geq$1 lines in at least one core. 
Typically, only one core per clump shows detectable $\varv_\mathrm{t}$$\geq$1 emission, but in some cases, multiple cores are detected: two cores in G335.78, G337.92, G301.14, and G343.13, and three cores in G019.88.

Figure\,\ref{fig:Figlines} shows a portion of the observed spectrum for a representative subsample of cores showing the progressive detection of $\varv_\mathrm{t}$$\geq$1 transitions across different $\Delta E/k_\mathrm{B}$ selected to cover a range of evolutionary stages. 
The sequence begins with G030.89~MM1 (classified as quiescent), which only shows low-$\Delta E/k_\mathrm{B}$ emission. 
In contrast, G019.88~MM3 exhibits low- and mid-$\Delta E/k_\mathrm{B}$ emission, but lacks high-$\Delta E/k_\mathrm{B}$ lines. 
Farther along the sequence, G335.78~MM1 displays emission across all three $\Delta E/k_\mathrm{B}$.
G337.40 ($\mathrm{H\,II}$) provides an example of a spectrum from more evolved regions, where the emission lines are generally brighter, reflecting a more advanced excitation than in earlier sources.
This progression illustrates the increasing complexity of the torsionally excited emission across the evolutionary classification of the clumps.

\begin{table*}
    \centering
    \caption{
    Detection of $\mathrm{CH_3OH}$ $7_\mathrm{k}$ -- $6_\mathrm{k}$ $\varv_\mathrm{t}$$\geq$1 transitions in the observed cores. } \label{tab:cores_energy}
    \begin{tabular}{lcccccc}
    \hline\hline
    Source & Core   & \multicolumn{2}{c}{Position (ICRS)}    & low-$E_\mathrm{u}/k_\mathrm{B}$ & mid-$E_\mathrm{u}/k_\mathrm{B}$        & high-$E_\mathrm{u}/k_\mathrm{B}$ \\
    Name   &        & R.A. (J2000)      & Dec. (J2000)       & (356--500)\,$\mathrm{K}$  & (500--700)\,$\mathrm{K}$ & $\geq$700\,$\mathrm{K}$\\
           &        &($\mathrm{h:m:s}$) & ($\mathrm{d:m:s}$) & $\sim$(29--40)\,$\mathrm{\mu m}$ & $\sim$ (19--27)\,$\mathrm{\mu m}$ & $\sim$ (14--19)\,$\mathrm{\mu m}$\\
    \hline
    G030.89 & MM1 & 18:47:13.66 & $-$01:45:03.26 & Y & N & N \\
    G014.19 & MM1 & 18:16:58.75 & $-$16:42:14.21 & Y & Y & Y \\ 
    G008.68 & MM1 & 18:06:23.48 & $-$21:37:10.53 & Y & Y & Y \\ 
    G023.21 & MM1 & 18:34:55.20 & $-$08:49:14.86 & Y & Y & Y \\
    G335.78 & MM1 & 16:29:47.33 & $-$48:15:52.24 & Y & Y & Y \\
    G335.78 & MM2 & 16:29:46.13 & $-$48:15:50.03 & Y & Y & Y \\
    G019.88 & MM1 & 18:29:14.36 & $-$11:50:22.52 & Y & Y & Y \\
    G019.88 & MM2 & 18:29:14.42 & $-$11:50:24.44 & Y & Y & Y \\
    G019.88 & MM3 & 18:29:14.34 & $-$11:50:20.11 & Y & Y & N \\
    G337.92 & MM1 & 16:41:10.39 & $-$47:08:03.13 & Y & Y & Y \\
    G337.92 & MM2 & 16:41:10.47 & $-$47:08:01.72 & Y & Y & Y \\
    G305.21 & MM1 & 13:11:13.72 & $-$62:34:41.27 & Y & Y & Y \\
    G301.14 & MM1 & 12:35:35.12 & $-$63:02:31.89 & Y & Y & Y \\
    G301.14 & MM2 & 12:35:34.99 & $-$63:02:30.39 & Y & Y & Y \\
    G337.40 & MM1 & 16:38:50.47 & $-$47:28:00.46 & Y & Y & Y \\
    G343.13 & MM1 & 16:58:17.22 & $-$42:52:07.51 & Y & Y & Y \\
    G343.13 & MM2 & 16:58:17.12 & $-$42:52:06.67 & Y & Y & N \\
    \hline
    \end{tabular}
    \tablefoot{The table lists the detection of
    $\mathrm{CH_3OH}$ $7_\mathrm{k}$ -- $6_\mathrm{k}$ $\varv_\mathrm{t}$$\geq$1
    transitions in each core, grouped by upper-energy level ($\Delta E/k_\mathrm{B}$). 
    The columns for low-, mid-, and high-$E_\mathrm{u}/k_\mathrm{B}$ levels ($E_\mathrm{u}/k_\mathrm{B}\,\leq\,500\,\mathrm{K}$, $500\,\mathrm{K}\,<\,E_\mathrm{u}/k_\mathrm{B}\,<\,750\,\mathrm{K}$, and $E_\mathrm{u}/k_\mathrm{B}\,\geq\,750\,\mathrm{K}$, respectively) indicate whether transitions were detected (Y) or not (N) at $\geq 3\sigma$. 
    Approximate wavelength intervals for each energy range are also provided: low 29--40\,$\mathrm{\mu m}$, mid 19--27\,$\mathrm{\mu m}$, and high 14--19\,$\mathrm{\mu m}$.}
\end{table*}

\section{Analysis} \label{sec:analysis}

\subsection{Distribution of ALMA emission and mid-IR counterparts} \label{sec:distrib}

Figs.\,\ref{fig:maps_b1} -- \ref{fig:maps_b2} show the ALMA 890\,$\mathrm{\mu m}$ dust continuum emission overlaid with the moment-0 contours of the torsionally excited $\mathrm{CH_3OH}$ lines for all 12 clumps. 
The $\varv_\mathrm{t}$$\geq$1 emission is generally compact (about one to two synthesized beams) and becomes more extended in more evolved regions (e.g., YSO and $\mathrm{H\,II}$).
In most cases, the $\varv_\mathrm{t}$$\geq$1 emission coincides with a single bright 890\,$\mathrm{\mu m}$ continuum core, although some clumps show multiple cores (e.g., G019.88 and G335.78), as described in Section\,\ref{sec:lines}.

Since the energy gaps between the torsionally excited levels and the ground state correspond to photons in the $\sim$14--40\,$\mathrm{\mu m}$ range, we compared the ALMA line emission with the closest available mid-IR 24\,$\mathrm{\mu m}$ images from MIPSGAL.
Despite the significant difference in angular resolution between ALMA ($\simeq$0.6$\arcsec$) and Spitzer ($\simeq$6$\arcsec$), a consistent association is observed between the $\varv_\mathrm{t}$$\geq$1 emission (core scale) and the mid-IR emission (clump scale), as shown in Figure\,\ref{fig:maps_ir}.

The only source in the sample without detectable $\varv_\mathrm{t}$$\geq$1 emission is G014.49, which also lacks a 24\,$\mathrm{\mu m}$ counterpart.
All other clumps exhibit $\varv_\mathrm{t}$$\geq$1 emission in all three $\Delta E/k_\mathrm{B}$, except for the cores G030.89~MM1 and G019.88~MM3, where high-$E_\mathrm{u}/k_\mathrm{B}$ transitions are not detected.

The spatial association between the line emission and the 24\,$\mathrm{\mu m}$ images is clear for six sources: G030.89, G014.19, G008.68, G023.21, G335.78, and G019.88. 
The further analysis showed a coincidence with mid-IR compact sources in the MIPSGAL catalog \citep{Gutermuth2015}, for which positions and fluxes are provided.
We identified two sources with faint 24\,$\mathrm{\mu m}$ emission coincident with compact $\varv_\mathrm{t}$$\geq$1 emission from two ALMA cores (G030.89~MM1 and G335.78~MM2), for which we estimated the fluxes through aperture photometry.
The comparison also revealed multiple compact $\varv_\mathrm{t}$$\geq$1 cores, such as in G019.88, which coincide with a single extended emission feature at 24\,$\mathrm{\mu m}$.

In contrast, five clumps (G305.21, G337.92, G301.14, G337.40, and G343.13) are saturated in MIPS images, and the direct comparison between ALMA cores and the 24\,$\mathrm{\mu m}$ emission required additional analysis (see Appendix\,\ref{apx:IR_fluxes}). These clumps represent the most evolved phases in our sample (one YSO and four $\mathrm{H\,II}$).
To constrain their  mid-IR fluxes, we supplemented this analysis with WISE and GLIMPSE images to explore IR emission at complementary wavelengths. 
More details are provided in Appendix\,\ref{apx:IR_fluxes}.

\subsection{Correlation between mid-IR and line luminosities} \label{sec:ll_diag}

We investigated the line luminosities of $\varv_\mathrm{t}$$\geq$1 methanol lines in three $\Delta E/k_\mathrm{B}$ regimes in embedded YSOs and examined their relation to the 24\,$\mathrm{\mu m}$ mid-IR emission. 
The methods we used to calculate luminosities are described in Appendix\,\ref{sec:lum_def}.

To account for differences in initial chemical conditions of each source, we normalized the line luminosities by the methanol abundance, $X_\mathrm{CH_3OH}$, adopting values reported by \citet{Giannetti2017b} that range from $0.4\,\times\,10^{-8}$ to $10.6\,\times\,10^{-8}$, indicating that methanol is abundant even in cold environments.
For G305.21, where no specific abundance estimate is available, we adopted a median value of $2.3\,\times\,10^{-8}$, representative of intermediate-stage (IRb or YSO) clumps from the same study. 
Although it is not source-specific, this median value provides a reasonable first-order approximation and ensures a homogeneous scaling across the sample.

Due to their different angular resolutions, ALMA and mid-IR observations probe distinct spatial scales, from cores to clumps. 
In some regions, a single ALMA core corresponds to one MIPSGAL source (e.g., G014.19 or G008.68), while in others, multiple ALMA cores are associated with a single extended mid-IR source (e.g., G019.88). 
For saturated MIPS images, we derived conservative lower limits to the mid-IR luminosities (except for G343.13), with complementary WISE measurements providing upper limits (except for G337.92 and G343.13).
To ensure a consistent comparison, we summarize the line luminosities of all ALMA cores associated with each mid-IR source (see Table\,\ref{tab:IR_flx}).

To explore the relation between $\varv_\mathrm{t}$$\geq$1 and 24\,$\mathrm{\mu m}$ luminosities, we performed a power-law fit in log-log space,

\begin{equation}
\log_{10} \left( \frac{L (\Delta E/k_\mathrm{B})}{X_\mathrm{CH_3OH}} \right) = \alpha + \beta \log_{10} (L_\mathrm{MIR}),
\end{equation}

\noindent where $\alpha$ is the intercept, and $\beta$ is the slope, $L (\Delta E/k_\mathrm{B})$/$X_\mathrm{CH_3OH}$ represents the normalized $\varv_\mathrm{t}$$\geq$1 methanol line luminosity in the three $\Delta E/k_\mathrm{B}$ regimes, and $L_\mathrm{MIR}$ is the 24\,$\mathrm{\mu m}$ luminosity (Appendix\,\ref{sec:lum_def}).

We performed a Bayesian linear regression using the package PyMC\footnote{\url{https://www.pymc.io/}} \citep{Salvatier2016, Patil2010}, which includes propagated measurement uncertainties and an intrinsic scatter term.
Normal priors on the intercept and slope were centered on the initial linear fit estimates.
Censored data (lower and upper limits) were incorporated into the model using conservative lower limits from MIPS and upper limits from WISE. We included them to account for sources with saturated mid-IR fluxes.
We also performed a model comparison to assess whether a constant (slope = 0) or a linear (nonzero slope) model described the data better. 
The resulting odds ratio of $\sim$10 for low-energy transitions indicates moderate evidence in favor of the linear relation.

\begin{figure}
\centering
\includegraphics[width=\linewidth]{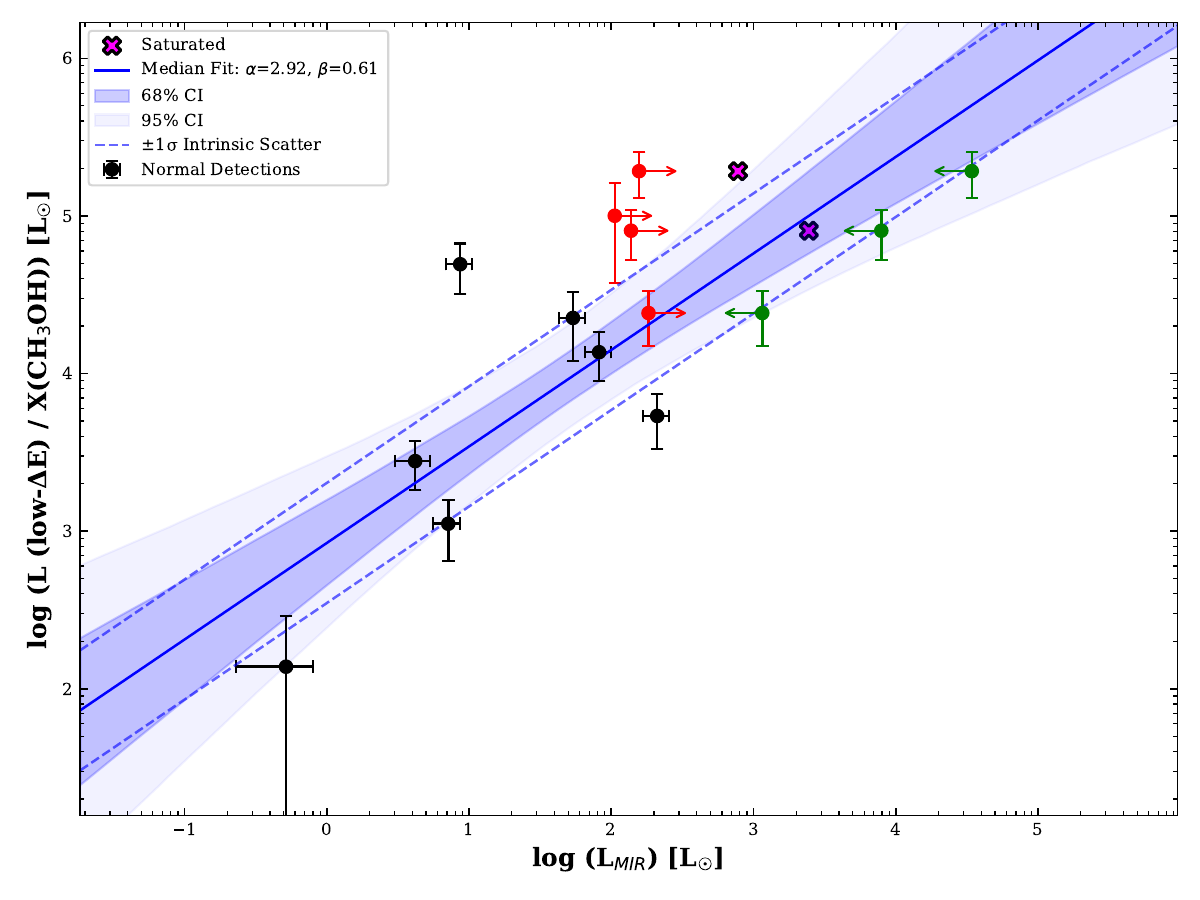}
\vspace{-2mm}
\includegraphics[width=\linewidth]{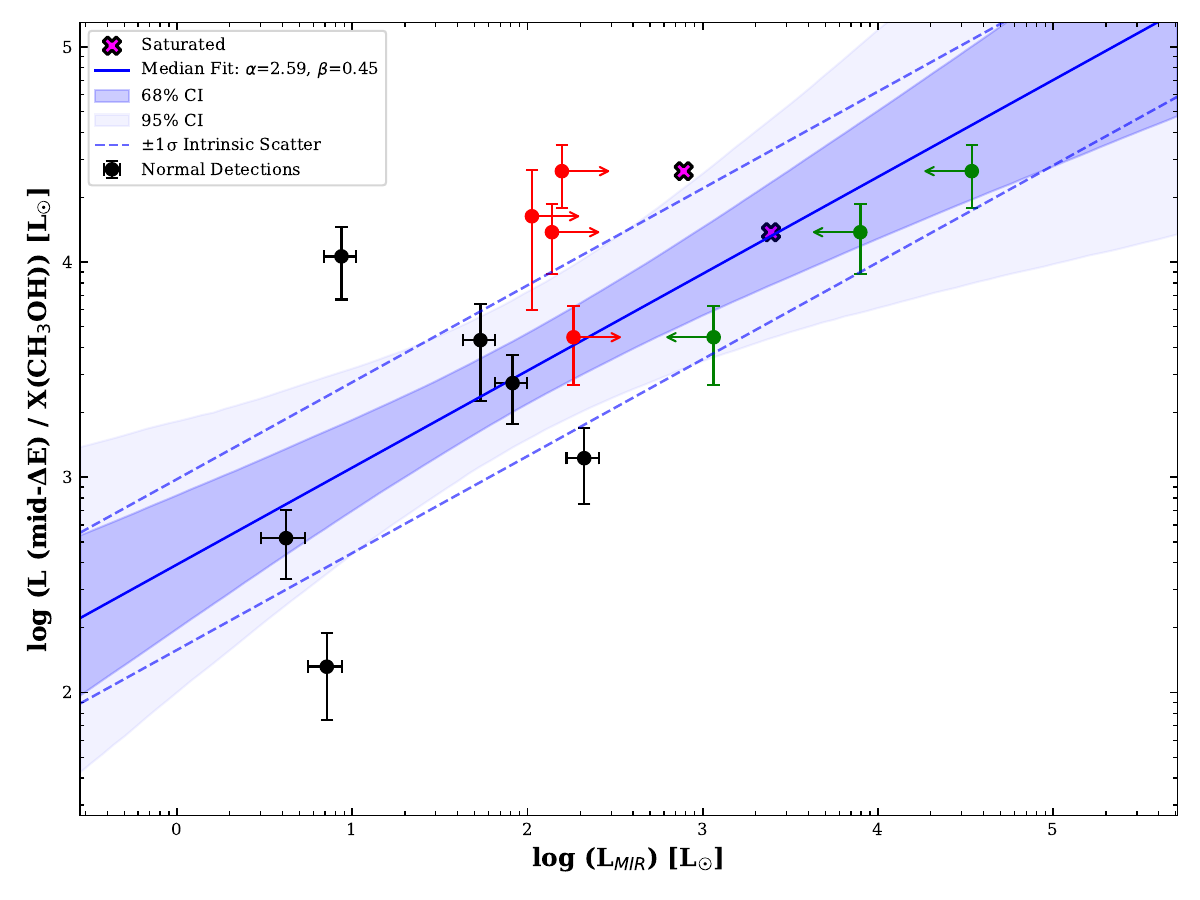}
\vspace{-2mm}
\includegraphics[width=\linewidth]{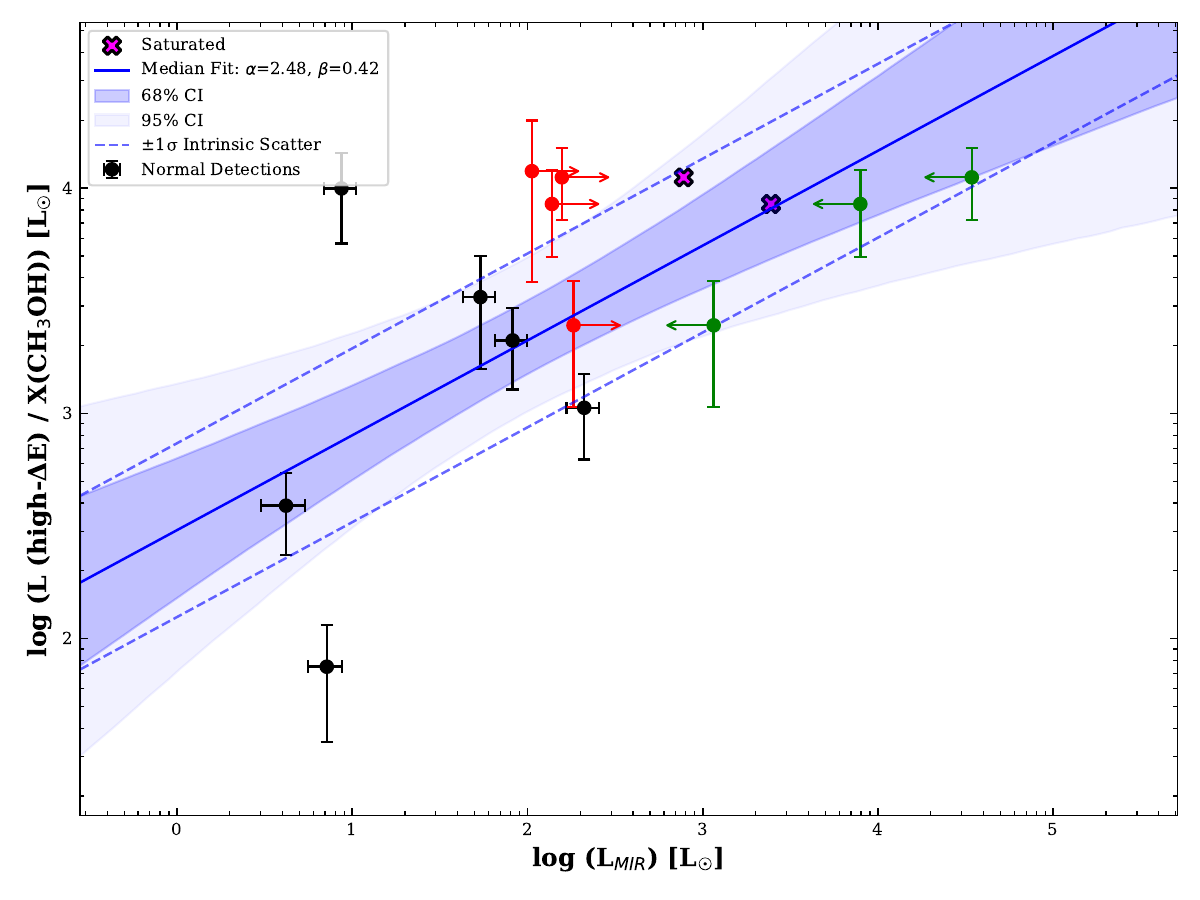}
\caption{
$\mathrm{log}L$ -- $\mathrm{log}L$ diagrams showing the relation between the mid-IR luminosity ($L_\mathrm{MIR}$) and the methanol $\varv_\mathrm{t}$$\geq$1 line luminosity normalized by the methanol abundance of the clump, $L(\Delta E/k_\mathrm{B})/X_\mathrm{CH_3OH}$, in the low (upper panel), mid (middle panel), and high (bottom panel) $\Delta E/k_\mathrm{B}$ regimes. 
The best-fit power-law relation (blue line) was obtained using Bayesian linear regression in log-log space. 
The dark and light shaded blue regions represent the 68\% and 95\% credible intervals (CI) of the regression, derived from the posterior distribution of the slope and intercept, accounting for their uncertainties.
The symbols represent sources with observed luminosities (black). For saturated sources, we show lower limits from MIPSGAL at 24\,$\mathrm{\mu m}$ (red), upper limits from WISE at 22\,$\mathrm{\mu m}$ (green), and, for visualization purposes only, expected fluxes scaled from JWST template ratios (magenta crosses; see Appendix\,\ref{sec:IR_def}).
}
\label{fig:LL_vts}
\end{figure}

The $\mathrm{log}L$ -- $\mathrm{log}L$ diagrams with the resulting fits are shown in Fig.\,\ref{fig:LL_vts} and the best-fit slopes and intercepts with their credible intervals (CI) are summarized in Table\,\ref{tab:coeff}.
In all three $\Delta E/k_\mathrm{B}$ regimes corresponding to mid-IR wavelengths of low 29--40\,$\mathrm{\mu m}$, mid 19--27\,$\mathrm{\mu m}$, and high 14--19\,$\mathrm{\mu m}$, the slopes are consistent within the uncertainties, ranging from $\sim$0.4 to 0.6, suggesting that the $\varv_\mathrm{t}$$\geq$1 methanol lines likely trace a common excitation mechanism despite differences in excitation energy.
Since these transitions are radiatively excited by mid-IR photons, the trend might also reflect the mid-IR radiation field of the sources.

For visualization purposes, Fig.\,\ref{fig:LL_vts} also indicates expected 24\,$\mathrm{\mu m}$ luminosities for saturated sources derived from GLIMPSE fluxes (see Appendix\,\ref{sec:IR_def}).

\begin{table*}
    \caption{Results of the Bayesian linear regression.} \label{tab:coeff}
    \centering
    \begin{tabular}{lcccccc}
    \hline\hline
    $\Delta E/k_\mathrm{B}$ & $\alpha$ & 68\% CI ($\alpha$) & 95\% CI ($\alpha$) & $\beta$ & 68\% CI ($\beta$) & 95\% CI ($\beta$) \\
    \hline
    low  & 2.92 & 2.65 -- 3.20 & 2.38 -- 3.47 & 0.61 & 0.49 -- 0.74 & 0.38 -- 0.89 \\
    mid  & 2.59 & 2.29 -- 2.92 & 2.00 -- 3.26 & 0.45 & 0.32 -- 0.58 & 0.18 -- 0.72 \\
    high & 2.48 & 2.18 -- 2.80 & 1.85 -- 3.13 & 0.42 & 0.29 -- 0.56 & 0.16 -- 0.71  \\
    \hline
    \end{tabular}
    \tablefoot{Best-fit parameters (median coefficient $\alpha$ intercept, $\beta$ slope) and their 68\% and 95\% credible intervals (CI) for the Bayesian power-law fit in log-log space between $L(\Delta E/k_\mathrm{B})/X_\mathrm{CH_3OH}$ and $L_\mathrm{MIR}$ at different $\Delta E/k_\mathrm{B}$.}

\end{table*}

\section{Discussion} \label{sec:discussion}

Our pilot sample was chosen to test the hypothesis that $\mathrm{CH_3OH}$ $\varv_\mathrm{t}$$\geq$1 lines are radiatively excited by mid-IR photons and can serve as indirect tracers of the mid-IR radiation fields around YSOs in different evolutionary stages.
Based on the source classification, we expected that mid-IR quiescent sources would not show $\varv_\mathrm{t}$$\geq$1 lines because either the absence of protostellar activity or its very early stages leaves the dust too cold to emit significant mid-IR radiation.
Conversely, more evolved sources are expected with stronger mid-IR fields capable of efficiently populating the torsional levels.
Consistent with these expectations, the only clump (G014.49) without $\varv_\mathrm{t}$$\geq$1 lines also lacks 24\,$\mathrm{\mu m}$ emission.
In contrast, G030.89, previously classified as quiescent, exhibits faint 24\,$\mathrm{\mu m}$ emission along with detectable $\varv_\mathrm{t}$$\geq$1 lines (Appendix\,\ref{sec:IR_def}). 
Based on these observations, we reclassified it as protostellar, further supporting the interpretation that these lines are sensitive tracers of embedded protostellar activity.

The power-law correlation between $L(\Delta E/k_\mathrm{B})/X_\mathrm{CH_3OH}$ and mid-IR luminosity across different $\Delta E/k_\mathrm{B}$ is promising (Fig.\,\ref{fig:LL_vts}). 
This relation appears largely independent of distance because all sources are within a relatively narrow range of $\sim$3--5\,$\mathrm{kpc}$.
Variations in the initial conditions are mitigated because the normalization by $X_\mathrm{CH_3OH}$ accounts for abundance differences.
However, several factors can limit the conclusions that can be drawn from the observed trend.
The sample is statistically small, the lowest-luminosity objects are underrepresented, and
they also tend to exhibit weak or undetected high-$E_\mathrm{u}/k_\mathrm{B}$ lines. This likely affects the slope steepness. 
In contrast, the brightest more evolved sources are saturated in the mid-IR, which only provides lower and upper limits on the slope.
Consequently, a larger sample, particularly of low-luminosity sources, is needed to refine the slope and further assess the dependences of the relation. 
\citet{Giannetti2017b} found that gas temperatures derived from $\mathrm{CH_3OH}$ $\varv_\mathrm{t}$$\geq$1 transitions correlate more closely with 22\,$\mathrm{\mu m}$ luminosities than with bolometric luminosities, suggesting a local excitation mechanism linked to internal protostellar heating at the clump scale. 
Our $\mathrm{log}L$ -- $\mathrm{log}L$ diagrams similarly show a linear trend between the luminosities of $\varv_\mathrm{t}$$\geq$1 lines and 24\,$\mathrm{\mu m}$ emission, confirming the sensitivity of these lines to the local mid-IR environment.
We verified that the trend persists across 22--70\,$\mathrm{\mu m}$ luminosities from available catalogs, whereas no correlation is observed with 8\,$\mathrm{\mu m}$, consistent with the idea that these lines indirectly trace mid-IR photons with longer wavelengths, which excite them in the local radiation field of MYSOs.
As a result, the power-law trend suggests that these lines might also serve as quantitative proxies for internal mid-IR radiation fields, and by extension, for the evolutionary stages of the sources. 
A similar approach was adopted by \citet{Aalto2015}, who used vibrationally excited $\mathrm{HCN}$ lines to infer mid-IR luminosities in extragalactic sources.

Our results are consistent with the radiative transfer modeling by \citet{Leurini2007a}, who suggested that these transitions are almost exclusively driven by radiative pumping, while collisional excitation remains inefficient even at very high densities ($10^{10}$--$10^{11}\,\mathrm{cm}^{-3}$).
The co-location of $\varv_\mathrm{t}$$\geq$1 emission with the 24\,$\mathrm{\mu m}$ sources and the 890\,$\mathrm{\mu m}$ ALMA continuum brightest peaks implies that excitation occurs in dense compact regions in the immediate vicinity of the protostar. 
This finding reinforces the interpretation that $\mathrm{CH_3OH}$ $\varv_\mathrm{t}$$\geq$1 transitions probe the innermost environment of protostars, where radiative feedback dominates collisional processes.
The $\varv_\mathrm{t}$$\geq$1 lines are indeed pumped by photons at energies similar to those in Class II methanol masers \citep{Menten1986}, which trace the inner gas close to massive YSOs and are found in outer regions of disk winds \citep[e.g.,][]{Sanna2015}. 
Similarly, \citet{Leurini2016} observed $\mathrm{CH_3OH}$ $\varv_\mathrm{t}$=2 lines in the low-mass protostar HH 212. 
In this source, the emission traces velocity gradients consistent with rotating and expanding gas, which was interpreted as a disk wind by the authors. 
In the high-mass protocluster IRAS 05358$+$3543, \citet{Leurini2007b} observed double-peaked $\varv_\mathrm{t}$$\geq$1 profiles, which are indicative of disks. 
These studies suggested that $\mathrm{CH_3OH}$ $\varv_\mathrm{t}$$\geq$1 transitions trace the innermost rotating structures at the base of outflows, likely launched near the disk surface, across a broad range of stellar masses. 
A simultaneous study of torsionally excited methanol lines and Class II masers at a comparable resolution might confirm the origin of the two emissions from the same gas and further support the idea that they are both pumped by the same mechanism.

The low angular resolution of Spitzer/MIPS limits our ability to distinguish whether the 24\,$\mathrm{\mu m}$ emission arises from a single dominant core or from multiple embedded cores.
Consequently, in the $\mathrm{log}L$ -- $\mathrm{log}L$ diagram, we compare the total $\varv_\mathrm{t}$$\geq$1 line luminosities from individual cores with a 24\,$\mathrm{\mu m}$ luminosity representing the entire clump. This introduces uncertainties in interpretation.
This pilot study highlights the potential of torsionally excited methanol lines as tracers of internal radiative feedback, but further ALMA and high-resolution mid-IR observations (e.g., with JWST) are needed to support this hypothesis. 
It is essential to extend the analysis to a larger statistically significant sample, particularly including low-luminosity sources to refine the slope of the $\mathrm{log}L$ -- $\mathrm{log}L$ diagram and to better understand the earliest stages of high-mass star formation.

\section{Summary and conclusions} \label{sec:conclusions}

We investigated the excitation and spatial distribution of $\mathrm{CH_3OH}$ $\varv_\mathrm{t}$$\geq$1 lines in a sample of high-mass star-forming regions by comparing ALMA spectral line and continuum observations at 890\,$\mathrm{\mu m}$ with mid-IR data.

Our results show that $\mathrm{CH_3OH}$ $\varv_\mathrm{t}$$\geq$1 emission is compact and systematically associated with ALMA continuum peaks and 24\,$\mathrm{\mu m}$ sources, indicating a close spatial association with embedded protostellar objects. 
The only clump without mid-IR counterparts shows no torsionally excited methanol emission, consistent with radiative pumping as the dominant excitation mechanism. 
In a few cases, weak 24\,$\mathrm{\mu m}$ emission coincides with weak $\varv_\mathrm{t}$$\geq$1 lines, suggesting that these transitions can reveal faint embedded protostars 
that are difficult to identify in mid-IR images due to confusion and limited angular resolution.

The correlation between methanol lines and 24\,$\mathrm{\mu m}$ luminosities, which persists across multiple excitation ranges and is largely independent of the methanol abundance variations, reinforces the interpretation that local mid-IR radiation from warm dust dominates the excitation of these transitions. 
This supports the hypothesis that methanol $\varv_\mathrm{t}$$\geq$1 lines are reliable probes of the internal radiation environment of YSOs at the core scale and, by extension, their evolutionary stage.

In summary, $\mathrm{CH_3OH}$ $\varv_\mathrm{t}$$\geq$1 transitions provide an effective tool for identifying and characterizing embedded star formation and probing internal radiative feedback. 
Future high-resolution mid-IR observations, for instance, with JWST, combined with statistically significant surveys, will be essential to refine this diagnostic, extend its application to diverse environments and mass regimes, and improve our understanding of the earliest stages of protostellar evolution.

\begin{acknowledgements}
We dedicate this work to Karl M. Menten, whose profound expertise in this field laid the foundations of this study: his commitment, insight, and unwavering enthusiasm continue to inspire us and guide our research.

We thank Prof. G. Malloci for his constructive comments and support during the development of this study.
We are grateful to R. Paladino for the useful support and discussion.
We thank the Italian ARC node and the staff for the support in data reduction.

We thank the anonymous referee for constructive comments that improved the clarity of this paper.

This work has been supported by the project PRIN-INAF 2016 "From yOuNg Star clusters to planETary systems (ONSET)".
This publication has received funding from the European Union’s Horizon 2020 research and innovation program under grant agreement No 101004719 (ORP).
C.S. acknowledges the support from the European Southern Observatory (ESO) Early-Career Scientific Visitor Programme Garching (Germany).

This paper makes use of the following ALMA data: ADS/JAO.ALMA\#2017.1.00377.S.
ALMA is a partnership of ESO (representing its member states), NSF (USA) and NINS (Japan), together with NRC (Canada), MOST and ASIAA (Taiwan) and KASI (Republic of Korea), in co-operation with the Republic of Chile. 
The Joint ALMA Observatory is operated by ESO, AUI/NRAO and NAOJ. 

This work used \texttt{Astropy}\footnote{\url{http://www.astropy.org}}, a community developed core Python package for Astronomy \citep{astropy:2013, astropy:2018, astropy:2022}; \texttt{APLPY} \citep{APLpy2012, APLpy2019}; \texttt{matplotlib} \citep{Hunter2007}; \texttt{NumPy} \citep{NumPy2020}, \texttt{SciPy} \citep{Virtanen2020}, and CARTA software \citep[][]{CARTA2021, CARTA2022}
This research used the VizieR catalog, operated at CDS, Strasbourg, France. 
This document was prepared using the Overleaf web platform\footnote{Overleaf is available at \url{https://www.overleaf.com}.}.

\end{acknowledgements}

\bibliography{references/master_biblio}

@ARTICLE{Aalto2015,
       author = {{Aalto}, S. and {Mart{\'\i}n}, S. and {Costagliola}, F. and {Gonz{\'a}lez-Alfonso}, E. and {Muller}, S. and {Sakamoto}, K. and {Fuller}, G.~A. and {Garc{\'\i}a-Burillo}, S. and {van der Werf}, P. and {Neri}, R. and {Spaans}, M. and {Combes}, F. and {Viti}, S. and {M{\"u}hle}, S. and {Armus}, L. and {Evans}, A. and {Sturm}, E. and {Cernicharo}, J. and {Henkel}, C. and {Greve}, T.~R.},
        title = "{Probing highly obscured, self-absorbed galaxy nuclei with vibrationally excited HCN}",
      journal = {\aap},
         year = 2015,
        month = dec,
       volume = {584},
          eid = {A42},
        pages = {A42},
          doi = {10.1051/0004-6361/201526410},
archivePrefix = {arXiv},
       eprint = {1504.06824},
 primaryClass = {astro-ph.GA},
       adsurl = {https://ui.adsabs.harvard.edu/abs/2015A&A...584A..42A}
}

@ARTICLE{Benjamin2003,
       author = {{Benjamin}, Robert A. and {Churchwell}, E. and {Babler}, Brian L. and {Bania}, T.~M. and {Clemens}, Dan P. and {Cohen}, Martin and {Dickey}, John M. and {Indebetouw}, R{\'e}my and {Jackson}, James M. and {Kobulnicky}, Henry A. and {Lazarian}, Alex and {Marston}, A.~P. and {Mathis}, John S. and {Meade}, Marilyn R. and {Seager}, Sara and {Stolovy}, S.~R. and {Watson}, C. and {Whitney}, Barbara A. and {Wolff}, Michael J. and {Wolfire}, Mark G.},
        title = "{GLIMPSE. I. An SIRTF Legacy Project to Map the Inner Galaxy}",
      journal = {\pasp},
         year = 2003,
        month = aug,
       volume = {115},
       number = {810},
        pages = {953-964},
          doi = {10.1086/376696},
archivePrefix = {arXiv},
       eprint = {astro-ph/0306274},
 primaryClass = {astro-ph},
       adsurl = {https://ui.adsabs.harvard.edu/abs/2003PASP..115..953B}
}

@misc{Bradley2022,
       author = {{Bradley}, Larry and {Sip{\H{o}}cz}, Brigitta and {Robitaille}, Thomas and {Tollerud}, Erik and {Vin{\'\i}cius}, Z{\'e} and {Deil}, Christoph and {Barbary}, Kyle and {Wilson}, Tom J and {Busko}, Ivo and {Donath}, Axel and {G{\"u}nther}, Hans Moritz and {Cara}, Mihai and {Lim}, P.~L. and {Me{\ss}linger}, Sebastian and {Conseil}, Simon and {Bostroem}, Azalee and {Droettboom}, Michael and {Bray}, E.~M. and {Andersen Bratholm}, Lars and {Barentsen}, Geert and {Craig}, Matt and {Ginsburg}, Adam and {Rathi}, Shivangee and {Pascual}, Sergio and {Perren}, Gabriel and {Georgiev}, Iskren Y. and {De Val-Borro}, Miguel and {Kerzendorf}, Wolfgang and {Bach}, Yoonsoo P. and {Quint}, Bruno},
        title = "{astropy/photutils: 1.6.0}",
         year = 2022,
        month = dec,
          eid = {10.5281/zenodo.7419741},
          doi = {10.5281/zenodo.7419741},
      version = {1.6.0},
    publisher = {Zenodo},
       adsurl = {https://ui.adsabs.harvard.edu/abs/2022zndo...7419741B}
}

@ARTICLE{Carey2009,
       author = {{Carey}, S.~J. and {Noriega-Crespo}, A. and {Mizuno}, D.~R. and {Shenoy}, S. and {Paladini}, R. and {Kraemer}, K.~E. and {Price}, S.~D. and {Flagey}, N. and {Ryan}, E. and {Ingalls}, J.~G. and {Kuchar}, T.~A. and {Pinheiro Gon{\c{c}}alves}, Daniela and {Indebetouw}, R. and {Billot}, N. and {Marleau}, F.~R. and {Padgett}, D.~L. and {Rebull}, L.~M. and {Bressert}, E. and {Ali}, Babar and {Molinari}, S. and {Martin}, P.~G. and {Berriman}, G.~B. and {Boulanger}, F. and {Latter}, W.~B. and {Miville-Deschenes}, M.~A. and {Shipman}, R. and {Testi}, L.},
        title = "{MIPSGAL: A Survey of the Inner Galactic Plane at 24 and 70 {\ensuremath{\mu}}m}",
      journal = {\pasp},
         year = 2009,
        month = jan,
       volume = {121},
       number = {875},
        pages = {76},
          doi = {10.1086/596581},
       adsurl = {https://ui.adsabs.harvard.edu/abs/2009PASP..121...76C}
}

@ARTICLE{CarrollGoldsmith1981,
       author = {{Carroll}, T.~J. and {Goldsmith}, P.~F.},
        title = "{Infrared pumping and rotational excitation of molecules in interstellar clouds}",
      journal = {\apj},
         year = 1981,
        month = may,
       volume = {245},
        pages = {891-897},
          doi = {10.1086/158865},
       adsurl = {https://ui.adsabs.harvard.edu/abs/1981ApJ...245..891C}
}

@ARTICLE{Churchwell2009, 
       author = {{Churchwell}, Ed and {Babler}, Brian L. and {Meade}, Marilyn R. and {Whitney}, Barbara A. and {Benjamin}, Robert and {Indebetouw}, Remy and {Cyganowski}, Claudia and {Robitaille}, Thomas P. and {Povich}, Matthew and {Watson}, Christer and {Bracker}, Steve},
        title = "{The Spitzer/GLIMPSE Surveys: A New View of the Milky Way}",
      journal = {\pasp},
         year = 2009,
        month = mar,
       volume = {121},
       number = {877},
        pages = {213},
          doi = {10.1086/597811},
       adsurl = {https://ui.adsabs.harvard.edu/abs/2009PASP..121..213C}
}

@ARTICLE{Giannetti2017b,
       author = {{Giannetti}, A. and {Leurini}, S. and {Wyrowski}, F. and {Urquhart}, J. and {Csengeri}, T. and {Menten}, K.~M. and {K{\"o}nig}, C. and {G{\"u}sten}, R.},
        title = "{ATLASGAL-selected massive clumps in the inner Galaxy. V. Temperature structure and evolution}",
      journal = {\aap},
         year = 2017,
        month = jul,
       volume = {603},
          eid = {A33},
        pages = {A33},
          doi = {10.1051/0004-6361/201630048},
archivePrefix = {arXiv},
       eprint = {1703.08485},
 primaryClass = {astro-ph.GA},
       adsurl = {https://ui.adsabs.harvard.edu/abs/2017A&A...603A..33G}
}

@ARTICLE{Gutermuth2015,
       author = {{Gutermuth}, Robert A. and {Heyer}, Mark},
        title = "{A 24{\,}{\ensuremath{\mu}}m Point Source Catalog of the Galactic Plane from Spitzer/MIPSGAL}",
      journal = {\aj},
         year = 2015,
        month = feb,
       volume = {149},
       number = {2},
          eid = {64},
        pages = {64},
          doi = {10.1088/0004-6256/149/2/64},
archivePrefix = {arXiv},
       eprint = {1412.4751},
 primaryClass = {astro-ph.SR},
       adsurl = {https://ui.adsabs.harvard.edu/abs/2015AJ....149...64G}
}

@ARTICLE{Koenig2017,
       author = {{K{\"o}nig}, C. and {Urquhart}, J.~S. and {Csengeri}, T. and {Leurini}, S. and {Wyrowski}, F. and {Giannetti}, A. and {Wienen}, M. and {Pillai}, T. and {Kauffmann}, J. and {Menten}, K.~M. and {Schuller}, F.},
        title = "{ATLASGAL-selected massive clumps in the inner Galaxy. III. Dust continuum characterization of an evolutionary sample}",
      journal = {\aap},
         year = 2017,
        month = mar,
       volume = {599},
          eid = {A139},
        pages = {A139},
          doi = {10.1051/0004-6361/201526841},
archivePrefix = {arXiv},
       eprint = {1610.09055},
 primaryClass = {astro-ph.GA},
       adsurl = {https://ui.adsabs.harvard.edu/abs/2017A&A...599A.139K}
}

@ARTICLE{Leurini2007a,
       author = {{Leurini}, S. and {Schilke}, P. and {Wyrowski}, F. and {Menten}, K.~M.},
        title = "{Methanol as a diagnostic tool of interstellar clouds. II. Modelling high-mass protostellar objects}",
      journal = {\aap},
         year = 2007,
        month = apr,
       volume = {466},
       number = {1},
        pages = {215-228},
          doi = {10.1051/0004-6361:20054245},
       adsurl = {https://ui.adsabs.harvard.edu/abs/2007A&A...466..215L}
}

@ARTICLE{Leurini2007b,
       author = {{Leurini}, S. and {Beuther}, H. and {Schilke}, P. and {Wyrowski}, F. and {Zhang}, Q. and {Menten}, K.~M.},
        title = "{Multi-line (sub)millimetre observations of the high-mass proto cluster IRAS 05358+3543}",
      journal = {\aap},
         year = 2007,
        month = dec,
       volume = {475},
       number = {3},
        pages = {925-939},
          doi = {10.1051/0004-6361:20077977},
archivePrefix = {arXiv},
       eprint = {0710.4238},
 primaryClass = {astro-ph},
       adsurl = {https://ui.adsabs.harvard.edu/abs/2007A&A...475..925L}
}

@ARTICLE{Leurini2016,
       author = {{Leurini}, S. and {Codella}, C. and {Cabrit}, S. and {Gueth}, F. and {Giannetti}, A. and {Bacciotti}, F. and {Bachiller}, R. and {Ceccarelli}, C. and {Gusdorf}, A. and {Lefloch}, B. and {Podio}, L. and {Tafalla}, M.},
        title = "{Hot methanol from the inner region of the HH 212 protostellar system}",
      journal = {\aap},
         year = 2016,
        month = oct,
       volume = {595},
          eid = {L4},
        pages = {L4},
          doi = {10.1051/0004-6361/201629460},
archivePrefix = {arXiv},
       eprint = {1610.05322},
 primaryClass = {astro-ph.GA},
       adsurl = {https://ui.adsabs.harvard.edu/abs/2016A&A...595L...4L}
}

@ARTICLE{Molinari2010,
       author = {{Molinari}, S. and {Swinyard}, B. and {Bally}, J. and {Barlow}, M. and {Bernard}, J. -P. and {Martin}, P. and {Moore}, T. and {Noriega-Crespo}, A. and {Plume}, R. and {Testi}, L. and {Zavagno}, A. and {Abergel}, A. and {Ali}, B. and {Anderson}, L. and {Andr{\'e}}, P. and {Baluteau}, J. -P. and {Battersby}, C. and {Beltr{\'a}n}, M.~T. and {Benedettini}, M. and {Billot}, N. and {Blommaert}, J. and {Bontemps}, S. and {Boulanger}, F. and {Brand}, J. and {Brunt}, C. and {Burton}, M. and {Calzoletti}, L. and {Carey}, S. and {Caselli}, P. and {Cesaroni}, R. and {Cernicharo}, J. and {Chakrabarti}, S. and {Chrysostomou}, A. and {Cohen}, M. and {Compiegne}, M. and {de Bernardis}, P. and {de Gasperis}, G. and {di Giorgio}, A.~M. and {Elia}, D. and {Faustini}, F. and {Flagey}, N. and {Fukui}, Y. and {Fuller}, G.~A. and {Ganga}, K. and {Garcia-Lario}, P. and {Glenn}, J. and {Goldsmith}, P.~F. and {Griffin}, M. and {Hoare}, M. and {Huang}, M. and {Ikhenaode}, D. and {Joblin}, C. and {Joncas}, G. and {Juvela}, M. and {Kirk}, J.~M. and {Lagache}, G. and {Li}, J.~Z. and {Lim}, T.~L. and {Lord}, S.~D. and {Marengo}, M. and {Marshall}, D.~J. and {Masi}, S. and {Massi}, F. and {Matsuura}, M. and {Minier}, V. and {Miville-Desch{\^e}nes}, M. -A. and {Montier}, L.~A. and {Morgan}, L. and {Motte}, F. and {Mottram}, J.~C. and {M{\"u}ller}, T.~G. and {Natoli}, P. and {Neves}, J. and {Olmi}, L. and {Paladini}, R. and {Paradis}, D. and {Parsons}, H. and {Peretto}, N. and {Pestalozzi}, M. and {Pezzuto}, S. and {Piacentini}, F. and {Piazzo}, L. and {Polychroni}, D. and {Pomar{\`e}s}, M. and {Popescu}, C.~C. and {Reach}, W.~T. and {Ristorcelli}, I. and {Robitaille}, J. -F. and {Robitaille}, T. and {Rod{\'o}n}, J.~A. and {Roy}, A. and {Royer}, P. and {Russeil}, D. and {Saraceno}, P. and {Sauvage}, M. and {Schilke}, P. and {Schisano}, E. and {Schneider}, N. and {Schuller}, F. and {Schulz}, B. and {Sibthorpe}, B. and {Smith}, H.~A. and {Smith}, M.~D. and {Spinoglio}, L. and {Stamatellos}, D. and {Strafella}, F. and {Stringfellow}, G.~S. and {Sturm}, E. and {Taylor}, R. and {Thompson}, M.~A. and {Traficante}, A. and {Tuffs}, R.~J. and {Umana}, G. and {Valenziano}, L. and {Vavrek}, R. and {Veneziani}, M. and {Viti}, S. and {Waelkens}, C. and {Ward-Thompson}, D. and {White}, G. and {Wilcock}, L.~A. and {Wyrowski}, F. and {Yorke}, H.~W. and {Zhang}, Q.},
        title = "{Clouds, filaments, and protostars: The Herschel Hi-GAL Milky Way}",
      journal = {\aap},
         year = 2010,
        month = jul,
       volume = {518},
          eid = {L100},
        pages = {L100},
          doi = {10.1051/0004-6361/201014659},
archivePrefix = {arXiv},
       eprint = {1005.3317},
 primaryClass = {astro-ph.GA},
       adsurl = {https://ui.adsabs.harvard.edu/abs/2010A&A...518L.100M}
}

@misc{Robitaille2019,
       author = {{Robitaille}, Thomas and {Rice}, Tom and {Beaumont}, Chris and {Ginsburg}, Adam and {MacDonald}, Braden and {Rosolowsky}, Erik},
        title = "{astrodendro: Astronomical data dendrogram creator}",
 howpublished = {Astrophysics Source Code Library, record ascl:1907.016},
         year = 2019,
        month = jul,
          eid = {ascl:1907.016},
       adsurl = {https://ui.adsabs.harvard.edu/abs/2019ascl.soft07016R}
}

@ARTICLE{Rosolowsky2008,
       author = {{Rosolowsky}, E.~W. and {Pineda}, J.~E. and {Kauffmann}, J. and {Goodman}, A.~A.},
        title = "{Structural Analysis of Molecular Clouds: Dendrograms}",
      journal = {\apj},
         year = 2008,
        month = jun,
       volume = {679},
       number = {2},
        pages = {1338-1351},
          doi = {10.1086/587685},
archivePrefix = {arXiv},
       eprint = {0802.2944},
 primaryClass = {astro-ph},
       adsurl = {https://ui.adsabs.harvard.edu/abs/2008ApJ...679.1338R}
}

@ARTICLE{Sanna2015,
       author = {{Sanna}, A. and {Surcis}, G. and {Moscadelli}, L. and {Cesaroni}, R. and {Goddi}, C. and {Vlemmings}, W.~H.~T. and {Caratti o Garatti}, A.},
        title = "{Velocity and magnetic fields within 1000 AU of a massive YSO}",
      journal = {\aap},
         year = 2015,
        month = nov,
       volume = {583},
          eid = {L3},
        pages = {L3},
          doi = {10.1051/0004-6361/201526806},
archivePrefix = {arXiv},
       eprint = {1509.05428},
 primaryClass = {astro-ph.SR},
       adsurl = {https://ui.adsabs.harvard.edu/abs/2015A&A...583L...3S}
}

@ARTICLE{Sanchez2014,
       author = {{S{\'a}nchez-Monge}, {\'A}. and {Beltr{\'a}n}, M.~T. and {Cesaroni}, R. and {Etoka}, S. and {Galli}, D. and {Kumar}, M.~S.~N. and {Moscadelli}, L. and {Stanke}, T. and {van der Tak}, F.~F.~S. and {Vig}, S. and {Walmsley}, C.~M. and {Wang}, K. -S. and {Zinnecker}, H. and {Elia}, D. and {Molinari}, S. and {Schisano}, E.},
        title = "{A necklace of dense cores in the high-mass star forming region G35.20-0.74 N: ALMA observations}",
      journal = {\aap},
         year = 2014,
        month = sep,
       volume = {569},
          eid = {A11},
        pages = {A11},
          doi = {10.1051/0004-6361/201424032},
archivePrefix = {arXiv},
       eprint = {1406.4081},
 primaryClass = {astro-ph.GA},
       adsurl = {https://ui.adsabs.harvard.edu/abs/2014A&A...569A..11S}
}

@ARTICLE{Wienen2015,
       author = {{Wienen}, M. and {Wyrowski}, F. and {Menten}, K.~M. and {Urquhart}, J.~S. and {Csengeri}, T. and {Walmsley}, C.~M. and {Bontemps}, S. and {Russeil}, D. and {Bronfman}, L. and {Koribalski}, B.~S. and {Schuller}, F.},
        title = "{ATLASGAL - Kinematic distances and the dense gas mass distribution of the inner Galaxy}",
      journal = {\aap},
         year = 2015,
        month = jul,
       volume = {579},
          eid = {A91},
        pages = {A91},
          doi = {10.1051/0004-6361/201424802},
archivePrefix = {arXiv},
       eprint = {1503.00007},
 primaryClass = {astro-ph.SR},
       adsurl = {https://ui.adsabs.harvard.edu/abs/2015A&A...579A..91W}
}

@ARTICLE{Urquhart2018,
       author = {{Urquhart}, J.~S. and {K{\"o}nig}, C. and {Giannetti}, A. and {Leurini}, S. and {Moore}, T.~J.~T. and {Eden}, D.~J. and {Pillai}, T. and {Thompson}, M.~A. and {Braiding}, C. and {Burton}, M.~G. and {Csengeri}, T. and {Dempsey}, J.~T. and {Figura}, C. and {Froebrich}, D. and {Menten}, K.~M. and {Schuller}, F. and {Smith}, M.~D. and {Wyrowski}, F.},
        title = "{ATLASGAL - properties of a complete sample of Galactic clumps}",
      journal = {\mnras},
         year = 2018,
        month = jan,
       volume = {473},
       number = {1},
        pages = {1059-1102},
          doi = {10.1093/mnras/stx2258},
archivePrefix = {arXiv},
       eprint = {1709.00392},
 primaryClass = {astro-ph.GA},
       adsurl = {https://ui.adsabs.harvard.edu/abs/2018MNRAS.473.1059U}
}

@misc{GLIMPSE2009,
       author = {{Spitzer Science}, Center},
        title = "{VizieR Online Data Catalog: GLIMPSE Source Catalog (I + II + 3D) (IPAC 2008)}",
 howpublished = {VizieR On-line Data Catalog: II/293.  Originally published in: IPAC, Caltech (2008)},
         year = 2009,
        month = jun,
          eid = {II/293},
       adsurl = {https://ui.adsabs.harvard.edu/abs/2009yCat.2293....0S}
}

@ARTICLE{Muller2001,
       author = {{M{\"u}ller}, H.~S.~P. and {Thorwirth}, S. and {Roth}, D.~A. and {Winnewisser}, G.},
        title = "{The Cologne Database for Molecular Spectroscopy, CDMS}",
      journal = {\aap},
         year = 2001,
        month = apr,
       volume = {370},
        pages = {L49-L52},
          doi = {10.1051/0004-6361:20010367},
       adsurl = {https://ui.adsabs.harvard.edu/abs/2001A&A...370L..49M}
}

@ARTICLE{Pickett1998,
       author = {{Pickett}, H.~M. and {Poynter}, R.~L. and {Cohen}, E.~A. and {Delitsky}, M.~L. and {Pearson}, J.~C. and {M{\"u}ller}, H.~S.~P.},
        title = "{Submillimeter, millimeter and microwave spectral line catalog.}",
      journal = {\jqsrt},
         year = 1998,
        month = nov,
       volume = {60},
       number = {5},
        pages = {883-890},
          doi = {10.1016/S0022-4073(98)00091-0},
       adsurl = {https://ui.adsabs.harvard.edu/abs/1998JQSRT..60..883P}
}

@INPROCEEDINGS{Remijan2007,
       author = {{Remijan}, Anthony J. and {Markwick-Kemper}, A. and {ALMA Working Group on Spectral Line Frequencies}},
        title = "{Splatalogue: Database for Astronomical Spectroscopy}",
    booktitle = {American Astronomical Society Meeting Abstracts},
         year = 2007,
       series = {American Astronomical Society Meeting Abstracts},
       volume = {211},
        month = dec,
          eid = {132.11},
        pages = {132.11},
       adsurl = {https://ui.adsabs.harvard.edu/abs/2007AAS...21113211R}
}

@INPROCEEDINGS{Muller2005,
       author = {{M{\"u}ller}, Holger S.~P. and {Schl{\"o}der}, F. and {Stutzki}, J. and {Schlemmer}, S. and {Giesen}, T. and {Schilke}, P.},
        title = "{The Cologne Database for Molecular Spectroscopy, CDMS: A Tool for Astrochemists and Astrophysicists}",
    booktitle = {Astrochemistry: Recent Successes and Current Challenges},
         year = 2005,
       editor = {{Lis}, Dariusz C. and {Blake}, Geoffrey A. and {Herbst}, Eric},
       series = {IAU Symposium},
       volume = {231},
        month = jan,
        pages = {62},
       adsurl = {https://ui.adsabs.harvard.edu/abs/2005IAUS..231P..62M}
}

@INPROCEEDINGS{McMullin2007,
       author = {{McMullin}, J.~P. and {Waters}, B. and {Schiebel}, D. and {Young}, W. and {Golap}, K.},
        title = "{CASA Architecture and Applications}",
    booktitle = {Astronomical Data Analysis Software and Systems XVI},
         year = 2007,
       editor = {{Shaw}, R.~A. and {Hill}, F. and {Bell}, D.~J.},
       series = {Astronomical Society of the Pacific Conference Series},
       volume = {376},
        month = oct,
        pages = {127},
       adsurl = {https://ui.adsabs.harvard.edu/abs/2007ASPC..376..127M}
}

@article{astropy:2013,
  author    = {{Astropy Collaboration}},
  title     = {Astropy: A community Python package for astronomy},
  journal   = {A\&A},
  volume    = {558},
  pages     = {A33},
  year      = {2013}
}

@article{astropy:2018,
  author    = {{Astropy Collaboration}},
  title     = {The Astropy Project: Building an open-science project and status of the v2.0 core package},
  journal   = {AJ},
  volume    = {156},
  pages     = {123},
  year      = {2018}
}

@article{astropy:2022,
  author    = {{Astropy Collaboration}},
  title     = {The Astropy Project: Sustaining and growing a community-oriented open-source project and the latest major release (v5.0) of the core package},
  journal   = {AJ},
  volume    = {163},
  pages     = {286},
  year      = {2022}
}

@ARTICLE{Beuther2023,
       author = {{Beuther}, H. and {van Dishoeck}, E.~F. and {Tychoniec}, L. and {Gieser}, C. and {Kavanagh}, P.~J. and {Perotti}, G. and {van Gelder}, M.~L. and {Klaassen}, P. and {Caratti o Garatti}, A. and {Francis}, L. and {Rocha}, W.~R.~M. and {Slavicinska}, K. and {Ray}, T. and {Justtanont}, K. and {Linnartz}, H. and {Waelkens}, C. and {Colina}, L. and {Greve}, T. and {G{\"u}del}, M. and {Henning}, T. and {Lagage}, P. -O. and {Vandenbussche}, B. and {{\"O}stlin}, G. and {Wright}, G.},
        title = "{JWST Observations of Young protoStars (JOYS). Outflows and accretion in the high-mass star-forming region IRAS 23385+6053}",
      journal = {\aap},
         year = 2023,
        month = may,
       volume = {673},
          eid = {A121},
        pages = {A121},
          doi = {10.1051/0004-6361/202346167},
archivePrefix = {arXiv},
       eprint = {2303.13172},
 primaryClass = {astro-ph.SR},
       adsurl = {https://ui.adsabs.harvard.edu/abs/2023A&A...673A.121B}
}

@ARTICLE{KennicuttEvans2012,
       author = {{Kennicutt}, Robert C. and {Evans}, Neal J.},
        title = "{Star Formation in the Milky Way and Nearby Galaxies}",
      journal = {\araa},
         year = 2012,
        month = sep,
       volume = {50},
        pages = {531-608},
          doi = {10.1146/annurev-astro-081811-125610},
archivePrefix = {arXiv},
       eprint = {1204.3552},
 primaryClass = {astro-ph.GA},
       adsurl = {https://ui.adsabs.harvard.edu/abs/2012ARA&A..50..531K}
}

@ARTICLE{Molinari2008,
       author = {{Molinari}, S. and {Pezzuto}, S. and {Cesaroni}, R. and {Brand}, J. and {Faustini}, F. and {Testi}, L.},
        title = "{The evolution of the spectral energy distribution in massive young stellar objects}",
      journal = {\aap},
         year = 2008,
        month = apr,
       volume = {481},
       number = {2},
        pages = {345-365},
          doi = {10.1051/0004-6361:20078661},
       adsurl = {https://ui.adsabs.harvard.edu/abs/2008A&A...481..345M}
}

@ARTICLE{Molinari2016,
       author = {{Molinari}, S. and {Merello}, M. and {Elia}, D. and {Cesaroni}, R. and {Testi}, L. and {Robitaille}, T.},
        title = "{Calibration of Evolutionary Diagnostics in High-mass Star Formation}",
      journal = {\apjl},
         year = 2016,
        month = jul,
       volume = {826},
       number = {1},
          eid = {L8},
        pages = {L8},
          doi = {10.3847/2041-8205/826/1/L8},
archivePrefix = {arXiv},
       eprint = {1604.06192},
 primaryClass = {astro-ph.GA},
       adsurl = {https://ui.adsabs.harvard.edu/abs/2016ApJ...826L...8M}
}

@ARTICLE{Elia2017,
       author = {{Elia}, Davide and {Molinari}, S. and {Schisano}, E. and {Pestalozzi}, M. and {Pezzuto}, S. and {Merello}, M. and {Noriega-Crespo}, A. and {Moore}, T.~J.~T. and {Russeil}, D. and {Mottram}, J.~C. and {Paladini}, R. and {Strafella}, F. and {Benedettini}, M. and {Bernard}, J.~P. and {Di Giorgio}, A. and {Eden}, D.~J. and {Fukui}, Y. and {Plume}, R. and {Bally}, J. and {Martin}, P.~G. and {Ragan}, S.~E. and {Jaffa}, S.~E. and {Motte}, F. and {Olmi}, L. and {Schneider}, N. and {Testi}, L. and {Wyrowski}, F. and {Zavagno}, A. and {Calzoletti}, L. and {Faustini}, F. and {Natoli}, P. and {Palmeirim}, P. and {Piacentini}, F. and {Piazzo}, L. and {Pilbratt}, G.~L. and {Polychroni}, D. and {Baldeschi}, A. and {Beltr{\'a}n}, M.~T. and {Billot}, N. and {Cambr{\'e}sy}, L. and {Cesaroni}, R. and {Garc{\'\i}a-Lario}, P. and {Hoare}, M.~G. and {Huang}, M. and {Joncas}, G. and {Liu}, S.~J. and {Maiolo}, B.~M.~T. and {Marsh}, K.~A. and {Maruccia}, Y. and {M{\`e}ge}, P. and {Peretto}, N. and {Rygl}, K.~L.~J. and {Schilke}, P. and {Thompson}, M.~A. and {Traficante}, A. and {Umana}, G. and {Veneziani}, M. and {Ward-Thompson}, D. and {Whitworth}, A.~P. and {Arab}, H. and {Bandieramonte}, M. and {Becciani}, U. and {Brescia}, M. and {Buemi}, C. and {Bufano}, F. and {Butora}, R. and {Cavuoti}, S. and {Costa}, A. and {Fiorellino}, E. and {Hajnal}, A. and {Hayakawa}, T. and {Kacsuk}, P. and {Leto}, P. and {Li Causi}, G. and {Marchili}, N. and {Martinavarro-Armengol}, S. and {Mercurio}, A. and {Molinaro}, M. and {Riccio}, G. and {Sano}, H. and {Sciacca}, E. and {Tachihara}, K. and {Torii}, K. and {Trigilio}, C. and {Vitello}, F. and {Yamamoto}, H.},
        title = "{The Hi-GAL compact source catalogue - I. The physical properties of the clumps in the inner Galaxy (-71.0{\textdegree} < {\ensuremath{\ell}} < 67.0{\textdegree})}",
      journal = {\mnras},
         year = 2017,
        month = oct,
       volume = {471},
       number = {1},
        pages = {100-143},
          doi = {10.1093/mnras/stx1357},
archivePrefix = {arXiv},
       eprint = {1706.01046},
 primaryClass = {astro-ph.GA},
       adsurl = {https://ui.adsabs.harvard.edu/abs/2017MNRAS.471..100E}
}

@ARTICLE{Lada1984,
       author = {{Lada}, C.~J. and {Margulis}, M. and {Dearborn}, D.},
        title = "{The formation and early dynamical evolution of bound stellar systems.}",
      journal = {\apj},
         year = 1984,
        month = oct,
       volume = {285},
        pages = {141-152},
          doi = {10.1086/162485},
       adsurl = {https://ui.adsabs.harvard.edu/abs/1984ApJ...285..141L}
}

@ARTICLE{Andre1993,
       author = {{Andr{\'e}}, Philippe and {Ward-Thompson}, Derek and {Barsony}, Mary},
        title = "{Submillimeter Continuum Observations of rho Ophiuchi A: The Candidate Protostar VLA 1623 and Prestellar Clumps}",
      journal = {\apj},
         year = 1993,
        month = mar,
       volume = {406},
        pages = {122},
          doi = {10.1086/172425},
       adsurl = {https://ui.adsabs.harvard.edu/abs/1993ApJ...406..122A}
}

@ARTICLE{Evans2009,
       author = {{Evans}, II, Neal J. and {Dunham}, Michael M. and {J{\o}rgensen}, Jes K. and {Enoch}, Melissa L. and {Mer{\'\i}n}, Bruno and {van Dishoeck}, Ewine F. and {Alcal{\'a}}, Juan M. and {Myers}, Philip C. and {Stapelfeldt}, Karl R. and {Huard}, Tracy L. and {Allen}, Lori E. and {Harvey}, Paul M. and {van Kempen}, Tim and {Blake}, Geoffrey A. and {Koerner}, David W. and {Mundy}, Lee G. and {Padgett}, Deborah L. and {Sargent}, Anneila I.},
        title = "{The Spitzer c2d Legacy Results: Star-Formation Rates and Efficiencies; Evolution and Lifetimes}",
      journal = {\apjs},
         year = 2009,
        month = apr,
       volume = {181},
       number = {2},
        pages = {321-350},
          doi = {10.1088/0067-0049/181/2/321},
archivePrefix = {arXiv},
       eprint = {0811.1059},
 primaryClass = {astro-ph},
       adsurl = {https://ui.adsabs.harvard.edu/abs/2009ApJS..181..321E}
}

@ARTICLE{Motte2018,
       author = {{Motte}, Fr{\'e}d{\'e}rique and {Bontemps}, Sylvain and {Louvet}, Fabien},
        title = "{High-Mass Star and Massive Cluster Formation in the Milky Way}",
      journal = {\araa},
         year = 2018,
        month = sep,
       volume = {56},
        pages = {41-82},
          doi = {10.1146/annurev-astro-091916-055235},
archivePrefix = {arXiv},
       eprint = {1706.00118},
 primaryClass = {astro-ph.GA},
       adsurl = {https://ui.adsabs.harvard.edu/abs/2018ARA&A..56...41M}
}

@ARTICLE{ZinneckerYorke2007,
       author = {{Zinnecker}, Hans and {Yorke}, Harold W.},
        title = "{Toward Understanding Massive Star Formation}",
      journal = {\araa},
         year = 2007,
        month = sep,
       volume = {45},
       number = {1},
        pages = {481-563},
          doi = {10.1146/annurev.astro.44.051905.092549},
archivePrefix = {arXiv},
       eprint = {0707.1279},
 primaryClass = {astro-ph},
       adsurl = {https://ui.adsabs.harvard.edu/abs/2007ARA&A..45..481Z}
}

@ARTICLE{Saraceno1996,
       author = {{Saraceno}, P. and {Andr{\'e}}, P. and {Ceccarelli}, C. and {Griffin}, M. and {Molinari}, S.},
        title = "{An evolutionary diagram for young stellar objects.}",
      journal = {\aap},
         year = 1996,
        month = may,
       volume = {309},
        pages = {827-839},
       adsurl = {https://ui.adsabs.harvard.edu/abs/1996A&A...309..827S}
}

@ARTICLE{Chen2025,
       author = {{Chen}, Li and {Qin}, Sheng-Li and {Liu}, Tie and {Goldsmith}, Paul F. and {Liu}, Xunchuan and {Peng}, Yaping and {Tang}, Xindi and {Garay}, Guido and {Kou}, Zhiping and {Tang}, Mengyao and {Sanhueza}, Patricio and {Li}, Zi-Yang and {Gorai}, Prasanta and {Das}, Swagat R. and {Bronfman}, Leonardo and {Dewangan}, Lokesh and {Garc{\'\i}a}, Pablo and {Li}, Shanghuo and {Lee}, Chang Won and {Liu}, Hong-Li and {T{\'o}th}, L. Viktor and {Chibueze}, James O. and {Hwang}, Jihye and {Li}, Xiaohu and {Xu}, Fengwei and {Zou}, Jiahang and {Jiao}, Wenyu and {Zhang}, Zhenying and {Zhang}, Yong},
        title = "{The ALMA-ATOMS survey: Vibrationally excited HC$_{3}$N lines in hot cores}",
      journal = {\aap},
         year = 2025,
        month = feb,
       volume = {694},
          eid = {A166},
        pages = {A166},
          doi = {10.1051/0004-6361/202452598},
archivePrefix = {arXiv},
       eprint = {2412.12546},
 primaryClass = {astro-ph.GA},
       adsurl = {https://ui.adsabs.harvard.edu/abs/2025A&A...694A.166C}
}

@ARTICLE{Wyrowski1999,
       author = {{Wyrowski}, F. and {Schilke}, P. and {Walmsley}, C.~M.},
        title = "{Vibrationally excited HC\_3N toward hot cores}",
      journal = {\aap},
         year = 1999,
        month = jan,
       volume = {341},
        pages = {882-895},
       adsurl = {https://ui.adsabs.harvard.edu/abs/1999A&A...341..882W}
}

@ARTICLE{Giannetti2014,
       author = {{Giannetti}, A. and {Wyrowski}, F. and {Brand}, J. and {Csengeri}, T. and {Fontani}, F. and {Walmsley}, C.~M. and {Nguyen Luong}, Q. and {Beuther}, H. and {Schuller}, F. and {G{\"u}sten}, R. and {Menten}, K.~M.},
        title = "{ATLASGAL-selected massive clumps in the inner Galaxy. I. CO depletion and isotopic ratios}",
      journal = {\aap},
         year = 2014,
        month = oct,
       volume = {570},
          eid = {A65},
        pages = {A65},
          doi = {10.1051/0004-6361/201423692},
archivePrefix = {arXiv},
       eprint = {1407.2215},
 primaryClass = {astro-ph.GA},
       adsurl = {https://ui.adsabs.harvard.edu/abs/2014A&A...570A..65G}
}

@ARTICLE{Schuller2009,
       author = {{Schuller}, F. and {Menten}, K.~M. and {Contreras}, Y. and {Wyrowski}, F. and {Schilke}, P. and {Bronfman}, L. and {Henning}, T. and {Walmsley}, C.~M. and {Beuther}, H. and {Bontemps}, S. and {Cesaroni}, R. and {Deharveng}, L. and {Garay}, G. and {Herpin}, F. and {Lefloch}, B. and {Linz}, H. and {Mardones}, D. and {Minier}, V. and {Molinari}, S. and {Motte}, F. and {Nyman}, L. -{\r{A}}. and {Reveret}, V. and {Risacher}, C. and {Russeil}, D. and {Schneider}, N. and {Testi}, L. and {Troost}, T. and {Vasyunina}, T. and {Wienen}, M. and {Zavagno}, A. and {Kovacs}, A. and {Kreysa}, E. and {Siringo}, G. and {Wei{\ss}}, A.},
        title = "{ATLASGAL - The APEX telescope large area survey of the galaxy at 870 {\ensuremath{\mu}}m}",
      journal = {\aap},
         year = 2009,
        month = sep,
       volume = {504},
       number = {2},
        pages = {415-427},
          doi = {10.1051/0004-6361/200811568},
archivePrefix = {arXiv},
       eprint = {0903.1369},
 primaryClass = {astro-ph.GA},
       adsurl = {https://ui.adsabs.harvard.edu/abs/2009A&A...504..415S}
}

@ARTICLE{McElwain2023,
       author = {{McElwain}, Michael W. and {Feinberg}, Lee D. and {Perrin}, Marshall D. and {Clampin}, Mark and {Mountain}, C. Matt and {Lallo}, Matthew D. and {Lajoie}, Charles-Philippe and {Kimble}, Randy A. and {Bowers}, Charles W. and {Stark}, Christopher C. and {Acton}, D. Scott and {Atkinson}, Charles and {Barinek}, Beth and {Barto}, Allison and {Basinger}, Scott and {Beck}, Tracy and {Bergkoetter}, Matthew D. and {Bluth}, Marcel and {Boucarut}, Rene A. and {Brady}, Gregory R. and {Brooks}, Keira J. and {Brown}, Bob and {Byard}, John and {Carey}, Larkin and {Carrasquilla}, Maria and {Chae}, Dan and {Chaney}, David and {Chayer}, Pierre and {Chonis}, Taylor and {Cohen}, Lester and {Cole}, Helen J. and {Comeau}, Thomas M. and {Coon}, Matthew and {Coppock}, Eric and {Coyle}, Laura and {Dean}, Bruce H. and {Dziak}, Kenneth J. and {Eisenhower}, Michael and {Flagey}, Nicolas and {Franck}, Randy and {Gallagher}, Benjamin and {Gilman}, Larry and {Glassman}, Tiffany and {Green}, Joseph J. and {Grieco}, John and {Haase}, Shari and {Hadjimichael}, Theodore J. and {Hagopian}, John G. and {Hahn}, Walter G. and {Hartig}, George F. and {Havey}, Keith A. and {Hayden}, William L. and {Hellekson}, Robert and {Hicks}, Brian and {Holfeltz}, Sherie T. and {Howard}, Joseph M. and {Huguet}, Jesse A. and {Jahne}, Brian and {Johnson}, Leslie A. and {Johnston}, John D. and {Jurling}, Alden S. and {Kegley}, Jeffrey R. and {Kennard}, Scott and {Keski-Kuha}, Ritva A. and {Knight}, J. Scott and {Kulp}, Bernard A. and {Levi}, Joshua S. and {Levine}, Marie B. and {Lightsey}, Paul and {Luetgens}, Robert A. and {Mather}, John C. and {Matthews}, Gary W. and {McKay}, Andrew G. and {Mehalick}, Kimberly I. and {Mel{\'e}ndez}, Marcio and {Mosier}, Gary E. and {Murphy}, Jess and {Nelan}, Edmund P. and {Niedner}, Malcolm B. and {Nol}, Darin M. and {Ohara}, Catherine M. and {Ohl}, Raymond G. and {Olczak}, Eugene and {Osborne}, Shannon B. and {Park}, Sang and {Perrygo}, Charles and {Pueyo}, Laurent and {Redding}, David C. and {Regan}, Michael W. and {Reynolds}, Paul and {Rifelli}, Rich and {Rigby}, Jane R. and {Sabatke}, Derek and {Saif}, Babak N. and {Scorse}, Thomas R. and {Seo}, Byoung-Joon and {Shi}, Fang and {Sigrist}, Norbert and {Smith}, Koby and {Smith}, J. Scott and {Smith}, Erin C. and {Sohn}, Sangmo Tony and {Stahl}, H. Philip and {Telfer}, Randal and {Terlecki}, Todd and {Texter}, Scott C. and {Van Buren}, David and {Van Campen}, Julie M. and {Vila}, Bego{\~n}a and {Voyton}, Mark F. and {Waldman}, Mark and {Walker}, Chanda B. and {Weiser}, Nick and {Wells}, Conrad and {West}, Garrett and {Whitman}, Tony L. and {Wolf}, Erin and {Zielinski}, Thomas P.},
        title = "{The James Webb Space Telescope Mission: Optical Telescope Element Design, Development, and Performance}",
      journal = {\pasp},
         year = 2023,
        month = may,
       volume = {135},
       number = {1047},
          eid = {058001},
        pages = {058001},
          doi = {10.1088/1538-3873/acada0},
archivePrefix = {arXiv},
       eprint = {2301.01779},
 primaryClass = {astro-ph.IM},
       adsurl = {https://ui.adsabs.harvard.edu/abs/2023PASP..135e8001M}
}

@ARTICLE{Csengeri2016,
       author = {{Csengeri}, T. and {Leurini}, S. and {Wyrowski}, F. and {Urquhart}, J.~S. and {Menten}, K.~M. and {Walmsley}, M. and {Bontemps}, S. and {Wienen}, M. and {Beuther}, H. and {Motte}, F. and {Nguyen-Luong}, Q. and {Schilke}, P. and {Schuller}, F. and {Zavagno}, A. and {Sanna}, C.},
        title = "{ATLASGAL-selected massive clumps in the inner Galaxy. II. Characterisation of different evolutionary stages and their SiO emission}",
      journal = {\aap},
         year = 2016,
        month = feb,
       volume = {586},
          eid = {A149},
        pages = {A149},
          doi = {10.1051/0004-6361/201425404},
archivePrefix = {arXiv},
       eprint = {1511.05138},
 primaryClass = {astro-ph.GA},
       adsurl = {https://ui.adsabs.harvard.edu/abs/2016A&A...586A.149C}
}

@ARTICLE{Urquhart2022,
       author = {{Urquhart}, J.~S. and {Wells}, M.~R.~A. and {Pillai}, T. and {Leurini}, S. and {Giannetti}, A. and {Moore}, T.~J.~T. and {Thompson}, M.~A. and {Figura}, C. and {Colombo}, D. and {Yang}, A.~Y. and {K{\"o}nig}, C. and {Wyrowski}, F. and {Menten}, K.~M. and {Rigby}, A.~J. and {Eden}, D.~J. and {Ragan}, S.~E.},
        title = "{ATLASGAL - evolutionary trends in high-mass star formation}",
      journal = {\mnras},
         year = 2022,
        month = mar,
       volume = {510},
       number = {3},
        pages = {3389-3407},
          doi = {10.1093/mnras/stab3511},
archivePrefix = {arXiv},
       eprint = {2111.12816},
 primaryClass = {astro-ph.GA},
       adsurl = {https://ui.adsabs.harvard.edu/abs/2022MNRAS.510.3389U}
}

@ARTICLE{Menten1986,
       author = {{Menten}, K.~M. and {Walmsley}, C.~M. and {Henkel}, C. and {Wilson}, T.~L. and {Snyder}, L.~E. and {Hollis}, J.~M. and {Lovas}, F.~J.},
        title = "{Torsionally excited methanol in hot molecular cloud cores.}",
      journal = {\aap},
         year = 1986,
        month = nov,
       volume = {169},
        pages = {271-280},
       adsurl = {https://ui.adsabs.harvard.edu/abs/1986A&A...169..271M}
}

@INCOLLECTION{Bontemps1996,
       author = {{Bontemps}, S. and {Andr{\'e}}, P. and {Terebey}, S. and {Cabrit}, S.},
        title = "{Evolution of Outflow Activity Around Low Mass Embedded Young Stellar Objects}",
    booktitle = {Disks and Outflows Around Young Stars},
         year = 1996,
       editor = {{Beckwith}, Steven and {Staude}, Jakob and {Quetz}, Axel and {Natta}, Antonella},
       volume = {465},
        pages = {270},
        publisher = {Springer-Verlag Berlin Heidelberg New York},
          doi = {10.1007/BFb0102645},
       adsurl = {https://ui.adsabs.harvard.edu/abs/1996LNP...465..270B}
}

@INPROCEEDINGS{Andre2000,
       author = {{Andr{\'e}}, P. and {Ward-Thompson}, D. and {Barsony}, M.},
        title = "{From Prestellar Cores to Protostars: the Initial Conditions of Star Formation}",
    booktitle = {Protostars and Planets IV},
         year = 2000,
       editor = {{Mannings}, V. and {Boss}, A.~P. and {Russell}, S.~S.},
        month = may,
        pages = {59},
          doi = {10.48550/arXiv.astro-ph/9903284},
archivePrefix = {arXiv},
       eprint = {astro-ph/9903284},
 primaryClass = {astro-ph},
       adsurl = {https://ui.adsabs.harvard.edu/abs/2000prpl.conf...59A}
}

@ARTICLE{Andre2008,
       author = {{Andr{\'e}}, Ph. and {Minier}, V. and {Gallais}, P. and {Rev{\'e}ret}, V. and {Le Pennec}, J. and {Rodriguez}, L. and {Boulade}, O. and {Doumayrou}, E. and {Dubreuil}, D. and {Lortholary}, M. and {Martignac}, J. and {Talvard}, M. and {De Breuck}, C. and {Hamon}, G. and {Schneider}, N. and {Bontemps}, S. and {Lagage}, P.~O. and {Pantin}, E. and {Roussel}, H. and {Miller}, M. and {Purcell}, C.~R. and {Hill}, T. and {Stutzki}, J.},
        title = "{First 450 {\ensuremath{\mu}}m dust continuum mapping of the massive star-forming region NGC 3576 with the P-ArT{\'e}MiS bolometer camera}",
      journal = {\aap},
         year = 2008,
        month = nov,
       volume = {490},
       number = {3},
        pages = {L27-L30},
          doi = {10.1051/0004-6361:200810957},
archivePrefix = {arXiv},
       eprint = {0809.3968},
 primaryClass = {astro-ph},
       adsurl = {https://ui.adsabs.harvard.edu/abs/2008A&A...490L..27A}
}

@ARTICLE{Goldsmith1982,
       author = {{Goldsmith}, P.~F. and {Snell}, R.~L. and {Deguchi}, S. and {Krotkov}, R. and {Linke}, R.~A.},
        title = "{Vibrationally excited cyanoacetylene in the Orion molecular cloud.}",
      journal = {\apj},
         year = 1982,
        month = sep,
       volume = {260},
        pages = {147-158},
          doi = {10.1086/160242},
       adsurl = {https://ui.adsabs.harvard.edu/abs/1982ApJ...260..147G}
}

@ARTICLE{Goldsmith1983,
       author = {{Goldsmith}, P.~F. and {Krotkov}, R. and {Snell}, R.~L. and {Brown}, R.~D. and {Godfrey}, P.},
        title = "{Vibrationally excited CH3CN and HC3N in Orion.}",
      journal = {\apj},
         year = 1983,
        month = nov,
       volume = {274},
        pages = {184-194},
          doi = {10.1086/161436},
       adsurl = {https://ui.adsabs.harvard.edu/abs/1983ApJ...274..184G}
}

@ARTICLE{Sobolev1994,
       author = {{Sobolev}, A.~M. and {Deguchi}, S.},
        title = "{Pumping of Class II methanol masers. I. The 2\_0\_-3\_-1\_E transition.}",
      journal = {\aap},
         year = 1994,
        month = nov,
       volume = {291},
        pages = {569-576},
       adsurl = {https://ui.adsabs.harvard.edu/abs/1994A&A...291..569S}
}

@ARTICLE{Wienen2021,
       author = {{Wienen}, M. and {Wyrowski}, F. and {Walmsley}, C.~M. and {Csengeri}, T. and {Pillai}, T. and {Giannetti}, A. and {Menten}, K.~M.},
        title = "{ATLASGAL-selected massive clumps in the inner Galaxy. IX. Deuteration of ammonia}",
      journal = {\aap},
         year = 2021,
        month = may,
       volume = {649},
          eid = {A21},
        pages = {A21},
          doi = {10.1051/0004-6361/201731208},
archivePrefix = {arXiv},
       eprint = {2102.04478},
 primaryClass = {astro-ph.GA},
       adsurl = {https://ui.adsabs.harvard.edu/abs/2021A&A...649A..21W}
}

@ARTICLE{Price2001,
       author = {{Price}, Stephan D. and {Egan}, Michael P. and {Carey}, Sean J. and {Mizuno}, Donald R. and {Kuchar}, Thomas A.},
        title = "{Midcourse Space Experiment Survey of the Galactic Plane}",
      journal = {\aj},
         year = 2001,
        month = may,
       volume = {121},
       number = {5},
        pages = {2819-2842},
          doi = {10.1086/320404},
       adsurl = {https://ui.adsabs.harvard.edu/abs/2001AJ....121.2819P}
}

@ARTICLE{Hoare2012,
       author = {{Hoare}, M.~G. and {Purcell}, C.~R. and {Churchwell}, E.~B. and {Diamond}, P. and {Cotton}, W.~D. and {Chandler}, C.~J. and {Smethurst}, S. and {Kurtz}, S.~E. and {Mundy}, L.~G. and {Dougherty}, S.~M. and {Fender}, R.~P. and {Fuller}, G.~A. and {Jackson}, J.~M. and {Garrington}, S.~T. and {Gledhill}, T.~R. and {Goldsmith}, P.~F. and {Lumsden}, S.~L. and {Mart{\'\i}}, J. and {Moore}, T.~J.~T. and {Muxlow}, T.~W.~B. and {Oudmaijer}, R.~D. and {Pandian}, J.~D. and {Paredes}, J.~M. and {Shepherd}, D.~S. and {Spencer}, R.~E. and {Thompson}, M.~A. and {Umana}, G. and {Urquhart}, J.~S. and {Zijlstra}, A.~A.},
        title = "{The Coordinated Radio and Infrared Survey for High-Mass Star Formation (The CORNISH Survey). I. Survey Design}",
      journal = {\pasp},
         year = 2012,
        month = sep,
       volume = {124},
       number = {919},
        pages = {939},
          doi = {10.1086/668058},
archivePrefix = {arXiv},
       eprint = {1208.3351},
 primaryClass = {astro-ph.GA},
       adsurl = {https://ui.adsabs.harvard.edu/abs/2012PASP..124..939H}
}

@ARTICLE{Purcell2013,
       author = {{Purcell}, C.~R. and {Hoare}, M.~G. and {Cotton}, W.~D. and {Lumsden}, S.~L. and {Urquhart}, J.~S. and {Chandler}, C. and {Churchwell}, E.~B. and {Diamond}, P. and {Dougherty}, S.~M. and {Fender}, R.~P. and {Fuller}, G. and {Garrington}, S.~T. and {Gledhill}, T.~M. and {Goldsmith}, P.~F. and {Hindson}, L. and {Jackson}, J.~M. and {Kurtz}, S.~E. and {Mart{\'\i}}, J. and {Moore}, T.~J.~T. and {Mundy}, L.~G. and {Muxlow}, T.~W.~B. and {Oudmaijer}, R.~D. and {Pandian}, J.~D. and {Paredes}, J.~M. and {Shepherd}, D.~S. and {Smethurst}, S. and {Spencer}, R.~E. and {Thompson}, M.~A. and {Umana}, G. and {Zijlstra}, A.~A.},
        title = "{The Coordinated Radio and Infrared Survey for High-mass Star Formation. II. Source Catalog}",
      journal = {\apjs},
         year = 2013,
        month = mar,
       volume = {205},
       number = {1},
          eid = {1},
        pages = {1},
          doi = {10.1088/0067-0049/205/1/1},
archivePrefix = {arXiv},
       eprint = {1211.7116},
 primaryClass = {astro-ph.GA},
       adsurl = {https://ui.adsabs.harvard.edu/abs/2013ApJS..205....1P}
}

@ARTICLE{Urquhart2009,
       author = {{Urquhart}, J.~S. and {Hoare}, M.~G. and {Purcell}, C.~R. and {Lumsden}, S.~L. and {Oudmaijer}, R.~D. and {Moore}, T.~J.~T. and {Busfield}, A.~L. and {Mottram}, J.~C. and {Davies}, B.},
        title = "{The RMS survey. 6 cm continuum VLA observations towards candidate massive YSOs in the northern hemisphere}",
      journal = {\aap},
         year = 2009,
        month = jul,
       volume = {501},
       number = {2},
        pages = {539-551},
          doi = {10.1051/0004-6361/200912108},
archivePrefix = {arXiv},
       eprint = {0905.1174},
 primaryClass = {astro-ph.GA},
       adsurl = {https://ui.adsabs.harvard.edu/abs/2009A&A...501..539U}
}

@ARTICLE{Walsh1999,
       author = {{Walsh}, A.~J. and {Burton}, M.~G. and {Hyland}, A.~R. and {Robinson}, G.},
        title = "{Studies of ultracompact HII regions - III. Near-infrared survey of selected regions}",
      journal = {\mnras},
         year = 1999,
        month = nov,
       volume = {309},
       number = {4},
        pages = {905-922},
          doi = {10.1046/j.1365-8711.1999.02890.x},
       adsurl = {https://ui.adsabs.harvard.edu/abs/1999MNRAS.309..905W}
}

@ARTICLE{Urquhart2007,
       author = {{Urquhart}, J.~S. and {Busfield}, A.~L. and {Hoare}, M.~G. and {Lumsden}, S.~L. and {Clarke}, A.~J. and {Moore}, T.~J.~T. and {Mottram}, J.~C. and {Oudmaijer}, R.~D.},
        title = "{The RMS survey. Radio observations of candidate massive YSOs in the southern hemisphere}",
      journal = {\aap},
         year = 2007,
        month = jan,
       volume = {461},
       number = {1},
        pages = {11-23},
          doi = {10.1051/0004-6361:20065837},
archivePrefix = {arXiv},
       eprint = {astro-ph/0605738},
 primaryClass = {astro-ph},
       adsurl = {https://ui.adsabs.harvard.edu/abs/2007A&A...461...11U}
}

@ARTICLE{Beltran2014,
       author = {{Beltr{\'a}n}, M.~T. and {S{\'a}nchez-Monge}, {\'A}. and {Cesaroni}, R. and {Kumar}, M.~S.~N. and {Galli}, D. and {Walmsley}, C.~M. and {Etoka}, S. and {Furuya}, R.~S. and {Moscadelli}, L. and {Stanke}, T. and {van der Tak}, F.~F.~S. and {Vig}, S. and {Wang}, K.-S. and {Zinnecker}, H. and {Elia}, D. and {Schisano}, E.},
        title = "{Filamentary structure and Keplerian rotation in the high-mass star-forming region G35.03+0.35 imaged with ALMA}",
      journal = {\aap},
         year = 2014,
        month = nov,
       volume = {571},
          eid = {A52},
        pages = {A52},
          doi = {10.1051/0004-6361/201424031},
       adsurl = {https://ui.adsabs.harvard.edu/abs/2014A&A...571A..52B}
}

@ARTICLE{Motte2007,
       author = {{Motte}, F. and {Bontemps}, S. and {Schilke}, P. and {Schneider}, N. and {Menten}, K.~M. and {Brogui{\`e}re}, D.},
        title = "{The earliest phases of high-mass star formation: a 3 square degree millimeter continuum mapping of Cygnus X}",
      journal = {\aap},
         year = 2007,
        month = dec,
       volume = {476},
       number = {3},
        pages = {1243-1260},
          doi = {10.1051/0004-6361:20077843},
archivePrefix = {arXiv},
       eprint = {0708.2774},
 primaryClass = {astro-ph},
       adsurl = {https://ui.adsabs.harvard.edu/abs/2007A&A...476.1243M}
}

@ARTICLE{Russeil2010,
       author = {{Russeil}, D. and {Zavagno}, A. and {Motte}, F. and {Schneider}, N. and {Bontemps}, S. and {Walsh}, A.~J.},
        title = "{The earliest phases of high-mass star formation: the NGC 6334-NGC 6357 complex}",
      journal = {\aap},
         year = 2010,
        month = jun,
       volume = {515},
          eid = {A55},
        pages = {A55},
          doi = {10.1051/0004-6361/200913632},
       adsurl = {https://ui.adsabs.harvard.edu/abs/2010A&A...515A..55R}
}

@ARTICLE{CASA2022,
       author = {{CASA Team} and {Bean}, Ben and {Bhatnagar}, Sanjay and {Castro}, Sandra and {Donovan Meyer}, Jennifer and {Emonts}, Bjorn and {Garcia}, Enrique and {Garwood}, Robert and {Golap}, Kumar and {Gonzalez Villalba}, Justo and {Harris}, Pamela and {Hayashi}, Yohei and {Hoskins}, Josh and {Hsieh}, Mingyu and {Jagannathan}, Preshanth and {Kawasaki}, Wataru and {Keimpema}, Aard and {Kettenis}, Mark and {Lopez}, Jorge and {Marvil}, Joshua and {Masters}, Joseph and {McNichols}, Andrew and {Mehringer}, David and {Miel}, Renaud and {Moellenbrock}, George and {Montesino}, Federico and {Nakazato}, Takeshi and {Ott}, Juergen and {Petry}, Dirk and {Pokorny}, Martin and {Raba}, Ryan and {Rau}, Urvashi and {Schiebel}, Darrell and {Schweighart}, Neal and {Sekhar}, Srikrishna and {Shimada}, Kazuhiko and {Small}, Des and {Steeb}, Jan-Willem and {Sugimoto}, Kanako and {Suoranta}, Ville and {Tsutsumi}, Takahiro and {van Bemmel}, Ilse M. and {Verkouter}, Marjolein and {Wells}, Akeem and {Xiong}, Wei and {Szomoru}, Arpad and {Griffith}, Morgan and {Glendenning}, Brian and {Kern}, Jeff},
        title = "{CASA, the Common Astronomy Software Applications for Radio Astronomy}",
      journal = {\pasp},
         year = 2022,
        month = nov,
       volume = {134},
       number = {1041},
          eid = {114501},
        pages = {114501},
          doi = {10.1088/1538-3873/ac9642},
archivePrefix = {arXiv},
       eprint = {2210.02276},
 primaryClass = {astro-ph.IM},
       adsurl = {https://ui.adsabs.harvard.edu/abs/2022PASP..134k4501C}
}

@ARTICLE{Hunter2023,
       author = {{Hunter}, Todd R. and {Indebetouw}, Remy and {Brogan}, Crystal L. and {Berry}, Kristin and {Chang}, Chin-Shin and {Francke}, Harold and {Geers}, Vincent C. and {G{\'o}mez}, Laura and {Hibbard}, John E. and {Humphreys}, Elizabeth M. and {Kent}, Brian R. and {Kepley}, Amanda A. and {Kunneriath}, Devaky and {Lipnicky}, Andrew and {Loomis}, Ryan A. and {Mason}, Brian S. and {Masters}, Joseph S. and {Maud}, Luke T. and {Muders}, Dirk and {Sabater}, Jose and {Sugimoto}, Kanako and {Sz{\H{u}}cs}, L{\'a}szl{\'o} and {Vasiliev}, Eugene and {Videla}, Liza and {Villard}, Eric and {Williams}, Stewart J. and {Xue}, Rui and {Yoon}, Ilsang},
        title = "{The ALMA Interferometric Pipeline Heuristics}",
      journal = {\pasp},
         year = 2023,
        month = jul,
       volume = {135},
       number = {1049},
          eid = {074501},
        pages = {074501},
          doi = {10.1088/1538-3873/ace216},
archivePrefix = {arXiv},
       eprint = {2306.07420},
 primaryClass = {astro-ph.IM},
       adsurl = {https://ui.adsabs.harvard.edu/abs/2023PASP..135g4501H}
}

@ARTICLE{Francis2020,
       author = {{Francis}, Logan and {Johnstone}, Doug and {Herczeg}, Gregory and {Hunter}, Todd R. and {Harsono}, Daniel},
        title = "{On the Accuracy of the ALMA Flux Calibration in the Time Domain and across Spectral Windows}",
      journal = {\aj},
         year = 2020,
        month = dec,
       volume = {160},
       number = {6},
          eid = {270},
        pages = {270},
          doi = {10.3847/1538-3881/abbe1a},
archivePrefix = {arXiv},
       eprint = {2010.02186},
 primaryClass = {astro-ph.IM},
       adsurl = {https://ui.adsabs.harvard.edu/abs/2020AJ....160..270F}
}

@MISC{Remijan2019,
       author = {{Remijan}, A. and {Biggs}, A. and {Cortes}, P.~A. and {Dent}, B. and {Di Franceso}, J. and {Fomalont}, E. and {Hales}, A. and {Kameno}, S. and {Mason}, B. and {Philips}, N. and {Saini}, K. and {Vila Vilaro}, B. and {Villard}, E.},
        title = "{ALMA Technical Handbook,ALMA Doc. 7.3, ver. 1.1}",
 howpublished = {2019, ALMA Technical Handbook,ALMA Doc. 7.3, ver. 1.1ISBN 978-3-923524-66-2},
         year = 2019,
        month = jun,
          doi = {10.5281/zenodo.4511522},
       adsurl = {https://ui.adsabs.harvard.edu/abs/2019athb.rept.....R}
}

@manual{Braatz2020,
  author       = {Braatz, J.},
  title        = {{ALMA Cycle 8 Proposer's Guide}},
  year         = {2020},
  organization = {ALMA Observatory},
  type         = {Doc. 8.2 v1.0},
  url          = {https://almascience.nrao.edu/documents-and-tools/cycle8/alma-proposers-guide}
}

@ARTICLE{Nayak2024,
       author = {{Nayak}, Omnarayani and {Hirschauer}, Alec S. and {Kavanagh}, Patrick J. and {Meixner}, Margaret and {Chu}, Laurie and {Habel}, Nolan and {Jones}, Olivia C. and {Lenki{\'c}}, Laura and {Nally}, Conor and {Reiter}, Megan and {Robberto}, Massimo and {Sargent}, B.~A.},
        title = "{JWST Mid-infrared Spectroscopy Resolves Gas, Dust, and Ice in Young Stellar Objects in the Large Magellanic Cloud}",
      journal = {\apj},
         year = 2024,
        month = mar,
       volume = {963},
       number = {2},
          eid = {94},
        pages = {94},
          doi = {10.3847/1538-4357/ad18bc},
       adsurl = {https://ui.adsabs.harvard.edu/abs/2024ApJ...963...94N}
}

@ARTICLE{Elia2010,
       author = {{Elia}, D. and {Schisano}, E. and {Molinari}, S. and {Robitaille}, T. and {Angl{\'e}s-Alc{\'a}zar}, D. and {Bally}, J. and {Battersby}, C. and {Benedettini}, M. and {Billot}, N. and {Calzoletti}, L. and {di Giorgio}, A.~M. and {Faustini}, F. and {Li}, J.~Z. and {Martin}, P. and {Morgan}, L. and {Motte}, F. and {Mottram}, J.~C. and {Natoli}, P. and {Olmi}, L. and {Paladini}, R. and {Piacentini}, F. and {Pestalozzi}, M. and {Pezzuto}, S. and {Polychroni}, D. and {Smith}, M.~D. and {Strafella}, F. and {Stringfellow}, G.~S. and {Testi}, L. and {Thompson}, M.~A. and {Traficante}, A. and {Veneziani}, M.},
        title = "{A Herschel study of YSO evolutionary stages and formation timelines in two fields of the Hi-GAL survey}",
      journal = {\aap},
         year = 2010,
        month = jul,
       volume = {518},
          eid = {L97},
        pages = {L97},
          doi = {10.1051/0004-6361/201014651},
archivePrefix = {arXiv},
       eprint = {1005.1783},
 primaryClass = {astro-ph.GA},
       adsurl = {https://ui.adsabs.harvard.edu/abs/2010A&A...518L..97E}
}

@ARTICLE{Ma2013,
       author = {{Ma}, Bo and {Tan}, Jonathan C. and {Barnes}, Peter J.},
        title = "{The Galactic Census of High- and Medium-mass Protostars. II. Luminosities and Evolutionary States of a Complete Sample of Dense Gas Clumps}",
      journal = {\apj},
         year = 2013,
        month = dec,
       volume = {779},
       number = {1},
          eid = {79},
        pages = {79},
          doi = {10.1088/0004-637X/779/1/79},
archivePrefix = {arXiv},
       eprint = {1211.6492},
 primaryClass = {astro-ph.GA},
       adsurl = {https://ui.adsabs.harvard.edu/abs/2013ApJ...779...79M}
}

@misc{Carey2007,
  author = {S. J. Carey and others},
  title = {MIPSGAL v2.0 Data Delivery Description Document (16 October 2007)},
  year = {2007},
  url = {http://mipsgal.ipac.caltech.edu/docs/mipsgal_delivery_guide_v2_16oct07.pdf},
  note = {Accessed: 2025-11-28}
}

@ARTICLE{Urquhart2014,
       author = {{Urquhart}, J.~S. and {Csengeri}, T. and {Wyrowski}, F. and {Schuller}, F. and {Bontemps}, S. and {Bronfman}, L. and {Menten}, K.~M. and {Walmsley}, C.~M. and {Contreras}, Y. and {Beuther}, H. and {Wienen}, M. and {Linz}, H.},
        title = "{ATLASGAL - Complete compact source catalogue: 280{\textdegree}<{\ensuremath{\ell}}< 60{\textdegree}}",
      journal = {\aap},
         year = 2014,
        month = aug,
       volume = {568},
          eid = {A41},
        pages = {A41},
          doi = {10.1051/0004-6361/201424126},
archivePrefix = {arXiv},
       eprint = {1406.5741},
 primaryClass = {astro-ph.GA},
       adsurl = {https://ui.adsabs.harvard.edu/abs/2014A&A...568A..41U}
}

@ARTICLE{Coletta2025,
       author = {{Coletta}, A. and {Molinari}, S. and {Schisano}, E. and {Traficante}, A. and {Elia}, D. and {Benedettini}, M. and {Mininni}, C. and {Soler}, J.~D. and {S{\'a}nchez-Monge}, {\'A}. and {Schilke}, P. and {Battersby}, C. and {Fuller}, G.~A. and {Beuther}, H. and {Zhang}, Q. and {Beltr{\'a}n}, M.~T. and {Jones}, B. and {Klessen}, R.~S. and {Walch}, S. and {Fontani}, F. and {Avison}, A. and {Brogan}, C.~L. and {Clarke}, S.~D. and {Hatchfield}, P. and {Hennebelle}, P. and {Ho}, P.~T.~P. and {Hunter}, T.~R. and {Johnston}, K.~G. and {Klaassen}, P.~D. and {Koch}, P.~M. and {Kuiper}, R. and {Lis}, D.~C. and {Liu}, T. and {Lumsden}, S.~L. and {Maruccia}, Y. and {M{\"o}ller}, T. and {Moscadelli}, L. and {Nucara}, A. and {Rigby}, A.~J. and {Rygl}, K.~L.~J. and {Sanhueza}, P. and {van der Tak}, F. and {Wells}, M.~R.~A. and {Wyrowski}, F. and {De Angelis}, F. and {Liu}, S. and {Ahmadi}, A. and {Bronfman}, L. and {Liu}, S.-Y. and {Su}, Y.-N. and {Tang}, Y. and {Testi}, L. and {Zinnecker}, H.},
        title = "{ALMAGAL: III. Compact source catalog: Fragmentation statistics and physical evolution of the core population}",
      journal = {\aap},
         year = 2025,
        month = apr,
       volume = {696},
          eid = {A151},
        pages = {A151},
          doi = {10.1051/0004-6361/202452706},
archivePrefix = {arXiv},
       eprint = {2503.05663},
 primaryClass = {astro-ph.GA},
       adsurl = {https://ui.adsabs.harvard.edu/abs/2025A&A...696A.151C}
}

@ARTICLE{Urquhart2013a,
       author = {{Urquhart}, J.~S. and {Thompson}, M.~A. and {Moore}, T.~J.~T. and {Purcell}, C.~R. and {Hoare}, M.~G. and {Schuller}, F. and {Wyrowski}, F. and {Csengeri}, T. and {Menten}, K.~M. and {Lumsden}, S.~L. and {Kurtz}, S. and {Walmsley}, C.~M. and {Bronfman}, L. and {Morgan}, L.~K. and {Eden}, D.~J. and {Russeil}, D.},
        title = "{ATLASGAL - properties of compact H II regions and their natal clumps}",
      journal = {\mnras},
         year = 2013,
        month = oct,
       volume = {435},
       number = {1},
        pages = {400-428},
          doi = {10.1093/mnras/stt1310},
archivePrefix = {arXiv},
       eprint = {1307.4105},
 primaryClass = {astro-ph.GA},
       adsurl = {https://ui.adsabs.harvard.edu/abs/2013MNRAS.435..400U}
}

@ARTICLE{Pilbratt2010,
       author = {{Pilbratt}, G.~L. and {Riedinger}, J.~R. and {Passvogel}, T. and {Crone}, G. and {Doyle}, D. and {Gageur}, U. and {Heras}, A.~M. and {Jewell}, C. and {Metcalfe}, L. and {Ott}, S. and {Schmidt}, M.},
        title = "{Herschel Space Observatory. An ESA facility for far-infrared and submillimetre astronomy}",
      journal = {\aap},
         year = 2010,
        month = jul,
       volume = {518},
          eid = {L1},
        pages = {L1},
          doi = {10.1051/0004-6361/201014759},
archivePrefix = {arXiv},
       eprint = {1005.5331},
 primaryClass = {astro-ph.IM},
       adsurl = {https://ui.adsabs.harvard.edu/abs/2010A&A...518L...1P}
}

@ARTICLE{Motte2022,
       author = {{Motte}, F. and {Bontemps}, S. and {Csengeri}, T. and {Pouteau}, Y. and {Louvet}, F. and {Stutz}, A.~M. and {Cunningham}, N. and {L{\'o}pez-Sepulcre}, A. and {Brouillet}, N. and {Galv{\'a}n-Madrid}, R. and {Ginsburg}, A. and {Maud}, L. and {Men'shchikov}, A. and {Nakamura}, F. and {Nony}, T. and {Sanhueza}, P. and {{\'A}lvarez-Guti{\'e}rrez}, R.~H. and {Armante}, M. and {Baug}, T. and {Bonfand}, M. and {Busquet}, G. and {Chapillon}, E. and {D{\'\i}az-Gonz{\'a}lez}, D. and {Fern{\'a}ndez-L{\'o}pez}, M. and {Guzm{\'a}n}, A.~E. and {Herpin}, F. and {Liu}, H.-L. and {Olguin}, F. and {Towner}, A.~P.~M. and {Bally}, J. and {Battersby}, C. and {Braine}, J. and {Bronfman}, L. and {Chen}, H.-R.~V. and {Dell'Ova}, P. and {Di Francesco}, J. and {Gonz{\'a}lez}, M. and {Gusdorf}, A. and {Hennebelle}, P. and {Izumi}, N. and {Joncour}, I. and {Lee}, Y.-N. and {Lefloch}, B. and {Lesaffre}, P. and {Lu}, X. and {Menten}, K.~M. and {Mignon-Risse}, R. and {Molet}, J. and {Moraux}, E. and {Mundy}, L. and {Nguyen Luong}, Q. and {Reyes}, N. and {Reyes Reyes}, S.~D. and {Robitaille}, J.-F. and {Rosolowsky}, E. and {Sandoval-Garrido}, N.~A. and {Schuller}, F. and {Svoboda}, B. and {Tatematsu}, K. and {Thomasson}, B. and {Walker}, D. and {Wu}, B. and {Whitworth}, A.~P. and {Wyrowski}, F.},
        title = "{ALMA-IMF. I. Investigating the origin of stellar masses: Introduction to the Large Program and first results}",
      journal = {\aap},
         year = 2022,
        month = jun,
       volume = {662},
          eid = {A8},
        pages = {A8},
          doi = {10.1051/0004-6361/202141677},
archivePrefix = {arXiv},
       eprint = {2112.08182},
 primaryClass = {astro-ph.GA},
       adsurl = {https://ui.adsabs.harvard.edu/abs/2022A&A...662A...8M}
}

@ARTICLE{Wright2010,
       author = {{Wright}, Edward L. and {Eisenhardt}, Peter R.~M. and {Mainzer}, Amy K. and {Ressler}, Michael E. and {Cutri}, Roc M. and {Jarrett}, Thomas and {Kirkpatrick}, J. Davy and {Padgett}, Deborah and {McMillan}, Robert S. and {Skrutskie}, Michael and {Stanford}, S.~A. and {Cohen}, Martin and {Walker}, Russell G. and {Mather}, John C. and {Leisawitz}, David and {Gautier}, III, Thomas N. and {McLean}, Ian and {Benford}, Dominic and {Lonsdale}, Carol J. and {Blain}, Andrew and {Mendez}, Bryan and {Irace}, William R. and {Duval}, Valerie and {Liu}, Fengchuan and {Royer}, Don and {Heinrichsen}, Ingolf and {Howard}, Joan and {Shannon}, Mark and {Kendall}, Martha and {Walsh}, Amy L. and {Larsen}, Mark and {Cardon}, Joel G. and {Schick}, Scott and {Schwalm}, Mark and {Abid}, Mohamed and {Fabinsky}, Beth and {Naes}, Larry and {Tsai}, Chao-Wei},
        title = "{The Wide-field Infrared Survey Explorer (WISE): Mission Description and Initial On-orbit Performance}",
      journal = {\aj},
         year = 2010,
        month = dec,
       volume = {140},
       number = {6},
        pages = {1868-1881},
          doi = {10.1088/0004-6256/140/6/1868},
archivePrefix = {arXiv},
       eprint = {1008.0031},
 primaryClass = {astro-ph.IM},
       adsurl = {https://ui.adsabs.harvard.edu/abs/2010AJ....140.1868W}
}

@ARTICLE{Gusten2006,
       author = {{G{\"u}sten}, R. and {Nyman}, L. {\r{A}}. and {Schilke}, P. and {Menten}, K. and {Cesarsky}, C. and {Booth}, R.},
        title = "{The Atacama Pathfinder EXperiment (APEX) - a new submillimeter facility for southern skies -}",
      journal = {\aap},
         year = 2006,
        month = aug,
       volume = {454},
       number = {2},
        pages = {L13-L16},
          doi = {10.1051/0004-6361:20065420},
       adsurl = {https://ui.adsabs.harvard.edu/abs/2006A&A...454L..13G}
}

@ARTICLE{Contreras2013,
       author = {{Contreras}, Y. and {Schuller}, F. and {Urquhart}, J.~S. and {Csengeri}, T. and {Wyrowski}, F. and {Beuther}, H. and {Bontemps}, S. and {Bronfman}, L. and {Henning}, T. and {Menten}, K.~M. and {Schilke}, P. and {Walmsley}, C.~M. and {Wienen}, M. and {Tackenberg}, J. and {Linz}, H.},
        title = "{ATLASGAL - compact source catalogue: 330{\textdegree} < {\ensuremath{\ell}} < 21{\textdegree}}",
      journal = {\aap},
         year = 2013,
        month = jan,
       volume = {549},
          eid = {A45},
        pages = {A45},
          doi = {10.1051/0004-6361/201220155},
archivePrefix = {arXiv},
       eprint = {1211.0741},
 primaryClass = {astro-ph.GA},
       adsurl = {https://ui.adsabs.harvard.edu/abs/2013A&A...549A..45C}
}

@ARTICLE{Csengeri2014,
       author = {{Csengeri}, T. and {Urquhart}, J.~S. and {Schuller}, F. and {Motte}, F. and {Bontemps}, S. and {Wyrowski}, F. and {Menten}, K.~M. and {Bronfman}, L. and {Beuther}, H. and {Henning}, Th. and {Testi}, L. and {Zavagno}, A. and {Walmsley}, M.},
        title = "{The ATLASGAL survey: a catalog of dust condensations in the Galactic plane}",
      journal = {\aap},
         year = 2014,
        month = may,
       volume = {565},
          eid = {A75},
        pages = {A75},
          doi = {10.1051/0004-6361/201322434},
archivePrefix = {arXiv},
       eprint = {1312.0937},
 primaryClass = {astro-ph.GA},
       adsurl = {https://ui.adsabs.harvard.edu/abs/2014A&A...565A..75C}
}

@misc{Cutri2012,
       author = {{Cutri}, R.~M. and {Wright}, E.~L. and {Conrow}, T. and {Bauer}, J. and {Benford}, D. and {Brandenburg}, H. and {Dailey}, J. and {Eisenhardt}, P.~R.~M. and {Evans}, T. and {Fajardo-Acosta}, S. and {Fowler}, J. and {Gelino}, C. and {Grillmair}, C. and {Harbut}, M. and {Hoffman}, D. and {Jarrett}, T. and {Kirkpatrick}, J.~D. and {Leisawitz}, D. and {Liu}, W. and {Mainzer}, A. and {Marsh}, K. and {Masci}, F. and {McCallon}, H. and {Padgett}, D. and {Ressler}, M.~E. and {Royer}, D. and {Skrutskie}, M.~F. and {Stanford}, S.~A. and {Wyatt}, P.~L. and {Tholen}, D. and {Tsai}, C.~W. and {Wachter}, S. and {Wheelock}, S.~L. and {Yan}, L. and {Alles}, R. and {Beck}, R. and {Grav}, T. and {Masiero}, J. and {McCollum}, B. and {McGehee}, P. and {Papin}, M. and {Wittman}, M.},
        title = "{Explanatory Supplement to the WISE All-Sky Data Release Products}",
 howpublished = {Explanatory Supplement to the WISE All-Sky Data Release Products},
         year = 2012,
        month = mar,
        pages = {1},
       adsurl = {https://ui.adsabs.harvard.edu/abs/2012wise.rept....1C}
}

@misc{Salvatier2016,
       author = {{Salvatier}, John and {Wiecki}, Thomas V. and {Fonnesbeck}, Christopher},
        title = "{PyMC3: Python probabilistic programming framework}",
 howpublished = {Astrophysics Source Code Library, record ascl:1610.016},
         year = 2016,
        month = oct,
          eid = {ascl:1610.016},
archivePrefix = {ascl},
       eprint = {1610.016},
       adsurl = {https://ui.adsabs.harvard.edu/abs/2016ascl.soft10016S}
}

@ARTICLE{Billington2019,
       author = {{Billington}, S.~J. and {Urquhart}, J.~S. and {K{\"o}nig}, C. and {Moore}, T.~J.~T. and {Eden}, D.~J. and {Breen}, S.~L. and {Kim}, W.-J. and {Thompson}, M.~A. and {Ellingsen}, S.~P. and {Menten}, K.~M. and {Wyrowski}, F. and {Leurini}, S.},
        title = "{ATLASGAL - physical parameters of dust clumps associated with 6.7 GHz methanol masers}",
      journal = {\mnras},
         year = 2019,
        month = dec,
       volume = {490},
       number = {2},
        pages = {2779-2798},
          doi = {10.1093/mnras/stz2691},
archivePrefix = {arXiv},
       eprint = {1907.00564},
 primaryClass = {astro-ph.GA},
       adsurl = {https://ui.adsabs.harvard.edu/abs/2019MNRAS.490.2779B}
}

@ARTICLE{Hunter2007,
       author = {{Hunter}, John D.},
        title = "{Matplotlib: A 2D Graphics Environment}",
      journal = {Computing in Science and Engineering},
         year = 2007,
        month = jan,
       volume = {9},
       number = {3},
        pages = {90-95},
          doi = {10.1109/MCSE.2007.55},
       adsurl = {https://ui.adsabs.harvard.edu/abs/2007CSE.....9...90H}
}

@misc{Virtanen2020,
       author = {{Virtanen}, Pauli and {Gommers}, Ralf and {Burovski}, Evgeni and {Oliphant}, Travis E. and {Weckesser}, Warren and {Cournapeau}, David and {Alexbrc} and {Peterson}, Pearu and {Reddy}, Tyler and {Haberland}, Matt and {Wilson}, Josh and {Nelson}, Andrew and {Endolith} and {Mayorov}, Nikolay and {Van Der Walt}, Stefan and {Laxalde}, Denis and {Polat}, Ilhan and {Brett}, Matthew and {Larson}, Eric and {Millman}, Jarrod and {Lars} and {Van Mulbregt}, Paul and {Eric-Jones} and {Carey}, CJ and {Moore}, Eric and {Kern}, Robert and {Leslie}, Tim and {Perktold}, Josef and {Striega}, Kai and {Feng}, Yu},
        title = "{scipy/scipy: SciPy 1.6.0}",
         year = 2020,
        month = dec,
          eid = {10.5281/zenodo.4406806},
          doi = {10.5281/zenodo.4406806},
      version = {v1.6.0},
    publisher = {Zenodo},
       adsurl = {https://ui.adsabs.harvard.edu/abs/2020zndo...4406806V}
}

@INPROCEEDINGS{CARTA2022,
       author = {{Ott}, Juergen and {Raba}, Ryan and {Hibbard}, John},
        title = "{CARTA: Cube Analysis and Rendering Tool for Astronomy 2.0 {\textrightarrow} 3.0}",
    booktitle = {American Astronomical Society Meeting \#240},
         year = 2022,
       series = {American Astronomical Society Meeting Abstracts},
       volume = {240},
        month = jun,
          eid = {215.05},
        pages = {215.05},
       adsurl = {https://ui.adsabs.harvard.edu/abs/2022AAS...24021505O}
}

@misc{APLpy2012,
       author = {{Robitaille}, Thomas and {Bressert}, Eli},
        title = "{APLpy: Astronomical Plotting Library in Python}",
 howpublished = {Astrophysics Source Code Library, record ascl:1208.017},
         year = 2012,
        month = aug,
          eid = {ascl:1208.017},
archivePrefix = {ascl},
       eprint = {1208.017},
       adsurl = {https://ui.adsabs.harvard.edu/abs/2012ascl.soft08017R}
}

@misc{APLpy2019,
       author = {{Robitaille}, Thomas},
        title = "{APLpy v2.0: The Astronomical Plotting Library in Python}",
         year = 2019,
        month = feb,
          eid = {10.5281/zenodo.2567476},
          doi = {10.5281/zenodo.2567476},
      version = {2.0},
    publisher = {Zenodo},
       adsurl = {https://ui.adsabs.harvard.edu/abs/2019zndo...2567476R}
}

@ARTICLE{NumPy2020,
       author = {{Harris}, Charles R. and {Millman}, K. Jarrod and {van der Walt}, St{\'e}fan J. and {Gommers}, Ralf and {Virtanen}, Pauli and {Cournapeau}, David and {Wieser}, Eric and {Taylor}, Julian and {Berg}, Sebastian and {Smith}, Nathaniel J. and {Kern}, Robert and {Picus}, Matti and {Hoyer}, Stephan and {van Kerkwijk}, Marten H. and {Brett}, Matthew and {Haldane}, Allan and {del R{\'\i}o}, Jaime Fern{\'a}ndez and {Wiebe}, Mark and {Peterson}, Pearu and {G{\'e}rard-Marchant}, Pierre and {Sheppard}, Kevin and {Reddy}, Tyler and {Weckesser}, Warren and {Abbasi}, Hameer and {Gohlke}, Christoph and {Oliphant}, Travis E.},
        title = "{Array programming with NumPy}",
      journal = {\nat},
         year = 2020,
        month = sep,
       volume = {585},
       number = {7825},
        pages = {357-362},
          doi = {10.1038/s41586-020-2649-2},
archivePrefix = {arXiv},
       eprint = {2006.10256},
 primaryClass = {cs.MS},
       adsurl = {https://ui.adsabs.harvard.edu/abs/2020Natur.585..357H}
}

@misc{CARTA2021,
       author = {{Comrie}, Angus and {Wang}, Kuo-Song and {Hsu}, Shou-Chieh and {Moraghan}, Anthony and {Harris}, Pamela and {Pang}, Qi and {Pi{\'n}ska}, Adrianna and {Chiang}, Cheng-Chin and {Chang}, Tien-Hao and {Hwang}, Yu-Hsuan and {Jan}, Hengtai and {Lin}, Ming-Yi and {Simmonds}, Rob},
        title = "{CARTA: The Cube Analysis and Rendering Tool for Astronomy}",
         year = 2021,
        month = jun,
          eid = {10.5281/zenodo.3377984},
          doi = {10.5281/zenodo.3377984},
      version = {2.0.0},
    publisher = {Zenodo},
       adsurl = {https://ui.adsabs.harvard.edu/abs/2021zndo...3377984C}
}

@ARTICLE{Patil2010,
       author = {{Patil}, Anand and {Huard}, David and {Fonnesbeck}, Christopher J.},
        title = "{PyMC: Bayesian Stochastic Modelling in Python}",
      journal = {Journal of Statistical Software},
         year = 2010,
        month = jul,
       volume = {35},
        pages = {1},
          doi = {10.18637/jss.v035.i04},
       adsurl = {https://ui.adsabs.harvard.edu/abs/2010JSS....35....1P}
}

@ARTICLE{Molinari2025,
       author = {{Molinari}, S. and {Schilke}, P. and {Battersby}, C. and {Ho}, P.~T.~P. and {S{\'a}nchez-Monge}, {\'A}. and {Traficante}, A. and {Jones}, B. and {Beltr{\'a}n}, M.~T. and {Beuther}, H. and {Fuller}, G.~A. and {Zhang}, Q. and {Klessen}, R.~S. and {Walch}, S. and {Tang}, Y.-W. and {Benedettini}, M. and {Elia}, D. and {Coletta}, A. and {Mininni}, C. and {Schisano}, E. and {Avison}, A. and {Law}, C.~Y. and {Nucara}, A. and {Soler}, J.~D. and {Stroud}, G. and {Wallace}, J. and {Wells}, M.~R.~A. and {Ahmadi}, A. and {Brogan}, C.~L. and {Hunter}, T.~R. and {Liu}, S.-Y. and {Pezzuto}, S. and {Su}, Y.-N. and {Zimmermann}, B. and {Zhang}, T. and {Wyrowski}, F. and {De Angelis}, F. and {Liu}, S. and {Clarke}, S.~D. and {Fontani}, F. and {Klaassen}, P.~D. and {Koch}, P. and {Johnston}, K.~G. and {Lebreuilly}, U. and {Liu}, T. and {Lumsden}, S.~L. and {Moeller}, T. and {Moscadelli}, L. and {Kuiper}, R. and {Lis}, D. and {Peretto}, N. and {Pfalzner}, S. and {Rigby}, A.~J. and {Sanhueza}, P. and {Rygl}, K.~L.~J. and {van der Tak}, F. and {Zinnecker}, H. and {Amaral}, F. and {Bally}, J. and {Bronfman}, L. and {Cesaroni}, R. and {Goh}, K. and {Hoare}, M.~G. and {Hatchfield}, P. and {Hennebelle}, P. and {Henning}, T. and {Kim}, K.-T. and {Kim}, W.-J. and {Maud}, L. and {Merello}, M. and {Nakamura}, F. and {Plume}, R. and {Qin}, S.-L. and {Svoboda}, B. and {Testi}, L. and {Veena}, V.~S. and {Walker}, D.},
        title = "{ALMAGAL: I. The ALMA evolutionary study of high-mass protocluster formation in the Galaxy: Presentation of the survey and early results}",
      journal = {\aap},
         year = 2025,
        month = apr,
       volume = {696},
          eid = {A149},
        pages = {A149},
          doi = {10.1051/0004-6361/202452702},
archivePrefix = {arXiv},
       eprint = {2503.05555},
 primaryClass = {astro-ph.GA},
       adsurl = {https://ui.adsabs.harvard.edu/abs/2025A&A...696A.149M}
}

@ARTICLE{Beuther2025,
       author = {{Beuther}, H. and {Kuiper}, R. and {Tafalla}, M.},
        title = "{Star Formation from Low to High Mass: A Comparative View}",
      journal = {\araa},
         year = 2025,
        month = aug,
       volume = {63},
       number = {1},
        pages = {1-44},
          doi = {10.1146/annurev-astro-013125-122023},
archivePrefix = {arXiv},
       eprint = {2501.16866},
 primaryClass = {astro-ph.GA},
       adsurl = {https://ui.adsabs.harvard.edu/abs/2025ARA&A..63....1B}
}

@INPROCEEDINGS{Tan2014,
       author = {{Tan}, J.~C. and {Beltr{\'a}n}, M.~T. and {Caselli}, P. and {Fontani}, F. and {Fuente}, A. and {Krumholz}, M.~R. and {McKee}, C.~F. and {Stolte}, A.},
        title = "{Massive Star Formation}",
    booktitle = {Protostars and Planets VI},
         year = 2014,
       editor = {{Beuther}, Henrik and {Klessen}, Ralf S. and {Dullemond}, Cornelis P. and {Henning}, Thomas},
        month = jan,
        pages = {149-172},
          doi = {10.2458/azu_uapress_9780816531240-ch007},
archivePrefix = {arXiv},
       eprint = {1402.0919},
 primaryClass = {astro-ph.GA},
       adsurl = {https://ui.adsabs.harvard.edu/abs/2014prpl.conf..149T}
}

@ARTICLE{Hatchfield2020,
       author = {{Hatchfield}, H. Perry and {Battersby}, Cara and {Keto}, Eric and {Walker}, Daniel and {Barnes}, Ashley and {Callanan}, Daniel and {Ginsburg}, Adam and {Henshaw}, Jonathan D. and {Kauffmann}, Jens and {Kruijssen}, J.~M. Diederik and {Longmore}, Steve N. and {Lu}, Xing and {Mills}, Elisabeth A.~C. and {Pillai}, Thushara and {Zhang}, Qizhou and {Bally}, John and {Butterfield}, Natalie and {Contreras}, Yanett A. and {Ho}, Luis C. and {Ott}, J{\"u}rgen and {Patel}, Nimesh and {Tolls}, Volker},
        title = "{CMZoom. II. Catalog of Compact Submillimeter Dust Continuum Sources in the Milky Way's Central Molecular Zone}",
      journal = {\apjs},
         year = 2020,
        month = nov,
       volume = {251},
       number = {1},
          eid = {14},
        pages = {14},
          doi = {10.3847/1538-4365/abb610},
archivePrefix = {arXiv},
       eprint = {2009.05052},
 primaryClass = {astro-ph.GA},
       adsurl = {https://ui.adsabs.harvard.edu/abs/2020ApJS..251...14H}
}

@ARTICLE{Rigby2024,
       author = {{Rigby}, Andrew J. and {Peretto}, Nicolas and {Anderson}, Michael and {Ragan}, Sarah E. and {Priestley}, Felix D. and {Fuller}, Gary A. and {Thompson}, Mark A. and {Traficante}, Alessio and {Watkins}, Elizabeth J. and {Williams}, Gwenllian M.},
        title = "{The dynamic centres of infrared-dark clouds and the formation of cores}",
      journal = {\mnras},
         year = 2024,
        month = feb,
       volume = {528},
       number = {2},
        pages = {1172-1197},
          doi = {10.1093/mnras/stae030},
archivePrefix = {arXiv},
       eprint = {2401.04238},
 primaryClass = {astro-ph.GA},
       adsurl = {https://ui.adsabs.harvard.edu/abs/2024MNRAS.528.1172R}
}

@ARTICLE{Wallace2026,
       author = {{Wallace}, Jennifer and {Kolz}, Taevis and {Battersby}, Cara and {Kuznetsova}, Aleksandra and {S{\'a}nchez-Monge}, {\'A}lvaro and {Schisano}, Eugenio and {Coletta}, Alessandro and {Zhang}, Qizhou and {Molinari}, Sergio and {Schilke}, Peter and {Ho}, Paul T.~P. and {Kuiper}, Rolf and {Zhang}, Tianwei and {M{\"o}ller}, Thomas and {Klessen}, Ralf S. and {Beltr{\'a}n}, Maria T. and {van der Tak}, Floris and {Pezzuto}, Stefania and {Beuther}, Henrik and {Traficante}, Alessio and {Elia}, Davide and {Bronfman}, Leonardo and {Klaassen}, Pamela and {Lis}, Dariusz C. and {Moscadelli}, Luca and {Rygl}, Kazi and {Benedettini}, Milena and {Law}, Chi Yan and {Allande}, Jofre and {Nucara}, Alice and {Koch}, Patrick M. and {Kim}, Won-ju and {Sanhueza}, Patricio and {Fuller}, Gary and {Stroud}, Georgie and {Jones}, Beth and {Brogan}, Crystal and {Hunter}, Todd and {Ahmadi}, Aida and {Avison}, Adam and {Johnston}, Katharine and {Liu}, Sheng-Yuan and {Mininni}, Chiara and {Su}, Yu-Nung and {Zinnecker}, Hans},
        title = "{ALMAGAL. VII. Cataloging Hierarchical Continuum Structure from Cores to Clumps across the Galactic Disk}",
      journal = {\apj},
         year = 2026,
        month = feb,
       volume = {998},
       number = {2},
          eid = {302},
        pages = {302},
          doi = {10.3847/1538-4357/ae2fec},
archivePrefix = {arXiv},
       eprint = {2510.12892},
 primaryClass = {astro-ph.GA},
       adsurl = {https://ui.adsabs.harvard.edu/abs/2026ApJ...998..302W}
}

@ARTICLE{McKee2007,
       author = {{McKee}, Christopher F. and {Ostriker}, Eve C.},
        title = "{Theory of Star Formation}",
      journal = {\araa},
         year = 2007,
        month = sep,
       volume = {45},
       number = {1},
        pages = {565-687},
          doi = {10.1146/annurev.astro.45.051806.110602},
archivePrefix = {arXiv},
       eprint = {0707.3514},
 primaryClass = {astro-ph},
       adsurl = {https://ui.adsabs.harvard.edu/abs/2007ARA&A..45..565M}
}

@ARTICLE{Krumholz2014,
       author = {{Krumholz}, Mark R.},
        title = "{The big problems in star formation: The star formation rate, stellar clustering, and the initial mass function}",
      journal = {\physrep},
         year = 2014,
        month = jun,
       volume = {539},
        pages = {49-134},
          doi = {10.1016/j.physrep.2014.02.001},
archivePrefix = {arXiv},
       eprint = {1402.0867},
 primaryClass = {astro-ph.GA},
       adsurl = {https://ui.adsabs.harvard.edu/abs/2014PhR...539...49K}
}

@ARTICLE{Robitaille2006,
       author = {{Robitaille}, Thomas P. and {Whitney}, Barbara A. and {Indebetouw}, Remy and {Wood}, Kenneth and {Denzmore}, Pia},
        title = "{Interpreting Spectral Energy Distributions from Young Stellar Objects. I. A Grid of 200,000 YSO Model SEDs}",
      journal = {\apjs},
         year = 2006,
        month = dec,
       volume = {167},
       number = {2},
        pages = {256-285},
          doi = {10.1086/508424},
archivePrefix = {arXiv},
       eprint = {astro-ph/0608234},
 primaryClass = {astro-ph},
       adsurl = {https://ui.adsabs.harvard.edu/abs/2006ApJS..167..256R}
}

@INPROCEEDINGS{Alexov2005,
       author = {{Alexov}, A. and {Berriman}, G.~B. and {Chiu}, N.-M. and {Good}, J.~C. and {Jarrett}, T.~H. and {Kong}, M. and {Laity}, A.~C. and {Monkewitz}, S.~M. and {Tahir-Kheli}, N.~D. and {Norton}, S.~W. and {Zhang}, A.},
        title = "{The NASA/IPAC Infrared Science Archive (IRSA): The Demo}",
    booktitle = {Astronomical Data Analysis Software and Systems XIV},
         year = 2005,
       editor = {{Shopbell}, P. and {Britton}, M. and {Ebert}, R.},
       series = {Astronomical Society of the Pacific Conference Series},
       volume = {347},
        month = dec,
        pages = {7},
       adsurl = {https://ui.adsabs.harvard.edu/abs/2005ASPC..347....7A}
}

@INPROCEEDINGS{Teplitz2018,
       author = {{Teplitz}, Harry I. and {Howell}, Justin and {Desai}, Vandana and {IRSA Team}},
        title = "{The NASA/IPAC Infrared Science Archive}",
    booktitle = {The Cosmic Wheel and the Legacy of the AKARI Archive: From Galaxies and Stars to Planets and Life},
         year = 2018,
       editor = {{Ootsubo}, Takafumi and {Yamamura}, Issei and {Murata}, Kazumi and {Onaka}, Takashi},
        month = mar,
        pages = {25-28},
       adsurl = {https://ui.adsabs.harvard.edu/abs/2018cwla.conf...25T}
}

\begin{appendix}

\nolinenumbers

\section{Methods} \label{apx:IR_fluxes}

\subsection{Infrared images analysis} \label{sec:IR_def}

To analyze the distribution of the mid-IR emission, we first visually compared ALMA emission with MIPS 24\,$\mathrm{\mu m}$ images using the Cube Analysis and Rendering Tool for Astronomy software \citep[CARTA;][]{CARTA2021, CARTA2022}.
Among the sources, G014.49 is the only one without detectable 24\,$\mathrm{\mu m}$ emission.
We divided the sources into two categories: those with detectable 24\,$\mathrm{\mu m}$ emission (G030.89, G014.19, G008.68, G023.21, G335.78, G019.88), and those with saturated emission (G337.92, G305.21, G301.14, G337.40, G343.13).

We used the Vizier catalog \citep{Gutermuth2015} to extract MIPSGAL sources within a radius of $\sim$6$\arcsec$ from the ATLASGAL clump positions, obtaining their positions and fluxes.
Sources G014.19, G008.68, G023.21, and G335.78 were already cataloged.
While for two new 24\,$\mathrm{\mu m}$ detections (G030.89~MM1 and G335.78~MM2), we performed aperture photometry using the photutils Python package \citep{Bradley2022}, adopting the parameters\footnote{Aperture radius: 6.35$\arcsec$; background annulus: 7.62$\arcsec$--17.78$\arcsec$; aperture correction factor: 1.63.} recommended by \citet{Gutermuth2015}.

In detail, G030.89~MM1, previously classified as quiescent by \citet{Urquhart2022}, shows faint 24\,$\mathrm{\mu m}$ emission ($\sim$0.06\,$\mathrm{Jy}$), close to the 5$\sigma$ detection threshold of 1.7\,$\mathrm{Jy}$ \citep{Carey2007}, which likely explains why it was not detected in previous searches.
Associated with G335.78, a second compact mid-IR source G335.78~MM2 ($\sim$0.1\,$\mathrm{Jy}$), is detected 12$\arcsec$ from the brighter G335.78~MM1 (the cataloged MIPSGAL source).
We classified G030.89~MM1 and G335.78~MM2 as protostellar according to the criteria of \citet{Urquhart2022}.

Although the angular resolutions of ALMA and MIPSGAL are different, the ALMA emission generally matches the cataloged MIPSGAL centroids within the MIPS beam size ($\sim$$6\arcsec$). 
However, in the case of G019.88, a slight shift is observed between the ALMA cores peaks and the MIPSGAL centroid.
This source shows an asymmetric elliptical distribution at 24\,$\mathrm{\mu m}$, with the emission peak aligning with the ALMA data, while the MIPSGAL centroid is offset, likely due to the 2D Gaussian method for source identification and systematic astrometric corrections \citep{Gutermuth2015}.

For sources with severe saturation, where MIPSGAL images saturate at $\sim$1700\,$\mathrm{MJy\,sr^{-1}}$ \citep{Carey2009} (typically in bright YSOs or $\mathrm{H\,II}$ regions), we examined MIPS images and supplemented with WISE and GLIMPSE images.
We estimated conservative lower limits for their fluxes by replacing the saturated (NaN) pixels in MIPS images with the maximum intensity from nearby non-saturated regions and performing aperture photometry on the corrected maps, using the same parameters described for unsaturated sources. 
We applied this method to G305.21, G301.14, G337.40, and G343.13, where saturation affected only a limited region.
However, it was not applicable to G337.92, where most of the image was saturated.

We also examined the WISE images and catalog data \citep{Cutri2012} within a radius of $\sim$6$\arcsec$ from the ATLASGAL clump positions, and identified three sources: G301.14, G337.40, and G343.13. 
For the remaining saturated sources (G337.92 and G305.21), we could not determine fluxes due to diffuse and extended emission, and no aperture photometry was performed.
Similarly, we inspected IRAC images and the GLIMPSE catalog \citep{GLIMPSE2009} within a radius of $\sim$2$\arcsec$ from the ATLASGAL clump positions, identifying two sources: G301.14 and G337.40. 
Fluxes at 8\,$\mathrm{\mu m}$ were available only for G337.40 in the catalog.
For G301.14, we performed aperture photometry following the IRAC Instrument Handbook\footnote{Aperture radius: 2.4$\arcsec$; background annulus: 2.4$\arcsec$–7.2$\arcsec$; aperture correction factor: 1.568.}.
A summary table of all IR fluxes used in this work is provided in Table\,\ref{tab:IR_flx}.

For saturated sources with available 22\,$\mathrm{\mu m}$ and 8\,$\mathrm{\mu m}$ fluxes, we estimated the 24\,$\mathrm{\mu m}$ fluxes using scaling factors derived from JWST spectroscopic observations of massive protoclusters in our Galaxy \citep{Beuther2023} and in the Large Magellanic Cloud \citep{Nayak2024}, spanning the spectral range of $\sim$5--28\,$\mathrm{\mu m}$. 
Specifically, the ratio of 24\,$\mathrm{\mu m}$/22\,$\mathrm{\mu m}$ fluxes ranges from $\sim$0.9 to 1.4, with a mean value of 1.24. 
The ratio of 24\,$\mathrm{\mu m}$/8\,$\mathrm{\mu m}$ fluxes ranges from $\sim$28 to $\sim$220, with a mean value of 100.
In Fig.\,\ref{fig:LL_vts}, flux estimates derived from WISE are used as upper limits, while those from GLIMPSE are used only for visualization purposes.

\subsection{Lines and IR luminosities} \label{sec:lum_def}

This section describes the methodology adopted to compute the luminosities of the $\varv_\mathrm{t}$$\geq$1 methanol lines (list of transitions in Table\,\ref{tab:freq_blocks}) and the mid-IR continuum emission at 24\,$\mathrm{\mu m}$.

The total luminosity of each source for a given $\Delta E/k_\mathrm{B}$ is computed by summing the contributions from each 890\,$\mathrm{\mu m}$ continuum core, integrating over all $\varv_\mathrm{t}$$\geq$1 methanol lines within that specific $\Delta E/k_\mathrm{B}$ range.

\begin{equation}
L(\Delta E/k_\mathrm{B}) = 4\pi d^2 \sum_{i} \left( \int_{\Delta \nu_i} I_{\varv_\mathrm{t} \geq 1} \, d\nu \right) \quad [erg\,s^{-1}],
\end{equation}

\noindent where $d$ is the distance to the source (in $\mathrm{kpc}$), $\Delta \nu_i$ is the linewidth for the i-th methanol line (in $\mathrm{MHz}$), $I_\mathrm{\varv_\mathrm{t} \geq 1}$ is the intensity of the methanol line at a frequency $\nu$ (in $\mathrm{Jy}$), the sum is taken over all ALMA cores indexed by $i$.

For the mid-IR continuum luminosity at 24\,$\mathrm{\mu m}$, we used the following formula:

\begin{equation}
L_\mathrm{MIR} = 4\pi d^2 \, \nu f_\nu \quad [erg\,s^{-1}],
\end{equation}

\noindent where $\nu$ is the central frequency of the MIPS 24\,$\mathrm{\mu m}$ band (in $\mathrm{Hz}$) and $f_\nu$ is the flux density at 24\,$\mathrm{\mu m}$ (in $\mathrm{Jy}$) from the MIPSGAL catalog or derived from WISE or GLIMPSE (see details in Appendix\,\ref{sec:IR_def}).

For comparison, we converted the calculated luminosities into solar units ($L_\mathrm{\odot}$).

\twocolumn

\begin{table*}[ht!]
\section{Supplementary tables}
    \caption{List of torsionally excited $\mathrm{CH_3OH}$ transitions analyzed in this paper.}
    \label{tab:freq_blocks}
    \centering
    \begin{tabular}{llc}
    \hline
    Rest Frequency ($\mathrm{MHz}$) & $\mathrm{CH_3OH}$ transition & $E_\mathrm{u}/k_\mathrm{B}$ ($\mathrm{K}$)\\
    \hline
    337463.703 & $7_6$ -- $6_6$ $A^+$ $\varv_\mathrm{t} = 1$         & 533.0\\
    337463.703 & $7_6$ -- $6_6$ $A^-$ $\varv_\mathrm{t} = 1$         & 533.0\\
    337490.562 & $7_{-6}$ -- $6_{-6}$ $E$ $\varv_\mathrm{t} = 1$     & 558.2\\
    337519.138 & $7_{3}$ -- $6_{3}$ $E$ $\varv_\mathrm{t} = 1$       & 482.2\\
    337546.116 & $7_{5}$ -- $6_{5}$ $A^+$ $\varv_\mathrm{t} = 1$     & 485.3\\ 
    337546.116 & $7_{5}$ -- $6_{5}$ $A^-$ $\varv_\mathrm{t} = 1$     & 485.3\\
    337605.288 & $7_{-2}$ -- $6_{-2}$ $E$ $\varv_\mathrm{t} = 1$     & 429.4\\
    337610.661 & $7_{-3}$ -- $6_{-3}$ $E$ $\varv_\mathrm{t} = 1$     & 387.4\\
    337610.68 & $7_{6}$ -- $6_{6}$ $E$ $\varv_\mathrm{t} = 1$        & 657.1\\
    337625.753 & $7_{2}$ -- $6_{2}$ $A^+$ $\varv_\mathrm{t} = 1$     & 363.4\\
    337635.754 & $7_{2}$ -- $6_{2}$ $A^-$ $\varv_\mathrm{t} = 1$     & 363.4\\
    337642.478 & $7_{1}$ -- $6_{1}$ $E$ $\varv_\mathrm{t} = 1$       & 356.2\\
    337643.915 & $7_{0}$ -- $6_{0}$ $E$ $\varv_\mathrm{t} = 1$       & 365.4\\
    337646.042 & $7_{-4}$ -- $6_{-4}$ $E$ $\varv_\mathrm{t} = 1$     & 470.2\\
    337648.209 & $7_{-5}$ -- $6_{-5}$ $E$ $\varv_\mathrm{t} = 1$     & 610.9\\
    337655.199 & $7_{3}$ -- $6_{3}$ $A^+$ $\varv_\mathrm{t} = 1$     & 460.9\\
    337655.236 & $7_{3}$ -- $6_{3}$ $A^-$ $\varv_\mathrm{t} = 1$     & 460.9\\
    337671.238 & $7_{2}$ -- $6_{2}$ $E$ $\varv_\mathrm{t} = 1$       & 464.7\\
    337685.248 & $7_{5}$ -- $6_{5}$ $E$ $\varv_\mathrm{t} = 1$       & 493.9\\
    337685.614 & $7_{4}$ -- $6_{4}$ $A^{+-}$ $\varv_\mathrm{t} = 1$  & 545.8\\
    337707.568 & $7_{-1}$ -- $6_{-1}$ $E$ $\varv_\mathrm{t} = 1$     & 478.2\\
    337748.83 & $7_{0}$ -- $6_{0}$ $A$ $\varv_\mathrm{t} = 1$        & 488.4\\
    337969.438 & $7_{-1}$ -- $6_{-1}$ $A^-$ $\varv_\mathrm{t} = 1$   & 390.1\\
    336605.889 & $7_{1}$ -- $6_{1}$ $A^+$ $\varv_\mathrm{t} = 2$     & 747.4\\
    336970.183 & $7_{6}$ -- $6_{6}$ $A^{+-}$ $\varv_\mathrm{t} = 2$  & 1022.6\\
    337021.917 & $7_{-3}$ -- $6_{-3}$ $E$ $\varv_\mathrm{t} = 2$     & 979.7\\
    337029.573 & $7_{2}$ -- $6_{2}$ $E$ $\varv_\mathrm{t} = 2$       & 941.3\\
    337029.662 & $7_{-2}$ -- $6_{-2}$ $E$ $\varv_\mathrm{t} = 2$     & 941.3\\
    337098.918 & $7_{5}$ -- $6_{5}$ $A^+-$ $\varv_\mathrm{t} = 2$    & 935.1\\
    337113.868 & $7_{1}$ -- $6_{1}$ $E$ $\varv_\mathrm{t} = 2$       & 863.6\\
    337159.158 & $7_{6}$ -- $6_{6}$ $E$ $\varv_\mathrm{t} = 2$       & 762.8\\
    337175.097 & $7_{-4}$ -- $6_{-4}$ $E$ $\varv_\mathrm{t} = 2$     & 809.2\\
    337186.488 & $7_{0}$ -- $6_{0}$ $E$ $\varv_\mathrm{t} = 2$       & 798.9\\
    337252.172 & $7_{3}$ -- $6_{3}$ $A^-$ $\varv_\mathrm{t} = 2$     & 722.8\\
    337252.173 & $7_{3}$ -- $6_{3}$ $A^+$ $\varv_\mathrm{t} = 2$     & 722.8\\
    337273.561 & $7_{4}$ -- $6_{4}$ $A^+$ $\varv_\mathrm{t} = 2$     & 679.2\\
    337279.18  & $7_{-2}$ -- $6_{-2}$ $E$ $\varv_\mathrm{t} = 2$     & 709.6\\
    337284.32  & $7_{0}$ -- $6_{0}$ $A^+$ $\varv_\mathrm{t} = 2$     & 572.9\\
    337312.36  & $7_{-1}$ -- $6_{-1}$ $E$ $\varv_\mathrm{t} = 2$     & 596.7\\
\hline
\end{tabular}
\end{table*}

\begin{table*}[ht!]
    \caption{Infrared fluxes of the selected sources.} \label{tab:IR_flx}
    \centering
    \begin{tabular}{llccc}
    \hline\hline
    Name & Core   & Flux 24\,$\mathrm{\mu m}$ ($\mathrm{Jy}$) & Flux 22\,$\mathrm{\mu m}$ ($\mathrm{Jy}$) & Flux 8\,$\mathrm{\mu m}$ ($\mathrm{Jy}$)\\
    \hline
    G030.89                  & MM1              & 0.006\tablefootmark{b}    &                           & \\ 
    G014.19                  & MM1              & 0.188\tablefootmark{c}    & 0.174\tablefootmark{c}    & \\ 
    G008.68                  & MM1              & 0.703\tablefootmark{c}    & 0.5783\tablefootmark{c}   & \\
    G023.21                  & MM1              & 0.104\tablefootmark{c}    &                           & \\
    G335.78                  & MM1              & 1.91\tablefootmark{c}     & 1.4\tablefootmark{c}      & \\
    G335.78                  & MM2              & 0.097\tablefootmark{b}    &                           & \\
    G019.88                  & MM1 $+$ MM2 $+$ MM3  & 4.884\tablefootmark{c}    & 5.35\tablefootmark{c}     & 0.720\tablefootmark{c}\\
    G337.92\tablefootmark{a} & MM1 $+$ MM2        &                           &                           & \\
    G305.21\tablefootmark{a} & MM1              & $>$2.134\tablefootmark{d} &                           & \\
    G301.14\tablefootmark{a} & MM1 $+$ MM2        & $>$2.699\tablefootmark{d} & 301\tablefootmark{c}      & 0.107\tablefootmark{b}\\
    G337.40\tablefootmark{a} & MM1              & $>$4.932\tablefootmark{d} & 143\tablefootmark{c}      & 0.692\tablefootmark{c}\\
    G343.13\tablefootmark{a} & MM1 $+$ MM2        & $>$8.028\tablefootmark{d} & 25.7\tablefootmark{c}     & \\
    \hline
    \end{tabular}
    \tablefoot{The indices indicate: \tablefoottext{a}{saturated sources in MIPSGAL images;} \tablefoottext{b}{fluxes measured in this work via aperture photometry;} \tablefoottext{c}{fluxes taken directly from the MIPSGAL \citep[24\,$\mathrm{\mu m}$,][]{Gutermuth2015}, WISE \citep[22\,$\mathrm{\mu m}$,][]{Cutri2012} and GLIMPSE \citep[8\,$\mathrm{\mu m}$,][]{GLIMPSE2009} catalogs;} \tablefoottext{d}{conservative lower limits for saturated sources in the 24\,$\mathrm{\mu m}$ maps.}}
\end{table*}

\FloatBarrier
\twocolumn

\begin{figure*}[ht!]
\section{Supplementary figures} \label{apx:Figures}
\vspace{-2.5mm}
    \centering   
    \includegraphics[width=0.30\linewidth,keepaspectratio]{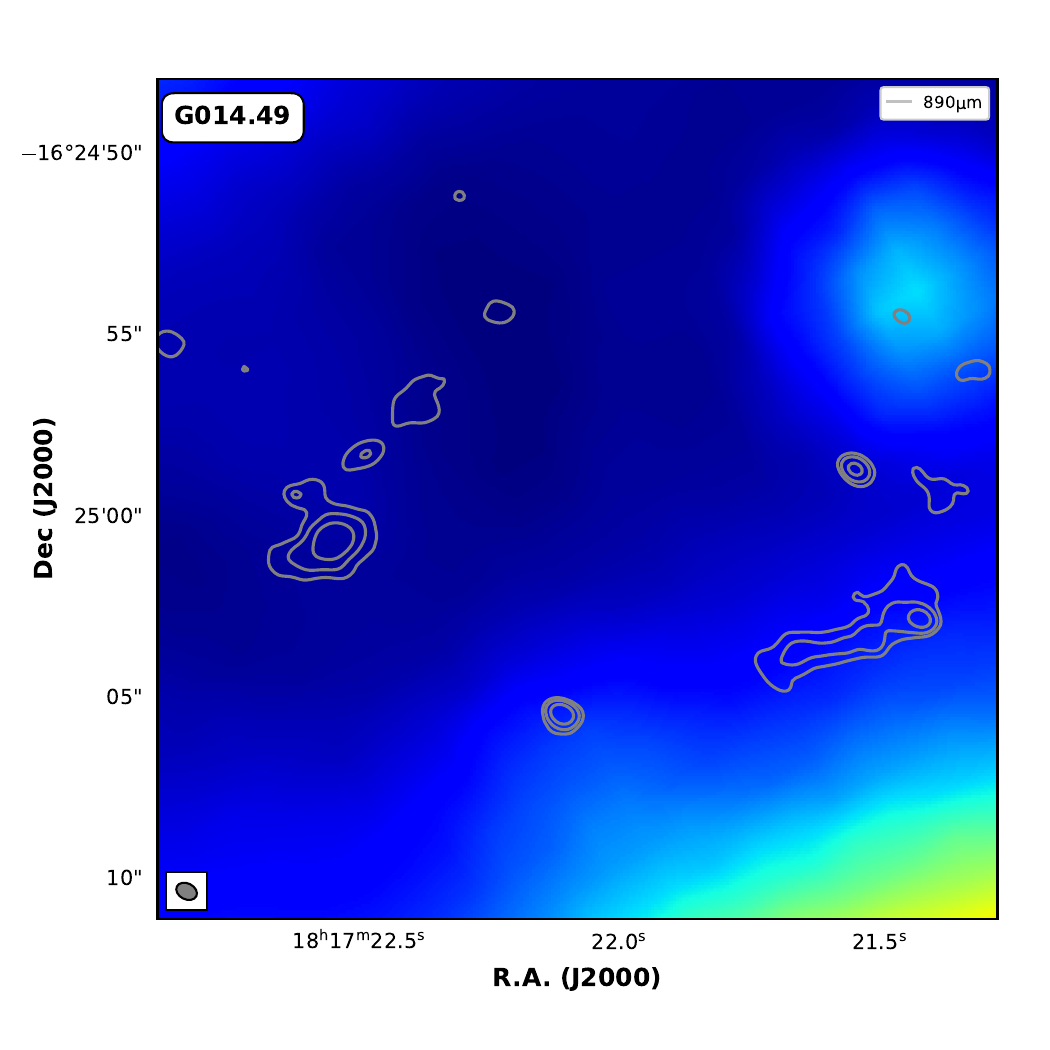}    
    \includegraphics[width=0.30\linewidth,keepaspectratio]{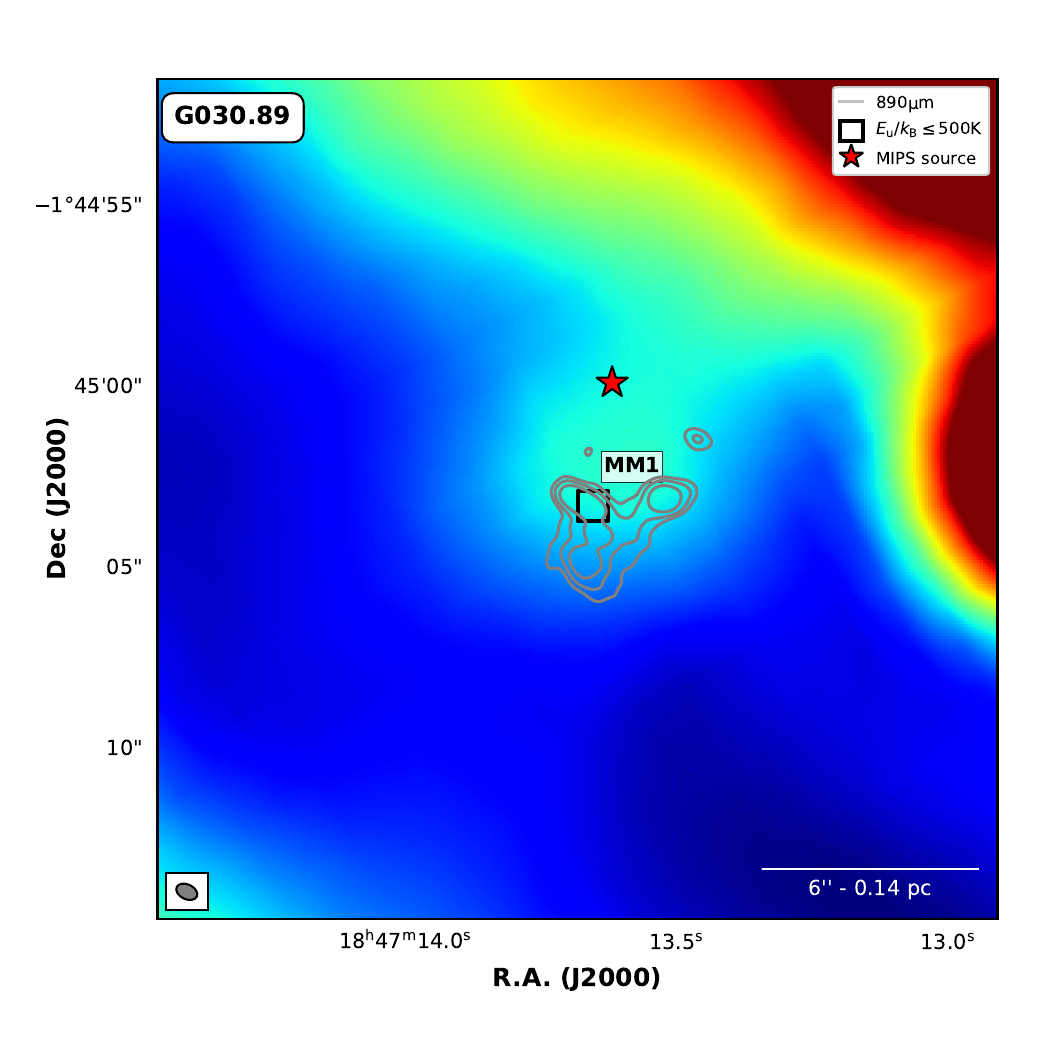}    
    \includegraphics[width=0.30\linewidth,keepaspectratio]{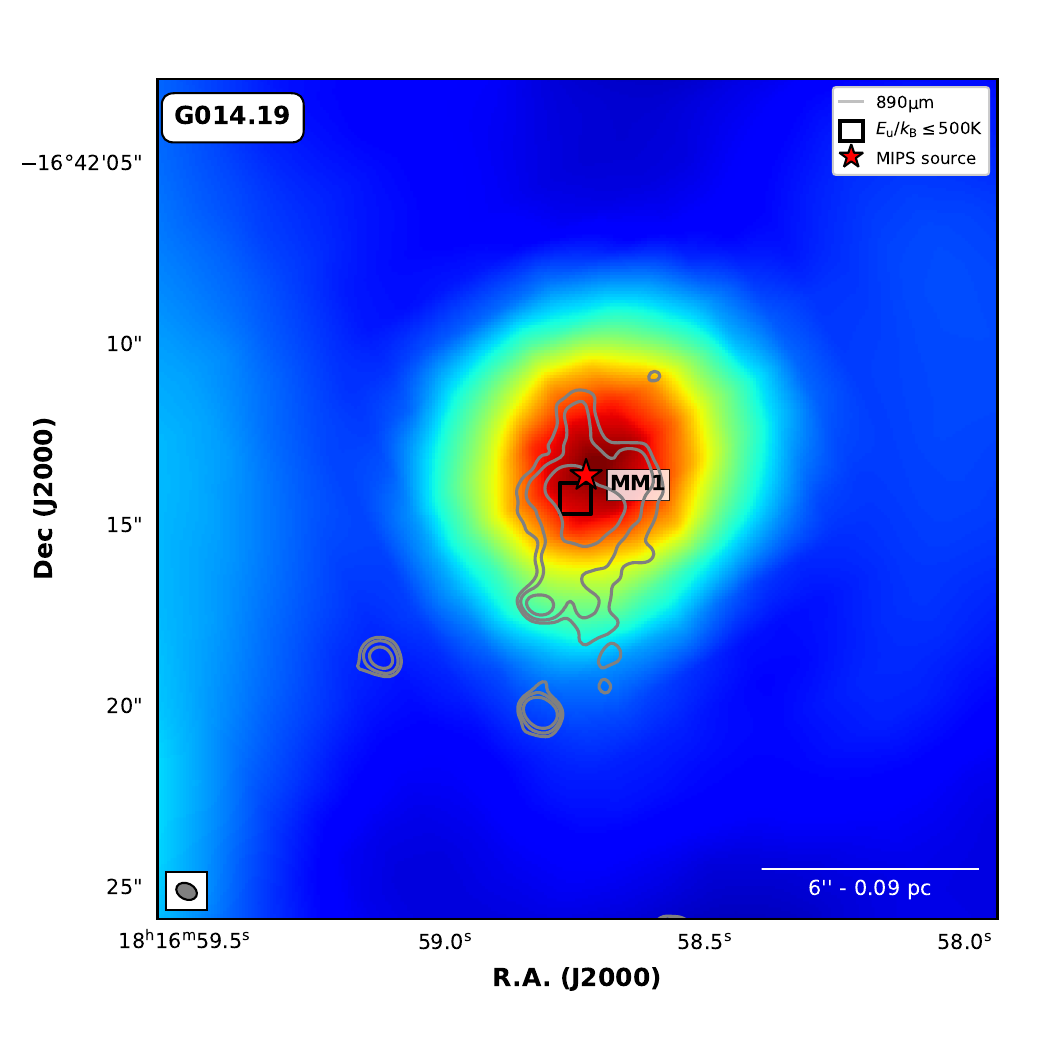}    
    \includegraphics[width=0.30\linewidth,keepaspectratio]{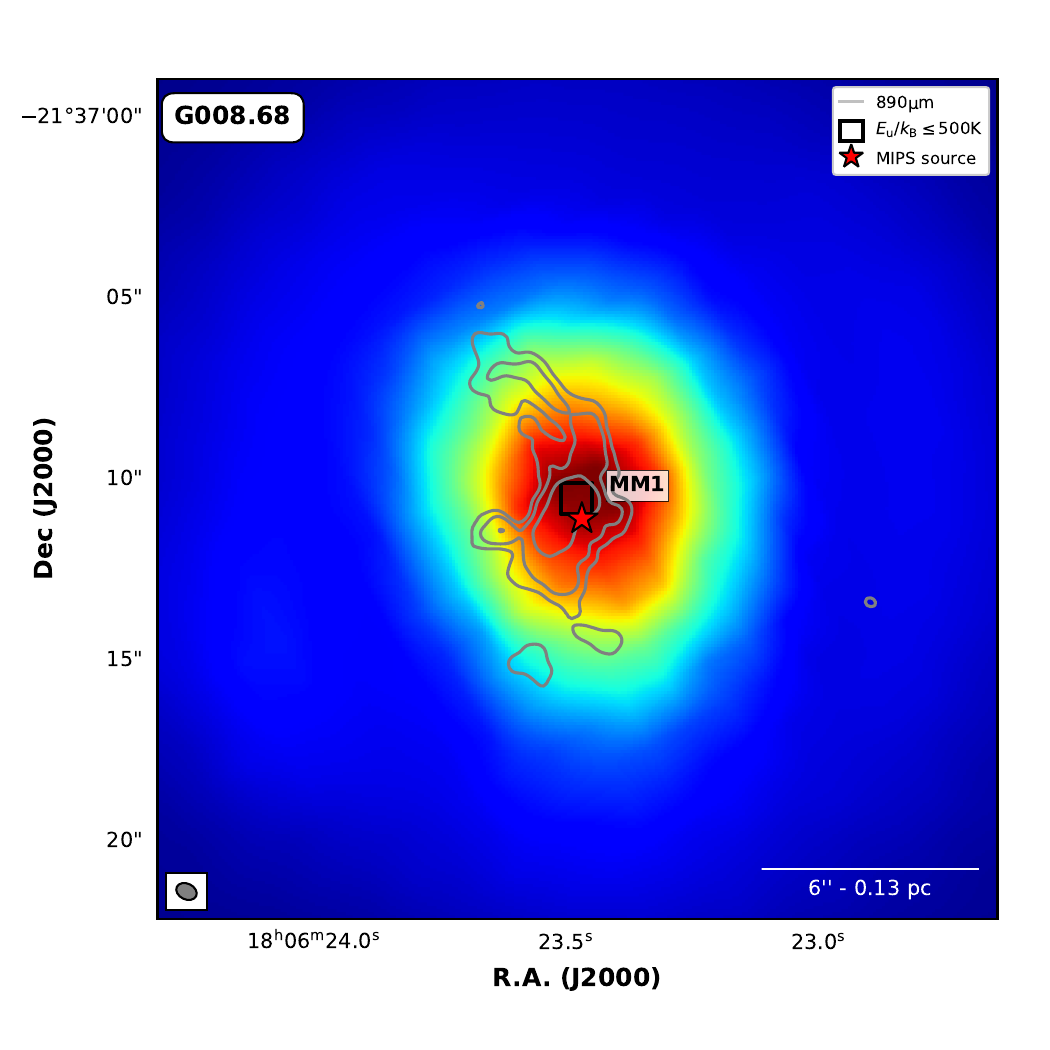}    
    \includegraphics[width=0.30\linewidth,keepaspectratio]{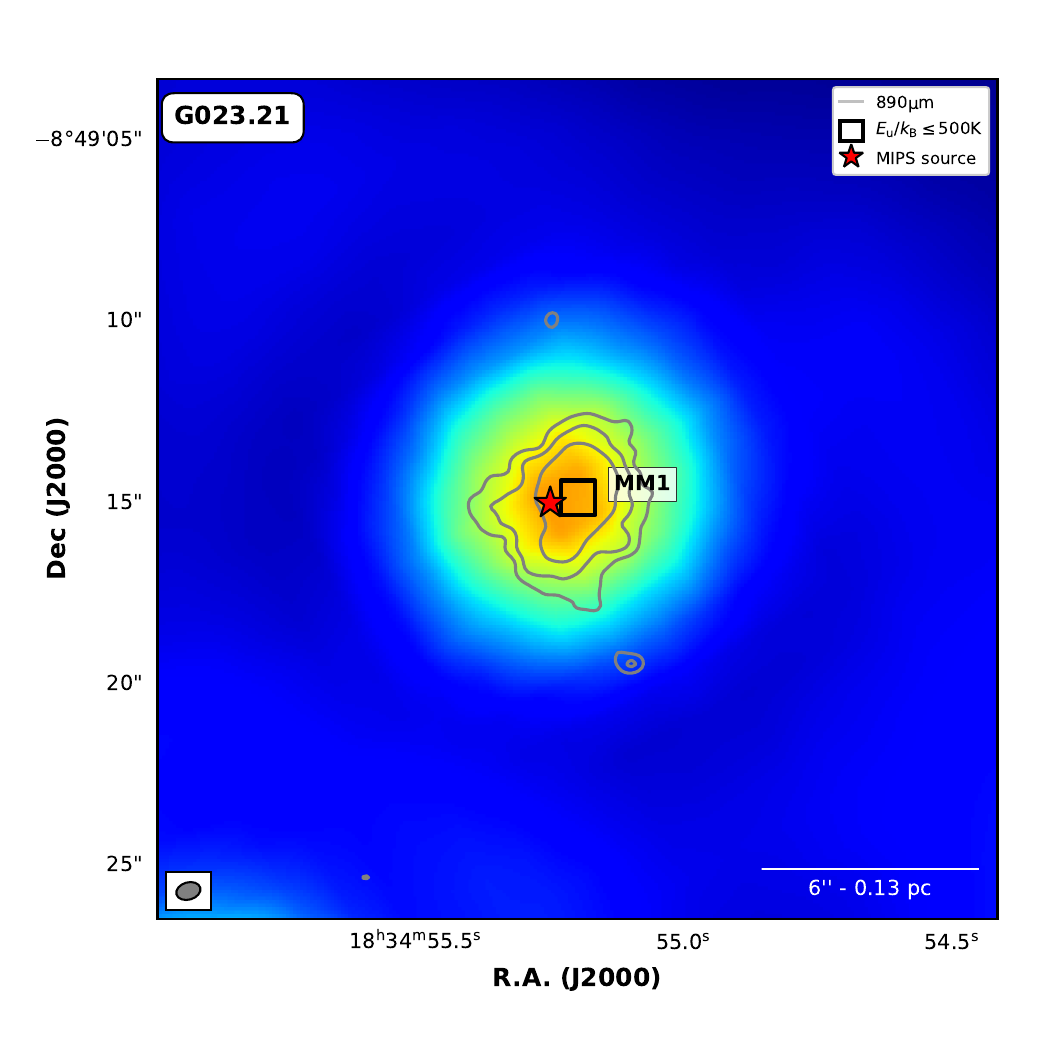}    
    \includegraphics[width=0.30\linewidth,keepaspectratio]{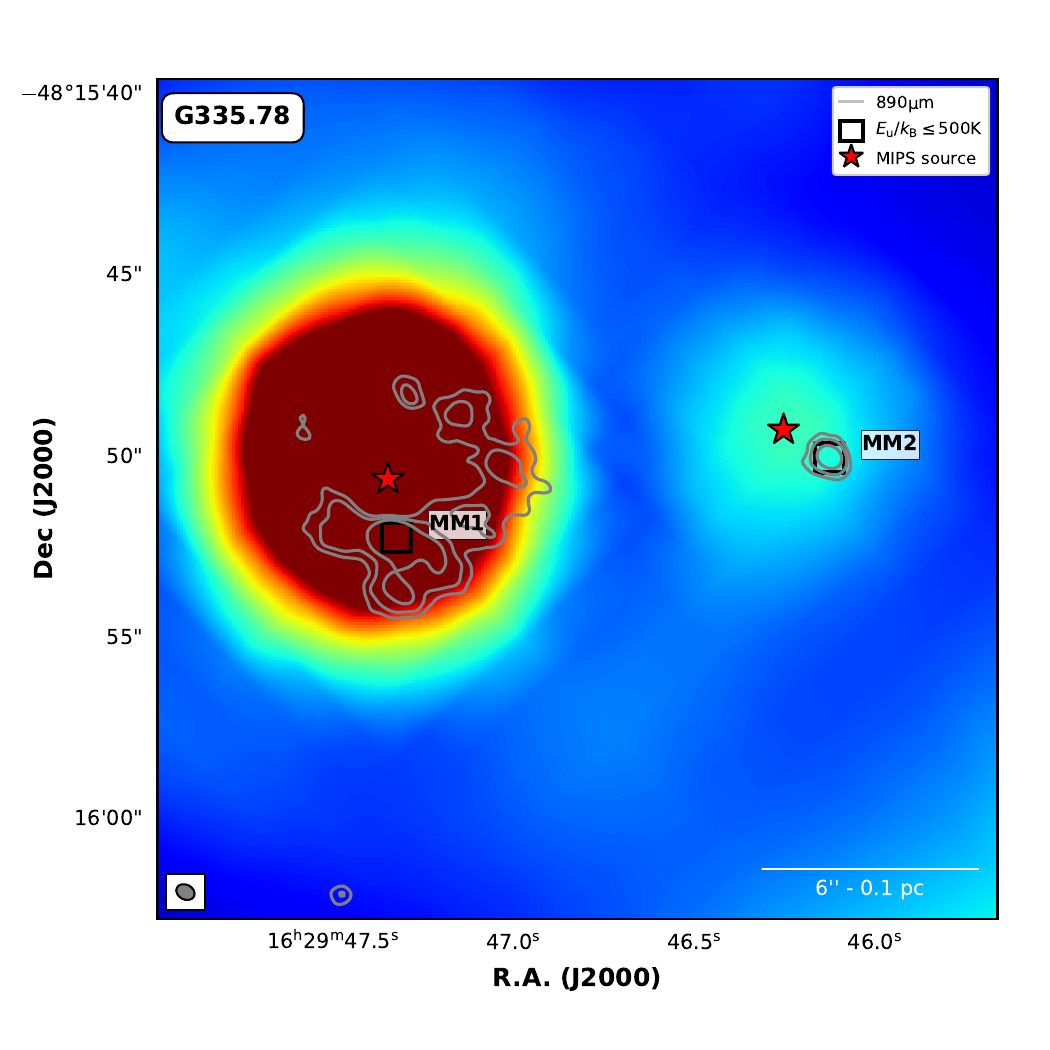} 
    
    \includegraphics[width=0.30\linewidth,keepaspectratio]{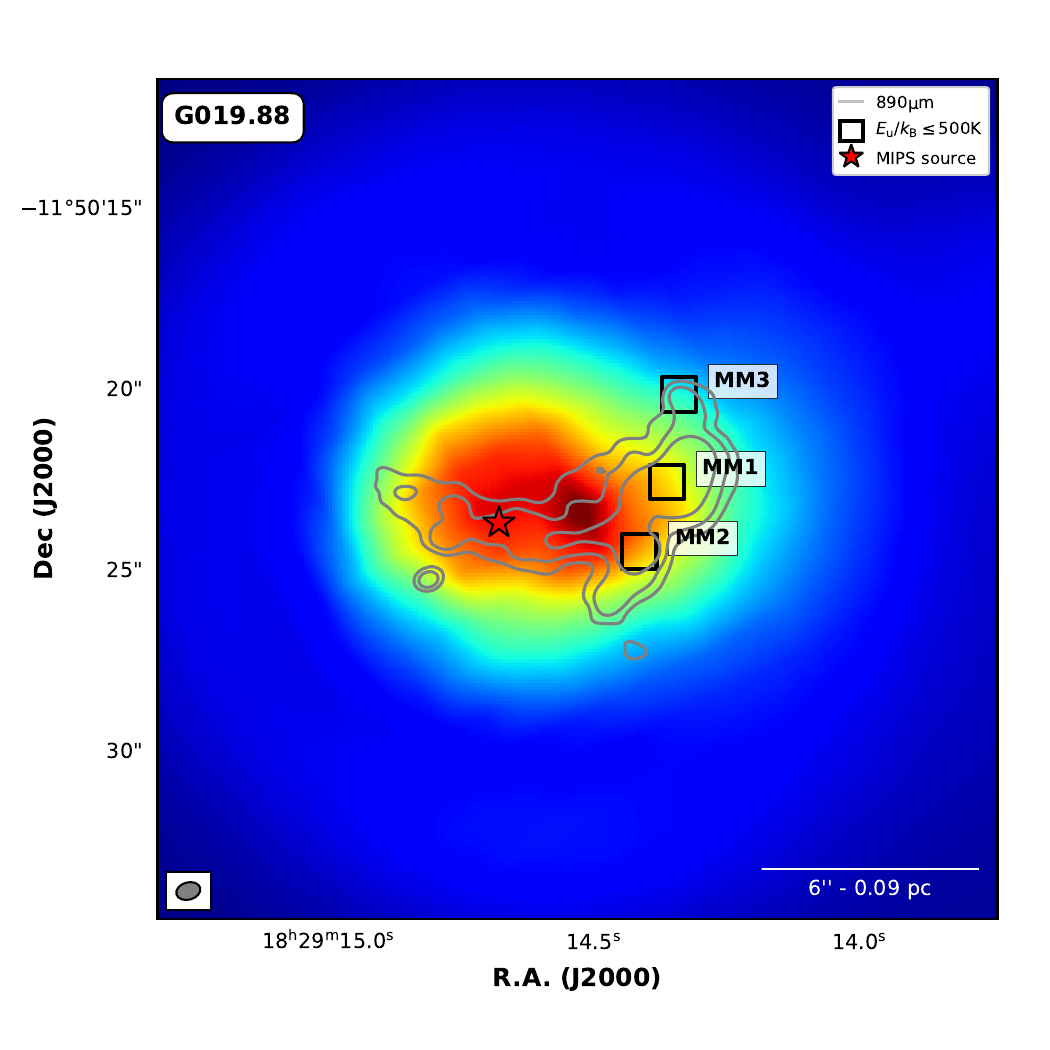}    
    \includegraphics[width=0.30\linewidth,keepaspectratio]{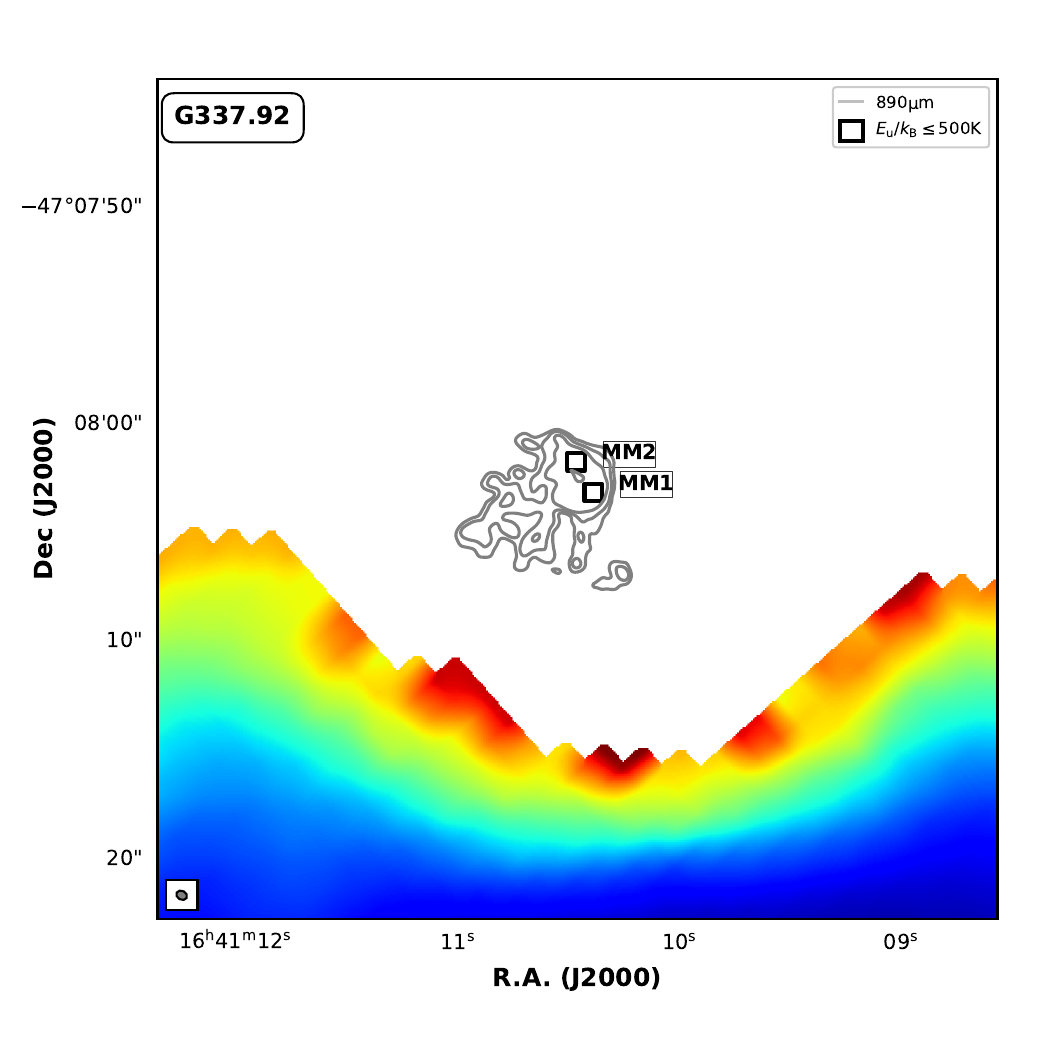}    
    \includegraphics[width=0.30\linewidth,keepaspectratio]{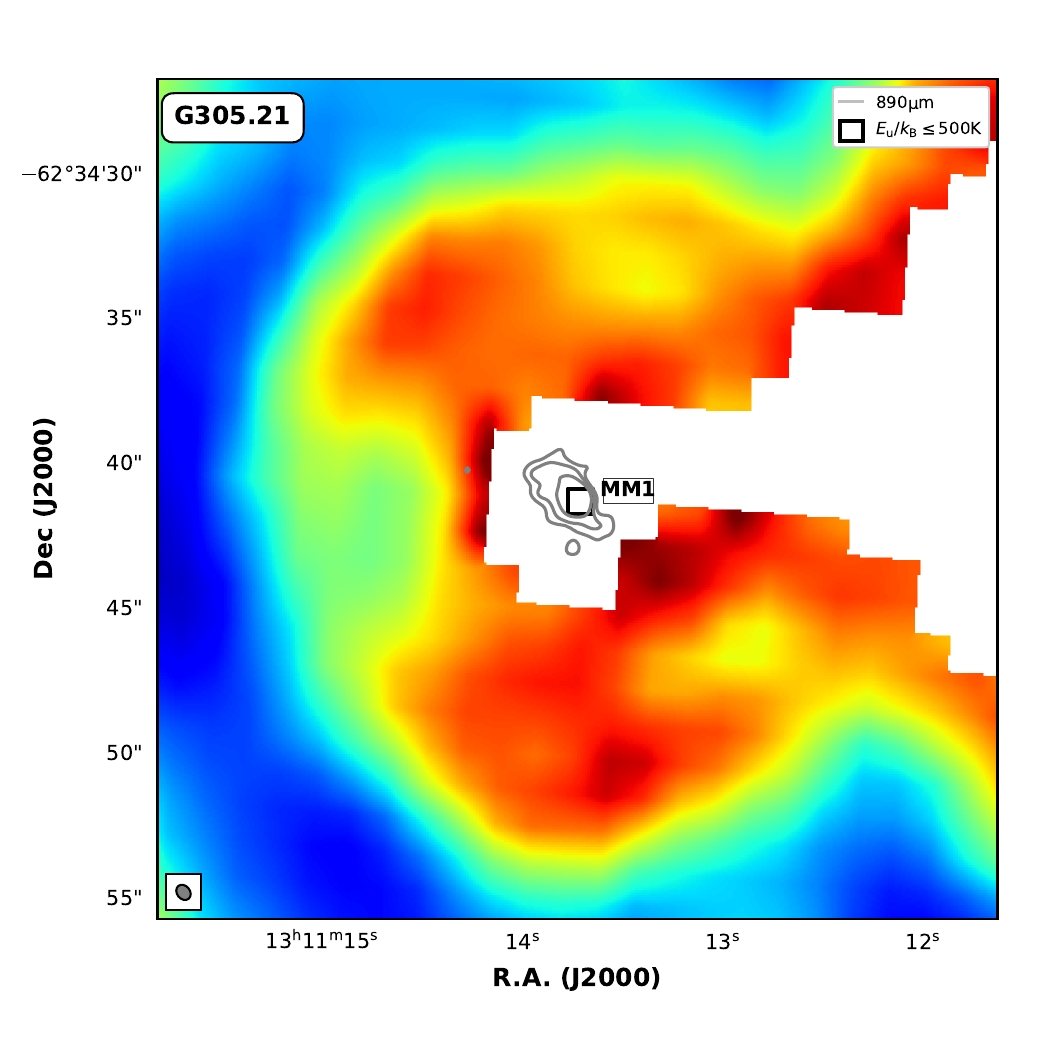}   
    \includegraphics[width=0.30\linewidth,keepaspectratio]{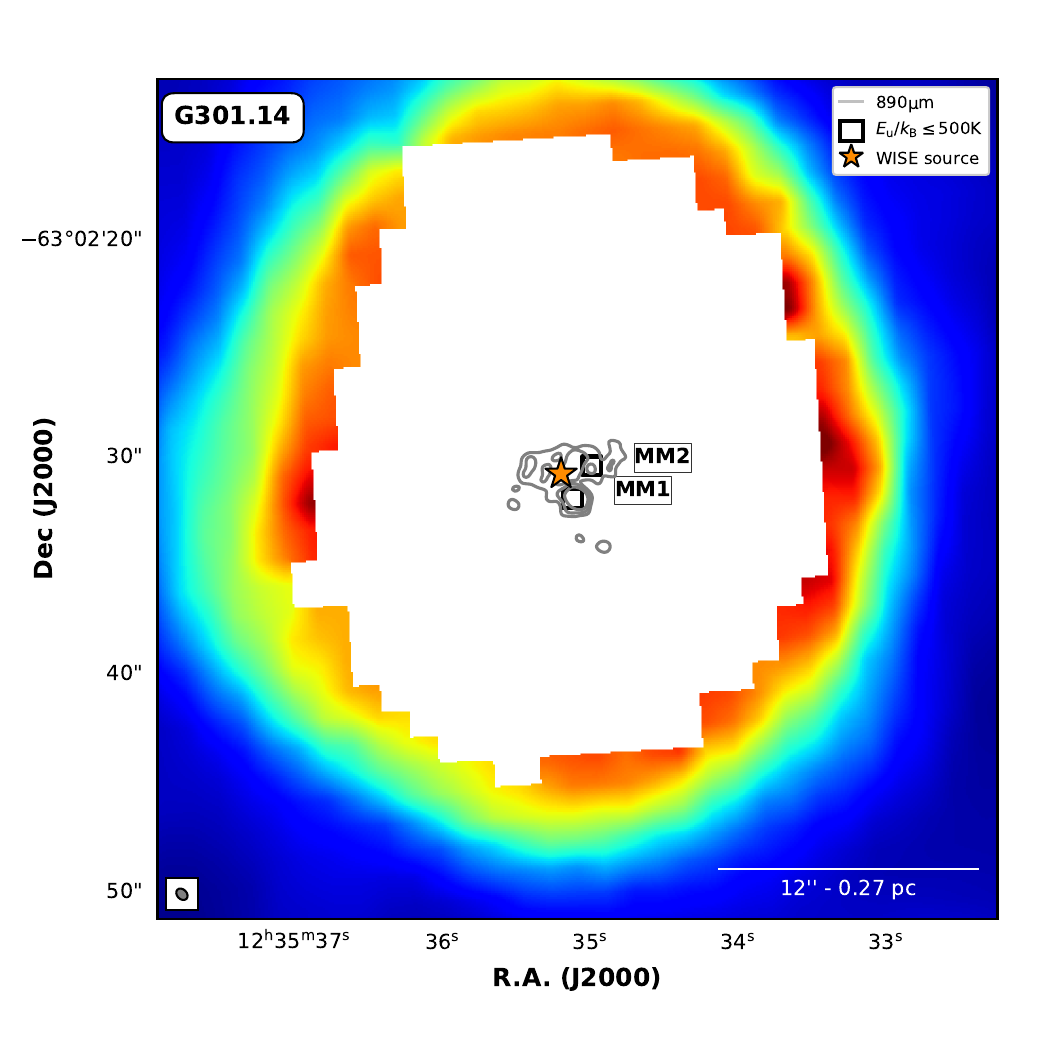}  
    \includegraphics[width=0.30\linewidth,keepaspectratio]{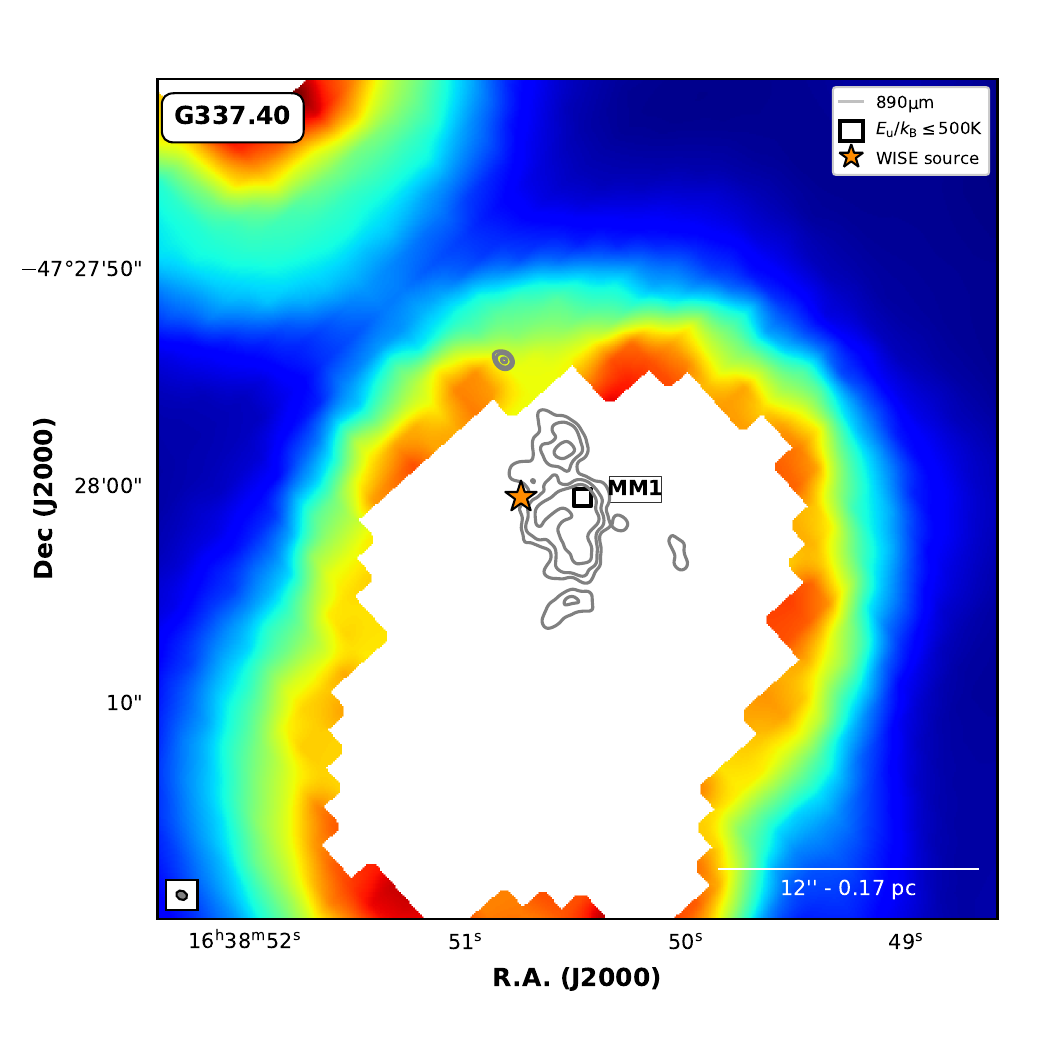}    
    \includegraphics[width=0.30\linewidth,keepaspectratio]{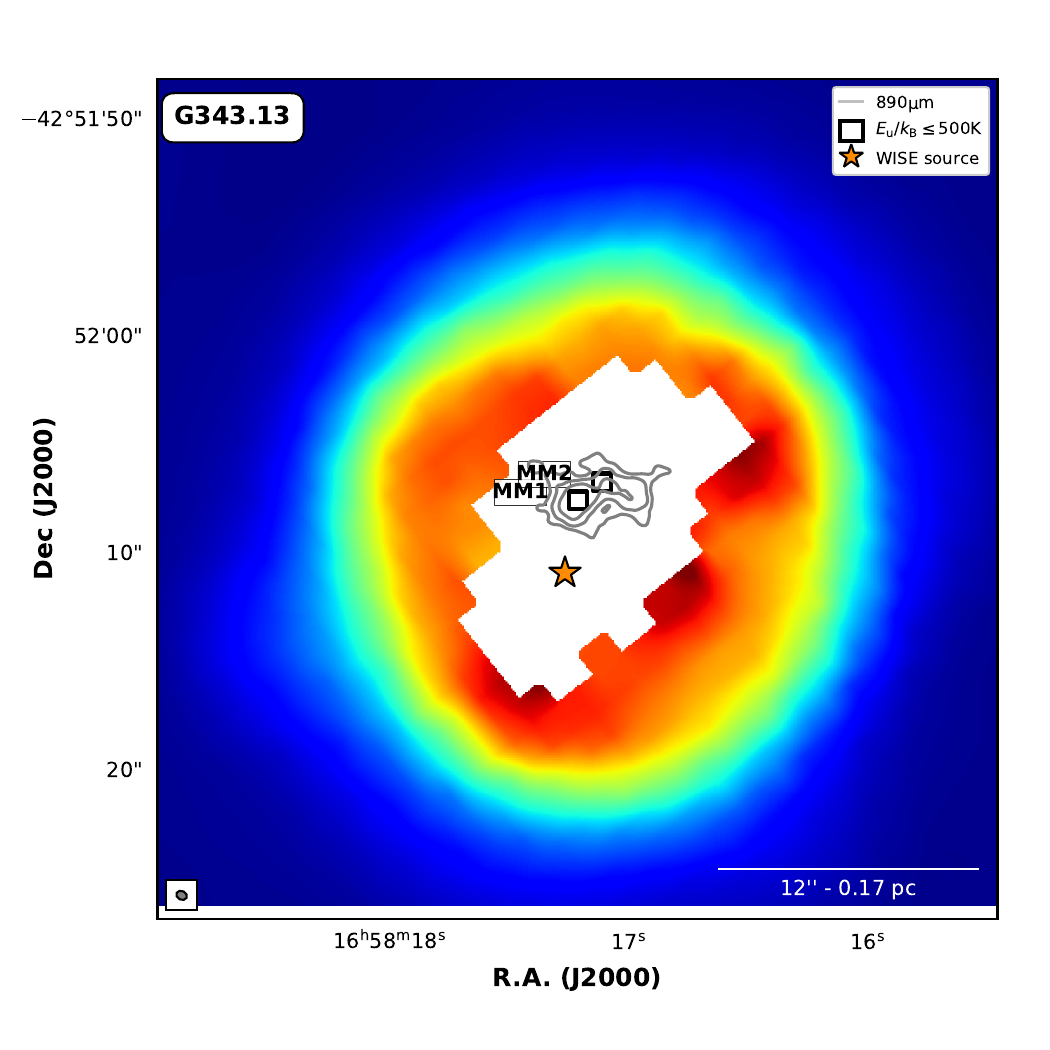}
    
    \caption{
    Panel showing the 24\,$\mathrm{\mu}m$ MIPS images (background) of all 12 sources in the sample.
    Silver contours show the continuum emission at [3, 5, 10]$\sigma$. 
    The black squares indicate the names and the positions of $\varv_\mathrm{t}$$\geq$$1$ cores with low-$\Delta E/k_\mathrm{B}$.
    For G014.49, the absence of lines contours and IR emission indicates a non-detection.
    The red star marks point sources in the MIPSGAL catalog \citep{Gutermuth2015}, with exceptions for G030.89 and secondary sources in G335.78, where MIPS centroids are determined in this work. 
    The orange stars indicate the WISE catalog centroid \citep{Cutri2012} for saturated MIPS sources: G337.92, G301.14, G337.40, and G343.13.
    The absence of stars (e.g., G305.21, G337.92) indicates complex, diffuse emission not in the catalogs.
    The ALMA beam size ($\sim$0.6$\arcsec$) is shown in the bottom-left corner, while the MIPSGAL ($\sim$6$\arcsec$) or the WISE ($\sim$12$\arcsec$) beam size is in the bottom-right.
    } 
    \label{fig:maps_ir}     
\end{figure*}

\end{appendix}

\end{document}